\documentclass{emulateapj}

\usepackage{float}

\usepackage{natbib}
\usepackage{color}
\newcommand{\comm}[1]{\textcolor{black}{{#1}}}
\shorttitle{Ages and heavy element abundances in the Sagittarius dwarf galaxy}
\shortauthors{C. J. Hansen et al.}

\slugcomment{Accepted for publication in Astrophysical Journal}

\begin{document}
\title{Ages and heavy element abundances from very metal-poor stars in the Sagittarius dwarf galaxy\altaffilmark{*}}

\author{
Camilla Juul Hansen\altaffilmark{1,2},  Mariam El-Souri\altaffilmark{1}, Lorenzo Monaco\altaffilmark{3}, Sandro Villanova\altaffilmark{4}, Piercarlo Bonifacio\altaffilmark{5}, Elisabetta Caffau\altaffilmark{5}, Luca Sbordone\altaffilmark{6}.
}
\altaffiltext{1}{University of Copenhagen, DARK Cosmology Centre, Juliane Maries Vej 30, 2100 Copenhagen Oe, Denmark}
\altaffiltext{2}{Max Planck Institute for Astronomy, Heidelberg, K\"onigstuhl 17, 69117 Heidelberg, Germany}
\altaffiltext{3}{Departamento de Ciencias Fisicas, Universidad Andres Bello, Fernandez Concha 700, Las Condes, Santiago, Chile}
\altaffiltext{4}{Departamento de Astronomia, Casilla 160, Universidad de Concepci{\'o}n, Chile\\}
\altaffiltext{5}{GEPI, Observatoire de Paris, PSL Research University, CNRS, Place Jules Janssen, 92195 Meudon, France}
\altaffiltext{6}{ESO - European Southern Observatory, Alonso de Cordova 3107, Vitacura, Santiago, Chile}
\altaffiltext{*}{Based on data obtained UVES/VLT ID: 083.B-0774, 075.B-0127.}
\begin{abstract}
Sagittarius (Sgr) is a massive disrupted dwarf spheroidal galaxy in the Milky Way halo that has undergone several stripping events. Previous chemical studies were restricted mainly to a few, metal-rich ([Fe/H]$\gtrapprox -1$) stars that suggested a top-light initial mass function (IMF).
Here we present the first high-resolution, very metal-poor ([Fe/H]=$-1$ to $-3$) sample of 13 giant stars in the main body of Sgr. We derive abundances of 13 elements namely C, Ca, Co, Fe, Sr, Ba, La, Ce, Nd, Eu, Dy, Pb, and Th which challenge the interpretation based on previous studies. Our abundances from Sgr mimic those of the metal-poor halo and our most metal-poor star ([Fe/H]$\sim -3$) indicates a pure r-process pollution. Abundances of Sr, Pb, and Th are presented for the first time in Sgr, allowing for age determination using nuclear cosmochronology. We calculate ages of $9\pm2.5$\,Gyr. Most of the sample stars have been enriched by a range of asymptotic giant branch (AGB) stars with masses between 1.3 and 5\,M$_{\odot}$. Sgr J190651.47-320147.23 shows a large overabundance of Pb (2.05\,dex) and a peculiar abundance pattern best fit by a 3\,M$_{\odot}$ AGB star.
Based on star-to-star scatter and observed abundance patterns a mixture of low- and high-mass AGB stars and supernovae (15-25\,M$_{\odot}$) are necessary to explain these patterns.
The high level (0.29$\pm$0.05\,dex) of Ca indicates that massive supernovae must have existed and polluted the early ISM of Sgr before it lost its gas. This result is in contrast with a top-light IMF with no massive stars polluting Sgr.
\end{abstract}

\section{Introduction}
The Sagittarius dwarf spheroidal (Sgr dSph) is the nearest \citep[26.3kpc;][]{Monaco2004}, massive dwarf galaxy  in the Milky Way (MW) and has been studied over two decades. It was discovered by \citet{Ibata1994} yet its most metal-poor component remains unexplored until now thereby limiting our past interpretation of the enrichment and formation of the MW and its satellites. Sgr is the third most massive satellite galaxy \citep[$2.1\cdot10^7$\,M$_{\odot}$ -- similar to Fornax dSph;][]{McConnachie2012} in the Local Group (LG) after the Large and Small Magellanic clouds (LMC and SMC, respectively). The Sgr system is currently undergoing tidal stripping from the interaction with the MW which has resulted in two large streams (an old, faint and a later stripped, brighter one; see, e.g., \citealt{Majewski2003, Belokurov2014, Koposov2015}). The old, faint stream was stripped when Sgr started falling into the MW $\sim9$\,Gyr ago, and the brighter one about $5-7$\,Gyr ago. The two streams are drawn from the main body of Sgr, which contains the massive globular cluster (GC) M54 as well as Ter7, Ter8, Arp2, and Pal12 \citep{Sbordone2007, Cohen2004}. Several studies have found a bimodal metallicity distribution of Sgr and M54 typically around [Fe/H] $= -1.5$ and $\sim -0.6$ \citep{Carretta2010,Bellazzini2008}. According to \citet{Carretta2010, deBoer2015}, Sgr shows an 'alpha-knee' at [Fe/H] = $-1.3$, which is consistent with the star formation history (SFH) of massive dwarf galaxies, and studying the SFH further they claim that the onset of the supernovae type Ia causing the occurrence of the knee happened 1-3\,Gyr after the initial star formation. 

Based on metallicity distribution functions, Sgr is found to have had an extended SFH which was terminated by the stripping of the brighter stream leaving the main body behind with no or little gas as we observe it today \citep{McWilliam2013}. Several globular clusters in the MW halo may have originated from Sgr \citep{Law2010} and they have different [Fe/H] indicating that Sgr had a complex evolution, chemical enrichment, and metallicity distribution. It also points towards a very efficient stripping of GCs into the MW, which indicates the importance of such stripped systems as building blocks of the MW halo.

Several spectroscopic studies have focused on the chemistry of Sgr and have shown it to be different and easily separable from that of the MW. The `classical' difference of dSphs showing lower [$\alpha$/Fe] than the MW is also seen in Sgr, at least in previous studies of Sgr stars more metal-rich than [Fe/H]$\geq -1.5$\,dex \citep[see, e.g., the recent large APOGEE study by][]{Hasselquist2017}. This is well explained by the poorer gas reservoir retained in the weaker gravitational potential of dSphs compared to larger galaxies, resulting in smaller molecular clouds and in turn in lower mass supernovae (SNe). Compared to lower mass SN the more massive ones produce and yield more $\alpha-$elements to enrich the next stellar generation \citep{Tinsley1979,Matteucci1990, Kobayashi2006}. The chemical composition of Sgr is however not limited to overall low $\alpha$ abundances; it has also been found to be underabundant in Fe-peak elements and conversely overabundant in slow neutron-capture ($s-$)process elements   \citep{Bonifacio2000, McWilliam2003,Monaco2005,Sbordone2007,McWilliam2013}.  Each of these studies analysed between 3 -- 12 stars belonging to the main body of Sgr, all of which have [Fe/H] $>-1.55$\,dex. The proposed explanation for this chemical enrichment pattern is that no massive supernovae have existed in Sgr, leaving the IMF steep and  top-light. The heavy element production was instead believed to be due to metal-poor, low-to-intermediate mass AGB stars 
confirmed by the large [La/Y] and [Ba/Y] ratio found in several studies \citep{Sbordone2007,McWilliam2013}. However, the large (235 stars belonging to M54 or the central nucleus of Sagittarius, Sgr, N), medium-resolution study by \citet{Mucciarelli2017} derived $\alpha-$abundances in stars spanning a broad range of metallicities reaching [Fe/H]$\sim -2$ and recovered for the first time an $\alpha-$enhancement.

Despite a number of high-resolution spectroscopic studies, the enrichment history of Sgr is yet not fully mapped, and open questions pertaining to the IMF and mass loss occurring during the tidal stripping and ram pressure exerted by the MW affect our interpretation of the chemical evolution of Sgr. The actual level of the Fe-peak elements is debated, as is the overall origin of the heavy elements (from both s- and r-processes) in various dwarf galaxies \citep{Venn2004}. Previous studies have placed constraints on the r-process sites in Sgr using e.g., Eu/O-ratios in small-number, metal-rich samples \citep{McWilliam2013}.
In spite of previous observations and analyses of Sgr, we still only know of one very and no extremely metal-poor stars in the main body of this system. This has prevented an in-depth investigation of the nature and enrichment of the very early stages of this galaxy.

The combination of $\alpha$, Fe-peak, and n-capture elements in our study allow for a re-assessment of the nature of both the AGB stars and supernovae that enriched the very to extremely metal-poor Sgr stars. Finally, Th abundances were derived for a few stars in the sample with the highest signal-to-noise ratio in their spectra. This combination of elements allowed us to discuss the early formation of the Fe-peak and heavy elements in greater detail than previously done in addition to determining ages of the main body of Sgr. This led us to revise the interpretation from the previous studies dominated by more metal-rich stars that suffered from observational biases and limitations. Our results point towards a very early generation of massive (15-20\,M$_{\odot}$) supernovae and AGB stars ($\sim5$\,M$_{\odot}$) polluting the early Sgr galaxy as seen in the MW and other massive dwarf galaxies \citep[e.g.,][]{Koch2008,Pompeia2008}. 

In Sect.~\ref{sec:sample} the sample and data reduction are described, Sect.~\ref{sec:stelpar} presents the stellar parameters and how they are derived, Sect.~\ref{sec:abun} the abundance analysis is outlined, and in Sect.~\ref{sec:results}, Sect.~\ref{sec:discussion}, and Sect.~\ref{sec:concl} the results, discussions, and conclusions can be found.

\section{Sample and data reduction}\label{sec:sample}
Observations were obtained using the high-resolution, cross-dispersed UV-Visual Echelle Spectrograph \citep[UVES;][]{Dekker2000} mounted at the unit 2 telescope (UT2/Keueyen) of the ESO Very Large Telescope (VLT) in Cerro Paranal, Chile.

Twelve out of 13 stars were observed with setup dic1 and central wavelengths of 390nm and 580nm, for the blue and red arms, respectively. We adopted a 1.4\arcsec ~ wide slit and 2x2 on-chip binning. All the stars were observed for $\sim$2400--13000s at an airmass between 1.0 and 1.3 in April and July, 2009. The very metal-poor star Sgr 2300225 was observed using a slightly different setup in an earlier run (in August 2005). Here a slit of 1.2\arcsec, 2x2 binning, and a dic1 setting centred on 390 and 564nm were used.  Details of the observations are provided in Table~\ref{tab:Obslog}. Throughout the paper we will refer to the stars using  their full ID except from in figures where we adopt S (for Sgr) plus their shortened coordinate identifiers. However, the IDs for stars 2300225, 3600436 and 2300275 do not follow the same convention and are not coordinates. Other targets IDs are from \citet{Giuffrida2010}.
The stars object of the present study
were selected for a high-resolution follow-up
of the metal-poor population of Sgr.
The selection was based on the metallicity derived
from FLAMES/Giraffe spectra \citep{Zaggia2004,Bonifacio2006,Giuffrida2010}.

\begin{table*}[htp]
\caption{Observations of our Sgr sample: ID, coordinates, total observing time,
distance to centre of M54, and heliocentric radial velocity.}
\begin{center}
\begin{tabular}{lcccrr}
\hline
Star (ID)   &  Ra   & Dec   & T$_{obs}$  & d~~~ &   RV$_{helio}$ \\
             &        &      & {[}s{]}  &  {[}arcmin{]}  & {[}km/s{]}\\
\hline
\object{Sgr J184323.07-290337.64} & 18 43 23.074  &  -29 03 37.650  &  9015  &  175.1  &  136.1 \\
\object{Sgr J184828.45-294929.70} & 18 48 28.460  &  -29 49 29.700  &  6010  &  94.2   &  127.6 \\
\object{Sgr J185211.31-311907.51} & 18 52 11.317  &  -31 19 07.510  &  6010  &  62.3   &  127.4 \\
\object{Sgr J185259.59-312135.11} & 18 52 59.590  &  -31 21 35.110  &  6010  &  59.0   &  154.9 \\
\object{Sgr J185533.85-300521.20} & 18 55 33.858  &  -30 05 21.200  &  6010  &  24.4    &  157.0 \\
\object{Sgr J185549.44-300349.30} & 18 55 49.444  &  -30 03 49.310  &  3005  &  26.9   &  135.7 \\ 
\object{Sgr J190039.06-310720.53} & 19 00 39.069  &  -31 07 20.540  &  6010  &  81.5   &  131.9 \\
\object{Sgr J190043.03-311704.33} & 19 00 43.035  &  -31 17 04.340  &  6010  &  87.2   &  153.9 \\
\object{Sgr J190638.43-315135.94} & 19 06 38.436  &  -31 51 35.950  &  2410  &  169.2  &  134.1 \\
\object{Sgr J190651.47-320147.23} & 19 06 51.471  &  -32 01 47.240  &  6010  &  176.6  &  165.5 \\ 
\object{Sgr 3600436}            & 18 53 35.657  &  -30 26 36.380  &  6010  &  19.0  &  138.7 \\  
\object{Sgr 2300225}            & 18 55 49.704  & -30 33 09.690   & 12685  &   10.9  &  167.2\\
\object{Sgr 2300275}            & 18 55 38.608  & -30 27 04.130   &  9915  &   7.8  & 137.0 \\
\hline
\end{tabular}
\end{center}
\label{tab:Obslog}
\end{table*}%

\begin{figure}[htbp]
\begin{center}
\includegraphics[scale=0.5]{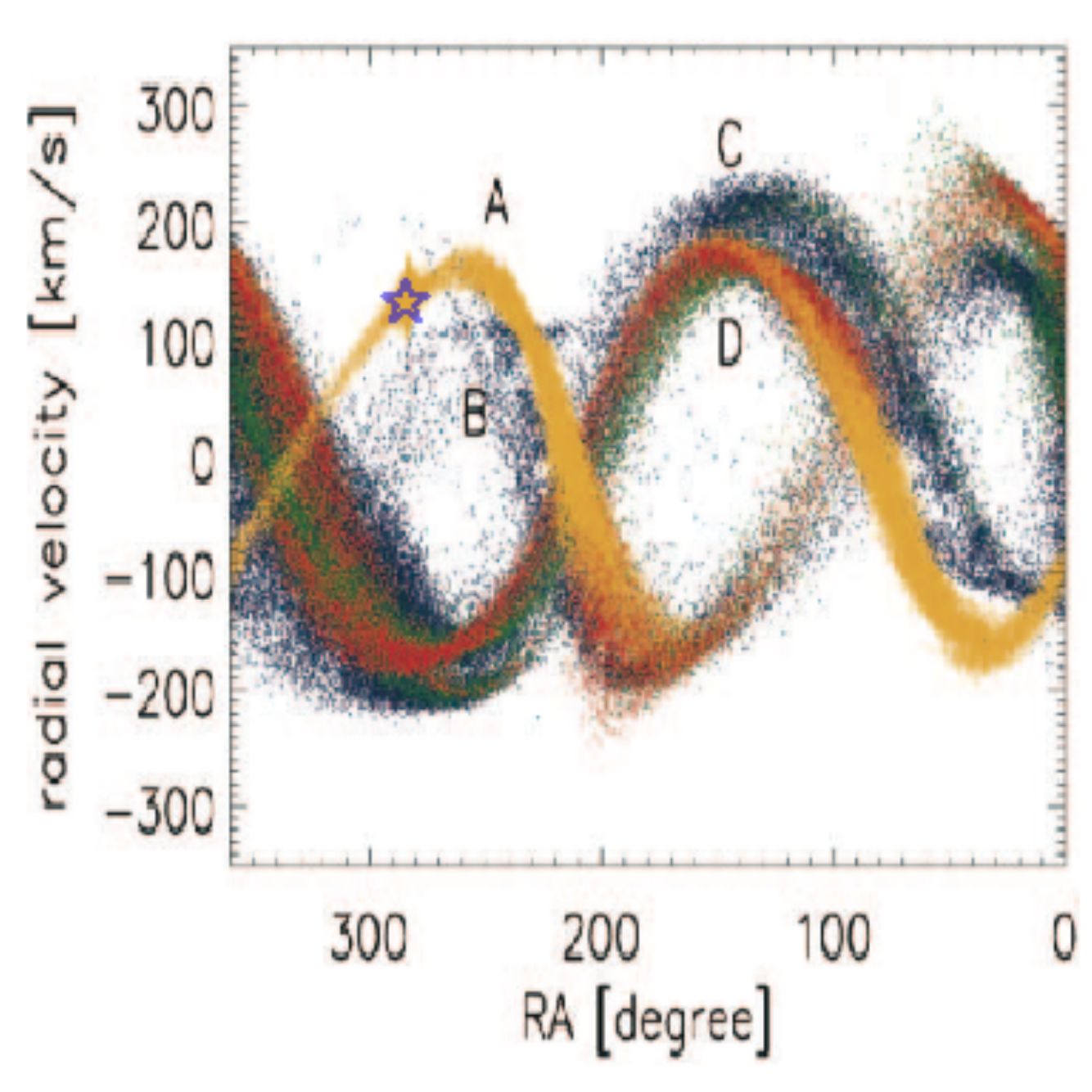}
\caption{Our sample (blue star) mean RV$_{hello}$ compared to the simulations of Sgr streams from \citet{Fellhauer2006}. The size of the star corresponds to the spread in RV and RA. The sample coincides perfectly with the location of Sgr main body.}
\label{Fig:Fellhauer}
\end{center}
\end{figure}

The distance to the centre of M54 is also listed in Table~\ref{tab:Obslog}, showing that all the stars are clearly outside the tidal radius of M54  \citep[7.4';][]{Trager1995} but centrally located in Sgr. Hence, the stars are not part of M54 but the main body of Sgr (see Fig.~\ref{Fig:Fellhauer}). The heliocentric radial velocities calculated using cross-correlation in IRAF are between 127 and 167 km/s for the individual stars. 
One to four frames were taken for each star, covering a maximum time span of about one day. Detected radial velocity variations between frames are generally small (below 0.7 km/s), being maximum for Sgr J190039.06-310720.53 and Sgr J190651.47-320147.23 with 1.0 km/s and 1.4 km/s, respectively, and we therefore consider them single. Observations over a longer baseline in time would be needed to truly probe the binary nature of these stars.

The sample's heliocentric corrected radial velocities are in good agreement with the average value of Sgr, N of $139.4\pm10.0$\,km/s \citep{Bellazzini2008}. Here we find an average of 143.6$\pm14.1$\,km/s and when excluding Sgr J190651.47-320147.23 an average and standard deviation of $141.8\pm 13.0$\,km/s is found.

\subsection*{Data reduction}
The data were reduced using the dedicated pipeline \footnote{see http://www.eso.org/sci/software/pipelines/}.
Data reduction includes bias subtraction, flat-field correction, wavelength calibration,
sky subtraction, and spectral rectification.
Radial velocities were measured by the {\it fxcor} package in IRAF \footnote{IRAF is distributed by the National
Optical Astronomy Observatory, which is operated by the Association of
Universities for Research in Astronomy, Inc., under cooperative agreement
with the National Science Foundation.}, using a synthetic spectrum of a typical giant star (T$_{\rm eff}$=4500 K, log(g)=2.0) as a template (see Table~\ref{tab:stelpar}). Finally, the median co-added spectra were normalised using the {\it continuum} package in IRAF.

\section{Stellar parameters}\label{sec:stelpar}
Initial atmospheric parameters were obtained in the following way.  
First, T$_{\rm eff}$ was derived from the 2MASS J-K color using the relation of
\citet{Ramirez2005}. Since this was only a first estimation, we did not adopt reddening correction. 
Surface gravities (log(g)) were obtained from the canonical equation:
$$ \log\left(\frac{g}{g_{\odot}}\right) =
         \log\left(\frac{M}{M_{\odot}}\right)
         + 4 \log\left(\frac{T_{\rm{eff}}}{T_{\odot}}\right)
         - \log\left(\frac{L}{L_{\odot}}\right). $$
where the mass M was assumed to be 1.0 M$_{\odot}$, and the
luminosity L/L$_{\odot}$ was obtained from the absolute magnitude, M$_{\rm V}$,
assuming an apparent distance modulus of the Sagittarius Dwarf Galaxy. The
bolometric correction (BC) was derived by adopting the relation 
BC-T$_{\rm eff}$ from \citet{Alonso1999}.
Finally, micro-turbulence velocity ($\xi$) was obtained from the
relation of \citet{Marino2008}.
Atmospheric models were calculated using ATLAS9 code \citep{Kurucz1970}
assuming our estimations of T$_{\rm eff}$, log(g), $\xi$, and assuming [Fe/H]=$-1.5$.\\ 
Then T$_{\rm eff}$, log(g), and $\xi$ were re-adjusted and new 
atmospheric models calculated in an interactive way in order to remove trends 
in excitation potential and reduced equivalent width (EW) versus abundance for T$_{\rm eff}$ and $\xi$, respectively, 
and to satisfy the ionization equilibrium for log(g). The [Fe/H] value of the model was changed at each 
iteration according to the output of the abundance analysis. 
The Local Thermodynamic Equilibrium (LTE) program MOOG \citep[][version 2014]{Sneden1973} was used
for the abundance analysis.
The final stellar parameters can be found in Table~\ref{tab:stelpar} and we adopt uncertainties on T$_{\rm eff}$/ log(g)/[Fe/H]/$\xi$ of 50\,K/0.2\,dex/0.2\,dex/0.1\,km/s.

\begin{table}[htp]
\caption{Stellar parameters: ID, effective temperature, gravity, [Fe/H], and microturbulence velocity ($\xi$).}
\begin{center}
\begin{tabular}{lcrcc}
\hline
Star (ID) &  T$_{eff}$ & log$g$ & $[$Fe/H$]$ & $\xi$ \\
  & [K] & [dex] & [dex] & [km/s] \\
\hline
\object{Sgr J184323.07-290337.64} &       4490 &       0.49 &	  $ -1.81$ &	 1.79 \\
\object{Sgr J184828.45-294929.70} &       4480 &       1.10 &	   $-1.44$ &	 1.54 \\
\object{Sgr J185211.31-311907.51} &       4825 &       2.00 &	  $ -1.07$ &	 1.34 \\
\object{Sgr J185259.59-312135.11} &       4595 &       1.10 &	   $-1.67$ &	 1.60 \\
\object{Sgr J185533.85-300521.20} &       4610 &       1.13 &	   $-1.46 $&	 1.56 \\
\object{Sgr J185549.44-300349.30} &       4320 &       0.03 &	  $ -1.43$ &	 1.68 \\
\object{Sgr J190039.06-310720.53} &       4660 &       1.03 &	   $-2.02$ &	 2.06 \\
\object{Sgr J190043.03-311704.33} &       4540 &       0.16 &	  $ -1.99$ &	 2.49 \\
\object{Sgr J190638.43-315135.94} &       4250 &       0.65 &	   $-1.47$ &	 1.72 \\
\object{Sgr J190651.47-320147.23} &       4500 &       0.81 &	   $-1.63$ &	 1.67 \\
\object{Sgr 3600436     }         &       4660 &       0.54 &	   $-1.63$&	 1.98 \\
\object{Sgr 2300225     }         &       4510 &       0.77 &	   $-2.56$ &	 1.56 \\
\object{Sgr 2300275     }         &       4975 &       1.90 &     $-2.96$  &     1.50\\
\hline
\end{tabular}
\end{center}
\label{tab:stelpar}
\end{table}%

\section{Abundance analysis}
\label{sec:abun}
Elemental abundances have been derived for C, Ca, Co, Sr, Ba, La, Ce, Nd, Eu, Dy, Pb, and Th. The details will be discussed below, and the line lists are provided in the online Table~\ref{tab:online}. All the stellar abundances have been derived using MOOG \citep[][version 2014]{Sneden1973} and the 1D ATLAS models with new opacity distribution functions \citep{Castelli2003} interpolated to the stellar parameters determined as explained above (Sect.~\ref{sec:stelpar}). The Solar abundances are from \citet{Asplund2009}.  
As most of the lines we employ for deriving heavy element abundances are located in the blue part of the spectrum, the absorption features are often blended and in some cases saturated. In these cases we include the lines in the final weighted mean, but with a lower weight. If the line is saturated and blended a weight of 0 or 0.3 is assigned, if the line is only blended or slightly saturated weight 0.5 is used, if the line is not saturated but slightly blended weight 0.7 - 0.8 is assigned. When the line is a clean and can easily be modelled (or even allow for equivalent width measurements) weight = 1.0 is used. Below the weight we used to calculate the mean is listed in parenthesis after each line. In general, the final average is not very different ($0.0 - 0.1$\,dex but mostly $0.02 - 0.03$\,dex) if we use a straight or a weighted mean, except for when we are dealing with upper or lower limits in the mean value. Hence, the weighted versus the straight mean value may differ significantly if we only have two measurements and one of them is a limit, however, this is the case for only eight out of the 159 abundance values presented here.

As a test of our weighting scheme, we calculate the difference between the straight mean and the weighted mean, to explore the impact on the calculated abundances. For all elements, the average differences are $<0.01$\,dex which are below the standard deviation (scatter). The best case is La, where the average difference amounts to only 0.003\,dex and the standard deviation 0.01\,dex. There are two exceptions, namely Sr, where the two lower limits have been included resulting in an averaged difference for the 13 stars of $0.04\pm0.06$\,dex, and Ca where the weighted mean and standard deviation are 0.29 and 0.17\,dex while the straight mean is 0.27 and its standard deviation 0.16\,dex. Within the uncertainty and standard deviation these numbers are the same, which confirms that our weights are acceptable.

\begin{table*}[ht]
\caption{Stellar abundances (weighted mean) and standard deviation for the 13 Sgr main body stars. }
\begin{center}
\begin{tabular}{l*{11}{r}}
\hline				 														    
\hline
Element   &Sgr J184323  &	$\sigma$  &  Sgr J184828	 & $\sigma$  &	 Sgr J185211&  $\sigma$	&     Sgr J185259 &$\sigma$	          & Sgr J1885533&   $\sigma$ \\
\hline				 															  
{[Fe/H]} &      $-1.81$  &    0.20	   &   $-1.44$ & 0.20	  &	 $-1.07$ &  0.20    &   $-1.67$ & 0.20            &   $-1.46$ &           0.20           \\			 
{[C/Fe]} &     $ -1.00$  &      0.25     &   $-0.75$ & 0.25 &	 $-0.55$  & 0.25    &   $-0.85$ &  0.25            &   $-0.90$ &       0.25          \\      
{[Ca/Fe]} &      0.31    &    0.27      &     0.38 &  0.25	  &	 0.49  &   0.05        &    0.28  &  0.19               &  	 0.17 &	 0.15         \\     
{[Co/Fe]} &    $ -0.03$  &    0.04     &  $-0.27$ &  0.35	  &	$-0.30$ &  --        &  $-0.28$ &   --          &   $-0.23$ &	 0.39          \\     
{[Sr/Fe]} &    $>-0.20$  &      --      &  $>0.10$ &   --	  &	 0.14  &  0.08         &    0.03  &  0.03               &  	 0.25 &	 0.02            \\     
{[Ba/Fe]} &    $ -0.07$  &    0.08      &	 0.43 &  0.09	  &	 0.22  &   0.28       &  $-0.19$ &  0.10              &   $-0.02$ &	0.03                  \\     
{[La/Fe]} &      0.13    &    0.11        &	 0.27 &  0.23	  &	 0.43  &   0.16        &   $-0.16$ &  0.13            &   $-0.04$ &	 0.12                  \\     
{[Ce/Fe]} &    $ -0.12$  &    0.06      &	 0.14 &  0.13	  &	 0.00  &   0.07       &  $ -0.18$ &  0.06             &   $-0.15$ &	0.03                    \\     
{[Nd/Fe]} &      0.15    &    0.13      &	 0.49 &  0.24	  &	 0.53  &   0.27       &    0.01   &  0.06               &  	 0.13 &	 0.08               \\     
{[Eu/Fe]} &      0.24    &    0.19      &	 0.27 &  0.23	  &	 0.29  &   0.06       &   $-0.17$ &  0.25              &  	0.06  &	 0.23                  \\     
{[Dy/Fe]} &      0.35    &     --         &	 0.80 &    --	  &	 0.70  &  --             &   0.10    &	 --                    &  	 0.40 &         --	            \\    
{[Pb/Fe]} &     $<0.50$ &     --       & $<0.10$&    --	  &   $<0.60$  &  --         &   $<0.10$ &	 --            &   $<0.20$ &        --	                  \\    
{[Th/Fe]} &      0.10    &     --         &	 ---    &    --	  &	 ---     &  --	 &    ---    &	 --                            &  	 ---  &           --	                         \\
\hline
             &  Sgr J185549 &                  &   Sgr J190039   &               &  Sgr J190043  &      	          & Sgr J190638 &          &Sgr J190651	&          	\\
\hline				        	        				 				     			     
{[Fe/H]}     &$-1.43$ &	0.20 &     $-2.02$&	   0.20  &   $-1.99$ & 0.20	          &   $-1.47$ &  0.20      &$-1.63$ &  0.20         \\  
{[C/Fe]}     &$-0.80$ &	0.25   &     $-0.90$ &      0.25   &	$-0.90$ & 0.25    &   $-0.90$ &  0.25       &   0.00 &  0.25                  	 \\  
{[Ca/Fe]}    &   0.54 &	0.35  &        0.15   &  0.08	    &	 0.23   & 0.08              & 	0.22 &  0.16             &   0.32 & 0.23	       \\  
{[Co/Fe]}   &  $-0.23$ &	0.25  &     $-0.40$ &  0.28	    &	$-0.08$ & --                 &  $-0.13$ &  0.14       &$-0.30$ & 0.21	      \\  
{[Sr/Fe]}    &   0.11 &	         0.03 &        0.10  &  --	    &	 0.42   & 0.18               & 	0.26 &  0.10           &   0.58 & 0.24	       \\  
{[Ba/Fe]}  &   0.13 &	0.05  &     $-0.32$ &  0.01	    &	$-0.09$ & 0.11              & 	0.05 &  0.01                   &   0.80 & 0.11	      \\  
{[La/Fe]}    &   0.27 &	0.06  &     $-0.04$ &  0.07	    &	 0.00   & 0.00               & 	0.01 &  0.13                  &   1.48 & 0.04	       \\  
{[Ce/Fe]}  &   0.02 &	0.16  &        0.11   &  0.11	    &	$-0.03$ & 0.12              &  $-0.25$ &  0.12              &   0.87 & 0.29	       \\  
{[Nd/Fe]}   &   0.55 &	0.13  &        0.15   &  0.16	    &	 0.09   & 0.05                & 	0.34 &  0.12                     &   1.24 & 0.20	        \\  
{[Eu/Fe]}  &   0.47 &	0.05  &        0.28   &  0.11	    &	 0.30   & 0.21                 & 	0.36 &  0.06                   &   0.81 & 0.08	        \\  
{[Dy/Fe]}   &   1.10 &   	--    &          0.50   &    --	    &	 0.25   &  --                   & 	0.70 &  --               &   0.85 &  --	               \\ 
{[Pb/Fe]}    &   0.60 &	         --    &         1.05    &  --	    &	$<1.10$ &  --      &         ---    &  --             &   2.05 &  --	              \\  
{[Th/Fe]}    &   0.65 &	--    &    $<-0.50$  &  --           &	 0.50   &  --                       & 	0.00 &  --                  &  $<0.9 $  &  --        	                \\
\hline
           &	  Sgr 3600436  &           &	Sgr 2300225  &    &       Sgr 2300275 & & &       \\
\hline					   											      
{[Fe/H]}   &  $-1.63$ &   0.20	 &	$-2.56$ &   0.20	& $-2.96$ & 0.2 & &    & &           \\  
{[C/Fe]}   &   $-0.80$ &  0.25 &      $-0.35$ &   0.25	& $-0.50$ &0.1 &&         & &     \\  
{[Ca/Fe]} &	0.28 &   0.23	 &	 0.48 &  0.28	        & $-0.1$ & --   &&     & &            \\  
{[Co/Fe]} &   $-0.28$ &   0.30	 &	$-0.15$ &  0.05        &   0.29   & 0.06 &&   & &         	 \\  
{[Sr/Fe]}  &	0.10 &   0.05	 &	 0.20 &  0.03	        & 0.60     & --      &&  & &               \\  
{[Ba/Fe]} &   $-0.19$ &   0.10	 &	$-0.70$ &  0.10	        & $-0.80$ & -- &&      & &           \\  
{[La/Fe]}   &   $-0.20$ &   0.03	 &	 0.00    &  --   	         & $<0.10$  & -- &&     & &              \\  
{[Ce/Fe]} &   $-0.04$ &   0.06	 &	 0.01 &  0.07	         &	---    &    --	   &&    & &            \\  
{[Nd/Fe]} &	0.10   &   0.08	 &	 0.20 &  0.20	         & $<0.40$ & -- &&       & &         \\  
{[Eu/Fe]} &	0.24   &   0.20	 &	$-0.12$ &  0.11         & $<0.20$ & -- &&       & &        	 \\  
{[Dy/Fe]}    &	0.50   &   --	&	0.00      &    --	 & ---    &    --	    &&          & &              \\ 
{[Pb/Fe]}   &	$<0.80$ &   --	&	$<1.25$ &    --	  & ---    &    --	    &&          & &                \\  
{[Th/Fe]}  &	0.65   &   --	&	---    &    --	          & ---    &    --	    &&          & &        \\
\hline
\hline
\end{tabular}
\end{center}
\label{tab:abun}
\end{table*}%

\subsection*{Carbon (Z=6)}
\begin{figure*}[htbp]
\begin{center}
\includegraphics[scale=0.54]{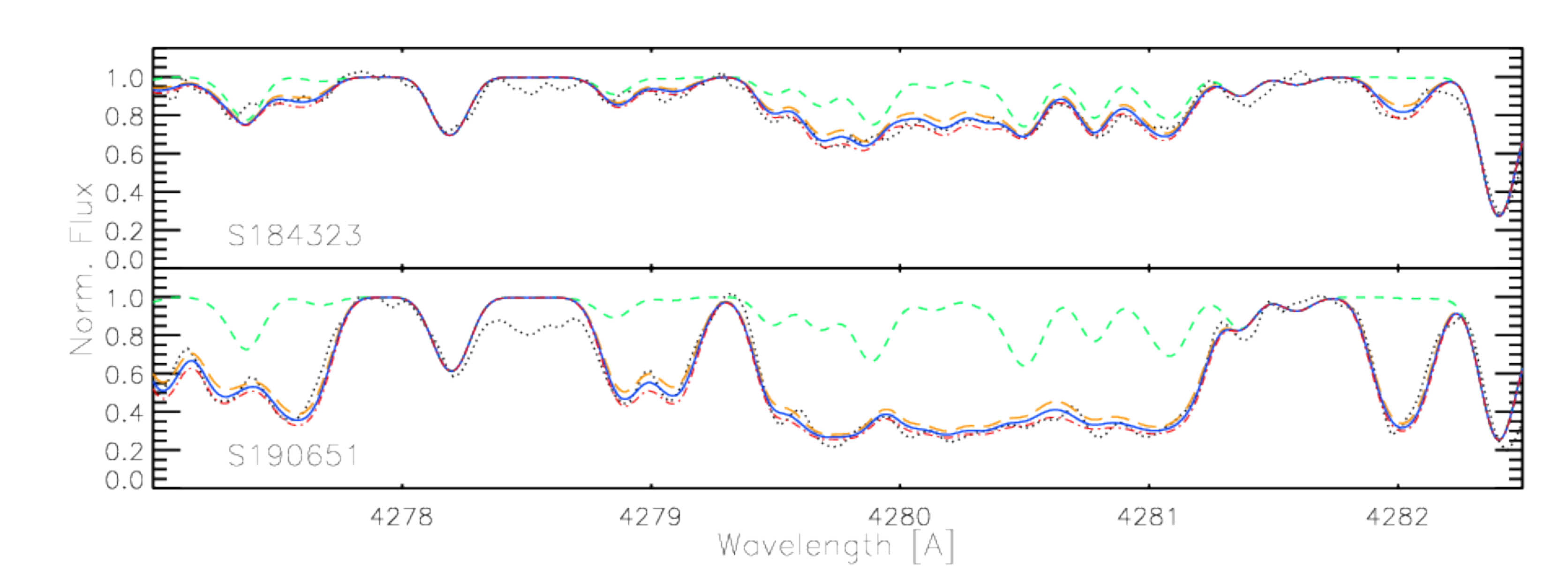}
\caption{Synthetic spectra fit to the observations of S184323 (top, black dotted line) in orange, blue, and red corresponding to [C/Fe]=$-1.1, -1.0, -0.9$ and to Sgr J190651.47-320147.23 (bottom, black dotted line) with [C/Fe]=$-0.1, 0.0, 0.1$ in the same colours. The green, dashed line corresponds to an abundance of $-5$ of the synthesised element.}
\label{Fig:Carbon}
\end{center}
\end{figure*}

The carbon abundances were derived by fitting synthetic spectra to the CH G-band at 4300\,\AA\,. We focused on the region 4280-4290\,\AA\ (w=1.0) as this region is mainly sensitive to C and there are only few atomic lines in this region (if the star is metal-poor -- see Fig.~\ref{Fig:Carbon}). For most of the stars a very low C abundance around [C/Fe]$\sim -0.71$\,dex is derived, except for Sgr J190651.47-320147.23 where we find [C/Fe]=0 (for more details see Sect.~\ref{sec:19065}). Our average carbon abundance is in good agreement with the [C/Fe] from the more metal-rich ([Fe/H]$>-1.2$) study by \citet{Hasselquist2017}. We note that the C abundances have been derived assuming molecular equilibrium (and Solar scaling all other elements like N and O), yet for these relatively cool  (and likely mixed) stars the CH abundance from the G-band could still be slightly off compared to what we would derive from other molecular C-bands like, e.g., CN, and we assign the [C/Fe] values a slightly larger uncertainty (of 0.25\,dex) for that reason as we cannot derive N abundances from our spectra. All the abundances are listed in Table~\ref{tab:abun}. 

\subsection*{Calcium (Z=20)}
Calcium is the only $\alpha-$element for which we present abundances here. A complete analysis of all lighter elements can be found in Monaco et al. 2017 (in prep.). Here we only present Ca abundances derived from two Ca lines located close to the Ba lines. For getting a rough estimate of the $\alpha-$abundance in these metal-poor stars, Ca was derived from the 5857.5\,\AA ~and 6493.8\,\AA~ lines (w=1.0 and 0.7, respectively). The reason for this is to trace the formation site of the heavy elements, where $\alpha-$elements provide insight into the nature (mass) of, e.g., the supernovae progenitor \citep{Kobayashi2006}. 
The average and standard deviation for the 13 stars studied here is $\langle[$Ca/Fe$]\rangle=0.29$, which is much higher than reported in previous studies \citep{McWilliam2013,Sbordone2007, Monaco2005} and in agreement with enhancements seen in more massive galaxies like the MW at similarly low metallicities ([Fe/H]$\leq-2$). This is in agreement with the medium-resolution study by \citet{Mucciarelli2017}.

\subsection*{Cobalt (Z=27)}
This is the only Fe-peak element we studied (again we refer to Monaco et al. 2017 in prep. for a detailed study of the lighter (Z$<30$) elements). The reason for studying Co I is due to the fact that it blends with our blue Th lines. 
In order to derive the Co abundances we used the 4121.3\,\AA\ line (w=1.0) combined with the wide Co line at 4020.9\,\AA\ (w=0.5), since this line is located just red wards of the 4019\,\AA\ Th line. However, the blue Co line is wide and complex, hence we assigned it the lower weight. For both lines hyperfine structure (HFS) was included in the line list\footnote{http://kurucz.harvard.edu/linelists.html}. The atomic data can be found in the online Table \ref{tab:online}.
We generally derive low (under abundant) Co abundances, and obtain an average of $-0.18$\,dex. This is in agreement with previous studies like \citet{Sbordone2007} and \citet{McWilliam2003,McWilliam2013} where they likewise find sub-Solar Co and Mn values, although at a higher metallicity. However, for our most metal-poor star ([Fe/H]$\sim-3$; Sgr 2300225) we find a super Solar Co abundance, [Co/Fe]$=0.29$.

\subsection*{Strontium (Z=38)}
We perform spectrum synthesis of two Sr II lines (4077.7, 4215.5\,\AA\,) in order to gain information about a light s-process element (for more details on, e.g., log$gf$ see \cite{Bergemann2012,Hansen2013}). The 4077\,\AA\ line is at the highest metallicities saturated resulting in lower limits, and we assign the values from this line a lower weight (0.5 -- see Fig.~\ref{Fig:SrBaEuspec}) while the 4215.5\,\AA\ is assigned full weight (1.0). The Sr abundances listed in Table~\ref{tab:abun} are thus weighted means. \comm{Due to the strong (sometimes saturated lines resulting in lower limits), the largest difference between straight and weighted mean is found for Sr (0.04\,dex).} The sample average and standard deviation of the Sr abundances are $0.21$ and 0.22, respectively, i.e., slightly above Solar.

\subsection*{Barium (Z=56)}
For barium we use two of the red lines 5853.7 and 6496.9\,\AA~  (w=0.8/1.0 and 1.0 -- see Fig.~\ref{Fig:SrBaEuspec}) in order to avoid the strong (easily saturating 4554.0\,\AA\ line as well as the heavily NLTE affected 6141.7\,\AA\ line; \citealt{Korotin2015}). However, the difference between assigning w=0.8 and w=1.0 is so small (\comm{$<0.01$\,}dex except for one case where it reaches 0.02\,dex) so we present a straight mean for Ba. Generally, a small line-to-line Ba abundance variation is found using these two Ba lines. We conduct both spectrum synthesis using the HFS from \cite{Gallagher2012}, and we measure EW to make sure that the Ba lines are not saturated (as reported in many other studies focusing on metal-rich, [Fe/H]$>-1$ stars, e.g., \citealt{McWilliam2013}).
The average Ba abundance is lower than what we measure for La and is just  below Solar $\langle$[Ba/Fe]$\rangle = -0.06$ with a standard deviation of\ 0.42\,dex (in part due to Sgr 2300225) indicating a fairly large star-to-star scatter for this s-process element ($\sim85\%$ in the Solar s-process distribution according to \citealt{Bisterzo2014}).

\subsection*{Lanthanum (Z=57)}
Lanthanum is another main s-process element \citep[75\% in the Solar System;][]{Bisterzo2014}, for which we derive abundances from two lines 4086.7 and 4123.2\,\AA{}, where the blue most one has Th as a blue wing blend. We therefore need to model the 4086-La line well, to make sure we separate the La contribution from the Th line. The difference between using equal weights or w=0.8 and 1.0, respectively, is very small ($\lesssim 0.01$\,dex\comm{, on average 0.003\,dex}). 
Spectrum syntheses of both La lines (including HFS -- \citealt{Lawler2001La}) result in an average La value of 0.17\,dex (standard deviation 0.43\comm{\,dex}), which is similar to the average we obtain for Sr. Lanthanum shows a slightly larger star-to-star abundance spread than Ba (somewhat driven by Sgr J190651.47-320147.23). With such a star-to-star abundance scatter, it is clear that the resulting heavy/light s-process ratio (HS/LS) must be discussed on a star to star basis to explore the formation site (see Sect.~\ref{sec:results}).

\subsection*{Cerium (Z=58)}
Five Ce lines were used with different weights (w = 1.0, 0.3, 0.8, 1.0, 0.5) owing to blends. The lines are:  4118.1, 4119.8, 4120.8, 4127.4, 4133.8\,\AA\ \citep[log$gf$ from][]{Lawler2009}. For this element we reinforce the weighted average since the lines are of varying quality. The abundance derived from synthesis vary as a consequence of a few unresolved blends as well as other heavy blends (e.g., from Nd). The weighted average and standard deviation are  $0.03$ and $0.29$\,dex. From this value we see that Ce behaves a lot like Ba, and we note that Ce like Ba is an even s-process element (84\% s). Further details will be discussed in the sections below.

\subsection*{Neodymium (Z=60)} 
The Nd abundances listed in Table~\ref{tab:abun} are based on seven lines: 4061.1, 4069.3, 4075.1, 4075.3, 4109.4, 4133.4, and 4135.3\,\AA\ with w = 1.0, 1.0, 0.8, 0.8, 0.8, 0.5, 0.7 where 4075.1 and 4075.3 lines blend together, and the 4133\,\AA\ line is a blended line resulting in this line often yielding upper limits. The  weighted mean is 0.34\,dex (standard devation 0.33\,dex) which is the highest average so far. Chronologically speaking, this is the first element where the r-process contributes by more than 20\%, and according to \citet{Bisterzo2014}  the s-/r-distribution is 57/43\%.

\begin{figure*}[htbp]
\begin{center}
\includegraphics[scale=0.6]{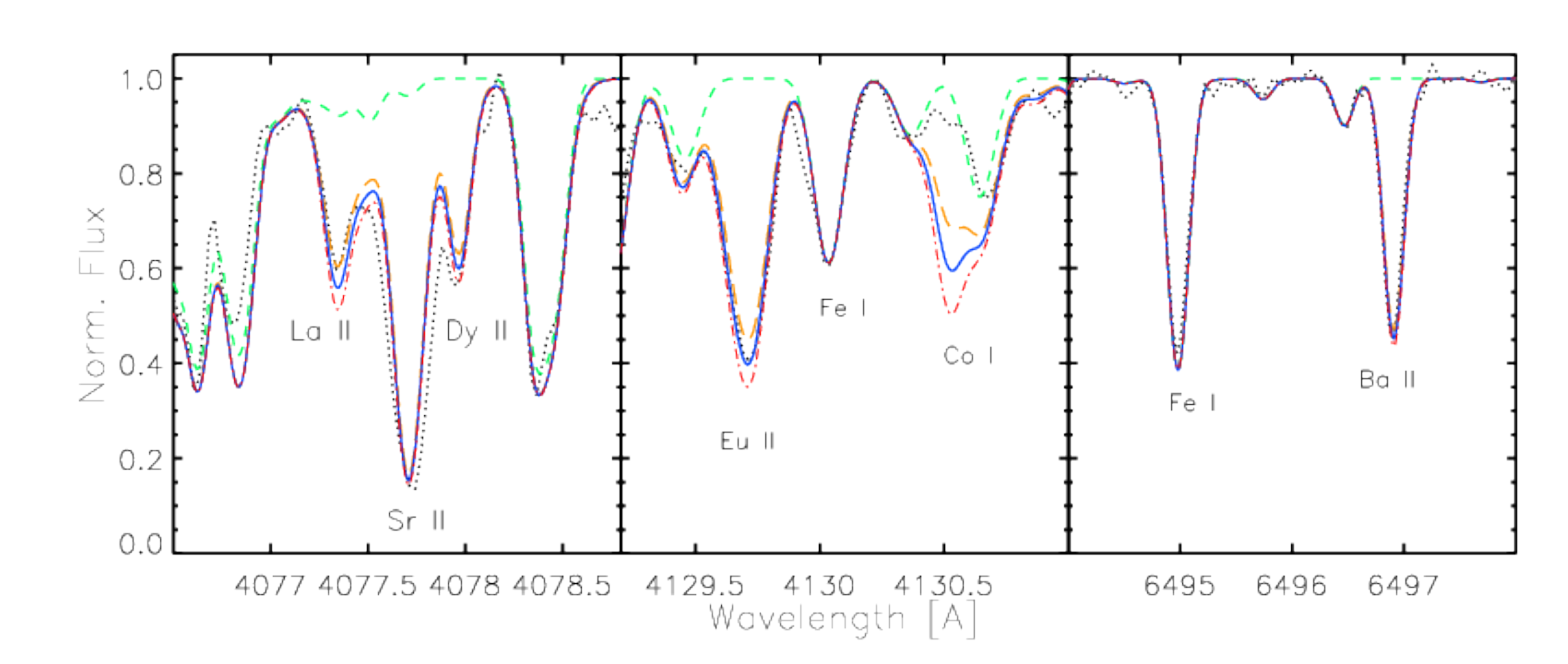}
\caption{Synthetic spectra fit to the observations of S185259 (black dotted line). The lines in orange (low), blue (intermediate), and red (high) are abundances are from left to right are as follows: [La/Fe]=$-0.16\pm0.1$, a saturated Sr line ([Sr/Fe]=$0.03\pm0.1$), [Dy/Fe]=$0.1\pm0.1$, [Eu/Fe]=$0.01\pm0.1$, Fe log$gf$ increased by 1\,dex, [Co/Fe]=$-0.5\pm0.2$, [Fe/H]=$-1.67$, [Ba/Fe]=$-0.1\pm0.1$. The green, dashed line corresponds to an abundance of $-5$ of the synthesised element.}
\label{Fig:SrBaEuspec}
\end{center}
\end{figure*}

\subsection*{Europium (Z=63)}
For Eu we used the two strongest lines, 4129.7 and 4205.1\,\AA\ where the blue most line is shown in Fig.~\ref{Fig:SrBaEuspec}. As seen from the spectra, several lines nearby or blending into the 4129\,\AA\ line have atomic data (oscillator strengths) that are poorly known. In \citet{Koch2002} fake Fe lines were introduced to obtain a better fit. In order to correctly reproduce the blending Fe line in the red Eu wing, the Fe log$gf$ had to be increased by 1\,dex for all stars (despite knowing the Fe-abundance to within 0.1-0.2\,dex accuracy). Despite the poorly constrained oscillator strengths of the surrounding Fe \comm{lines}, the Eu abundance is not affected by \comm{either} of the 4129.2 or 4130.0\,\AA\ lines and their log$gf$ values, so we decided assign both Eu lines full weight. Moreover, the Co line at 4130.5\,\AA\ can also not be reproduced with \comm{the} Co value derived from the two Co lines mentioned above even though we take HFS into account. \comm{Therefore, we} did not use this Co line in our study (see Fig.~\ref{Fig:SrBaEuspec}), and we note that it has no influence on the derived Eu abundances. However, this highlights the need for improved atomic data for a large number of lines in the blue (4100-4150\,\AA) region.

Europium is our best r-process tracer (94\% r-process in the Solar system; \citealt{Bisterzo2014}), which makes it the best r-process element for which we can derive stable abundances for a nuclear cosmochronometer (see Sect.~\ref{Sec:Th}). The average Eu abundance is 0.25\,dex and the standard deviation is 0.25\,dex.

\subsection*{Dysprosium (Z=66)}
We use one Dy line namely 4073.1 (w=1.0) since we find 4077.9 and 4103.3\,\AA\ to be too blended and we decided not to include them.  The 4103.3\,\AA\ line blends with H, La and Sr.  We have intentionally avoided the Dy line at 4077.9\,\AA~ which blends severely with Sr but also Nd.   
The average Dy abundance is 0.52\,dex. Dy is the next best r-process tracer after Eu with 85\% r-process material in the Solar system \citep{Bisterzo2014}.

\subsection*{Lead (Z=82)}
Lead is the heaviest s-process element we have studied here and we rely on the abundance synthesised from the 4057.8\,\AA\ line. We derived abundances for three stars and upper limits for seven stars. The rough average (treating limits and detections evenly in this estimate) is super Solar at 0.76\,dex which is somewhat biased by the high upper limits we derive. For comparison, the average abundance from the three Pb detections amounts to 1.23\,dex (see Table~\ref{tab:mean}) which is even higher and the values spread owing to Sgr J190651.47-320147.23 having a [Pb/Fe] = 2.05 (see Sect.~\ref{sec:19065}). However, since lead has not been investigated in Sgr dSph before, we choose to present these and lend the limits slightly more value.

\subsection*{Thorium (Z=90)}
Two Th lines were analysed, namely 4019.1 and 4086.5\,\AA\,. The 4086 line is very weak and yielded only upper limits while the 4019 line provided us with detections of Th in five stars and limits in two. The values listed in Table~\ref{tab:abun} are therefore only based on the 4019-Th line.

\begin{figure*}[htbp]
\begin{center}
\includegraphics[scale=0.5]{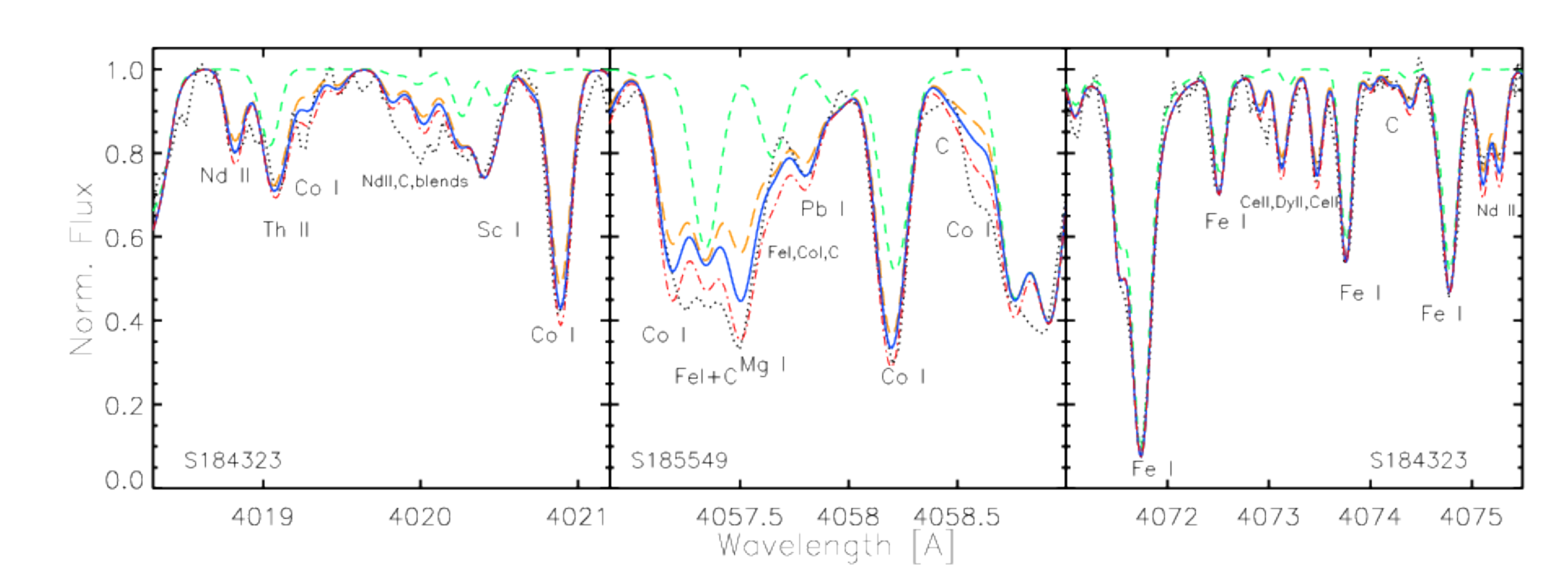}
\caption{Spectrum synthesis of S184323 with [Nd/Fe]=$0.15\pm0.1$, [Th/Fe]=0.1, [Co/Fe]=$-0.23, -0.03, 0.23$, [Sc/Fe]=$-0.25$. S185549 is shown in the mid panel with [Co/Fe]=$-0.23, -0.05, 0.15$, [Fe/H]=$-1.43,-1.23$, a blended Mg line ([Mg/Fe] = 0,0.5,1.0), [Pb/Fe]=$0.6\pm01$, and [C/Fe]=$-0.8\pm0.1$. Colours as in Fig.~\ref{Fig:SrBaEuspec}.} 
\label{Fig:specheavy}
\end{center}
\end{figure*}

This is a transition blended with not only atomic containments, but also CH lines. The first step is to establish the elements that blends. On the left hand/blue side the line is blended with Nd II at 4018.82\,\AA\ and Ni I at 4019.19\,\AA\,, whilst on the right/red side Th is blended with a CH line at 4019.13\,\AA\ and Co I at 4019.12\,\AA\,. In addition, Fe I and V II lines are blending into the Th line as pointed out by \citet{Caffau2008Th}. With Fe set by the metallicity and Solar scaled V and Ni, the first step in the analysis was to determine the abundance of Nd, Co and CH, respectively. The analysis of C, Co, and Nd was explained previously. The abundances of CH, Co, Nd, were added to the spectral synthesis including their uncertainties (see Fig.~\ref{Fig:specheavy}).
Very few stars have had Th detected in their spectra, and we stress that deriving Th in these stars is very demanding owing to the heavy line blending. In order to produce satisfactory synthetic spectra we had to update line lists both from VALD\footnote{Vienna Atomic Line Database - http://vald.astro.uu.se} and the recent compilation from \cite{Sneden2014} that includes molecular C lines. The final line list is appended in the online material (Table~\ref{tab:online}) and the average Th abundance and its standard deviation is 0.33 and 0.49\,dex, respectively (see Table~\ref{tab:mean}).

\section{Results}
\label{sec:results}

The chemical imprint of Sgr is known to be a mixture of high, heavy s-process, low $\alpha$, and low Fe-peak \citep{McWilliam2003,McWilliam2013,Hasselquist2017}. Our sample has revealed the first abundance enhancements with respect to Solar at very low metallicity in Sgr dSph. We find enhancements of both $\alpha-$ and s-process elements, and the most metal-poor star shows a (possible) pure r-process trace atypical for dwarf galaxies in particular for Sgr. Below we describe the results with increasing atomic numbers of the 12 elements studied.

\begin{figure}[!htbp]
\begin{center}
\includegraphics[scale=0.45]{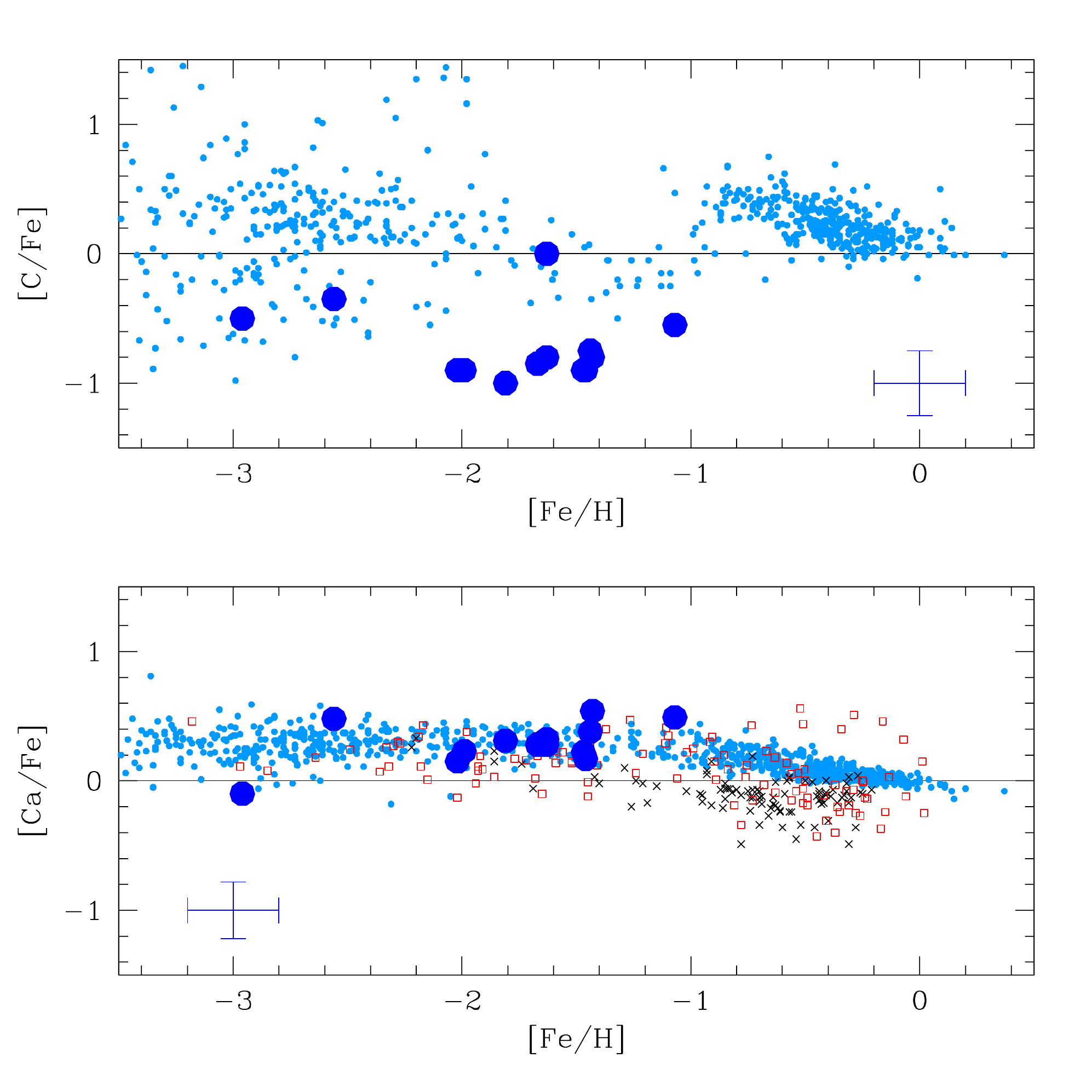}
\caption{Top: [C/Fe] vs [Fe/H] for our sample (filled circles) compared to literature studies of the MW \citep[gray filled squares]{Fulbright2000,Reddy2003,Reddy2006,Cayrel2004,Simmerer2004,Barklem2005,Francois2007,Johnson2014}. Bottom: [Ca/Fe] vs [Fe/H] -- symbols as above and additionally extra-galactic objects such as Magellanic
clouds \citep[blue filled squares]{Pompeia2008, Johnson2006, Mucciarelli2008, Mucciarelli2009},  
Draco, Sextans, Ursa Minor and Sagittarius dwarf galaxy and
the ultra-faint dwarf spheroidals Bo\"{o}tes I and Hercules
\citep[green filled squares]{Monaco2005,Sbordone2007,Shetrone2001,Ishigaki2014,Koch2008}.}
\label{Fig:CFe}
\end{center}
\end{figure}

Starting with our lightest studied element, carbon, we find a remarkable low [C/Fe] $\sim -0.71$ in all but one star (Sgr J190651.47-320147.23, which is Solar). So far carbon has not been studied in Sgr at high resolution and low metallicity ([Fe/H]$<-1.2$ -- see Figure~\ref{Fig:CFe}). 
The low C abundances were also shown by \citep{Hasselquist2017} albeit at higher metallicities. Our carbon abundances at low [Fe/H] are in good agreement with [C/Fe] in \citet{Hasselquist2017} which at their lowest [Fe/H]$\sim-1.2$ span a [C/Fe] from $-0.8$ to $-0.5$.  It should be borne in mind that our sample of stars, like those of \citet{Hasselquist2017} are very luminous, and therefore have very likely already undergone internal mixing. In the material that has been nuclearly processed and is mixed in the atmosphere, the C and the O  have been partly destroyed, to create N.
The approximate amount of reprocessed C can be estimated using the predictions from \citet{Placco2014b} and reading off the C-corrections from their Figure 15. This requires that we know the luminosity, which we can calculate using the distance modulus, the luminosity - absolute magnitude relation ($ \log(L/L_{\odot}) = 10^{-0.4(M_V - M_{V,\odot})} $) and assuming the distance to Sgr of 26.3\,kpc and $M_{V,\odot} = 4.83$. This results in $\log( L/L_{\odot})$ values between 2 and 2.6, which for the lowest gravity stars indicates a correction in [C/Fe] of $\sim0.2 - 0.4$\,dex and up to 0.6\,dex in one case (\object{Sgr J190039.06-310720.53}). Despite these fairly large corrections all our stars remain C-poor except for Sgr J190651.47$-$320147.23 (see Sect.~\ref{sec:19065}).
\citet{Hasselquist2017} have the abundances of also C and N, and can conclude that the ratio [(C+N+O)/Fe] is sub-Solar for the majority of their sample. They advocate a top-light initial mas function to explain this. We note that in their data there is a hint that this ratio increases at lower metallicity, although this is represented by very few stars. Our sample is ideal to check this trend at even lower metallicity, but we need to complement our C abundances with N abundances, at least. This is unfortunately not possible with the spectra currently available.

Our only $\alpha-$element, Ca, is generally found to be overabundant with respect to Solar, except for the most metal-poor star (Sgr 2300225) which has a remarkable low [Ca/Fe]$=-0.1$. The general Ca overabundance is in good agreement with \citet{Mucciarelli2017}, who down to [Fe/H]$\sim -2$ find Ca to be enhanced and match the Ca level of the MW.
\comm{For comparison to other $\alpha-$elements, we draw parallels to the sample of \citet{Hasselquist2017}. Their O and Mg abundances are deficient by $\sim0.1$\,dex compared to MW disk stars, and Si slightly less so (see their Figure 5). Their O and Ca trends in the same figure are seen to agree, and their Mg/Ca-ratio cluster around 0 ($\pm0.2$\,dex) as seen from their Figure 9. With this in mind, our trends and results from Ca should be representative of the $\alpha-$element behaviour in Sgr, even if Ca is slightly less mass dependent than Mg and O.}

\begin{figure}[htbp]
\begin{center}
\includegraphics[scale=0.30]{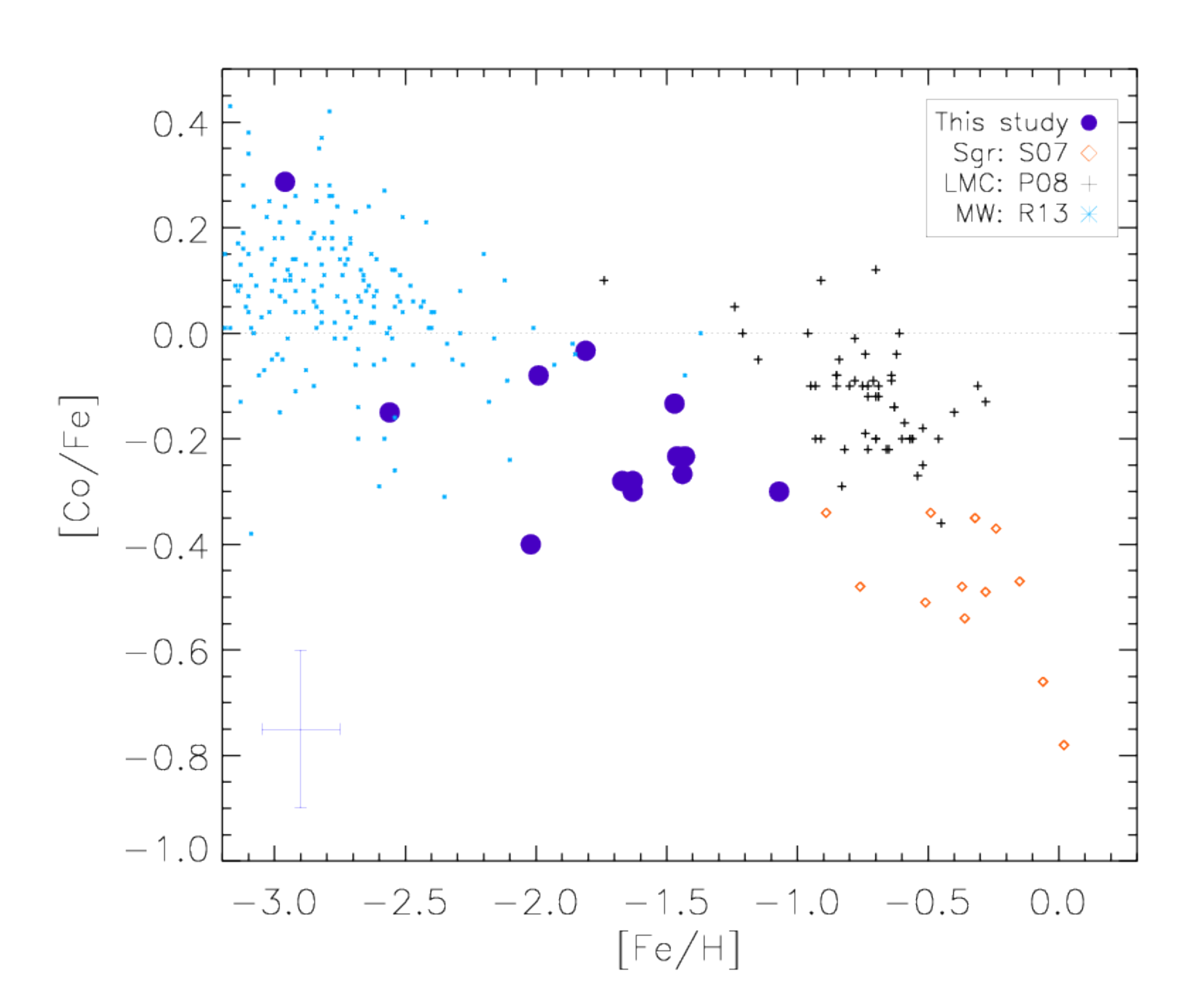}
\caption{[Co/Fe] vs [Fe/H] for our sample (filled blue circles) compared to literature studies of Sgr \citep[][S07, open, red diamonds]{Sbordone2007}, LMC \citep[][P08, black '+']{Pompeia2008}, and the MW halo \citep[][R13,turquoise dots]{Roederer2013}}
\label{Fig:CoFe}
\end{center}
\end{figure}

Except from Sgr 2300225, Fig.~\ref{Fig:CoFe} shows sub-Solar values of [Co/Fe] of $\sim -0.6$ at Solar metallicity which increases with decreasing [Fe/H] to around or just below [Co/Fe] = 0.0. This is also in agreement with the Sgr APOGEE data from \citet{Hasselquist2017} and the UVES/VLT data from \citet{Sbordone2007}  who showed low Co values ($-0.4$ down to $-0.8$) for their more metal-rich samples. Previous studies have drawn parallels between the formation and evolution of Sgr and the LMC \citep{Monaco2005, McWilliam2013} in that they both seem to have a top-light IMF and have lost gas early in their history. Therefore we compare our results to other studies of both Sgr and LMC to comment on this. 

Our results 1D, LTE abundances for Co agree well with the LMC \citep{Pompeia2008} and Sgr trends as well as some of the metal-poor, MW halo stars which also exhibit low Co values. Some of the LMC stars are even Solar or slightly above the [Co/Fe] Solar-scaled value. Since we rely on giant stars in this study, which may lower the Co abundances by $\sim 0.2$\,dex compared to dwarfs \citep[as noted in][]{Bonifacio2009}, the Co values are expected to be a bit lower than what dwarf stars in Sgr may exhibit. The 1D, LTE analysis may also be one of the reasons why we derive low Co abundances, as a 1D, NLTE analysis could increase the value by up 0.7\,dex in metal-poor dwarfs \citep[][]{Bergemann2010}. However, the final 3D, NLTE abundances would need to be calculated with our adopted stellar parameters to get the complete picture (which is beyond this paper's scope).  For this analysis we have used the recent Co I HFS loggf values from \citet{Lawler2015}, which for Co I yields values in good agreement with Co II \citep{Sneden2016} thereby reducing the need for the strong NLTE corrections. In summary, only slightly higher values would indeed be expected.

From a Galactic chemical evolution point of view, a large amount of Fe-peak elements is generally associated with supernovae type Ia. In Sgr these are expected to explode $1-3$\,Gyr after formation, and an increase in Fe-peak elements as a function of time. When SN Ia become more frequent and dilute the previous generation of SN II ($\alpha$-rich) material higher Fe-peak abundances and a decreasing [$\alpha$/Fe] are expected. Such a metallicity-dependent trend was clearly seen in \citet{McWilliam2003} for Mn. This is not the trend we find for our more metal-poor Sgr sample studying Co, and a metal-poor type Ia progenitor generation seem not to be able to explain our results. Our average $\langle$[Co/Fe]$\rangle = -0.18 \pm 0.05$  (see Table~\ref{tab:mean}), combined with the average $\alpha/$Fe-ratio obtained from Ca ($0.29\pm 0.05$ -- see Table~\ref{tab:mean}) results in an average [Ca/Co]$\sim0.47\pm0.07$. These are clearly higher values than the sub-Solar ones reported in previous, more metal-rich studies \citep[e.g., ][]{Monaco2005, Sbordone2007, McWilliam2013}. 

As the mass of the supernova tends to correlate with the amount of $\alpha-$elements ejected \citep{Kobayashi2006} our results indicate that more massive supernovae were indeed present and enriched the early composition of Sgr before it got accreted into the Milky Way and/or lost its gas (for more details on $\alpha-$ and lighter elements we refer to Monaco et al. in prep). To get an estimate of the SN mass, we compare our Ca/Co-ratio with yields from a supernova with [Fe/H]$\sim -2.4$. A good match (to within 0.16\,dex) for half the sample is found with the SN yields from a 20\,M$_{\odot}$ star \citep{Kobayashi2006}. For comparison a 13\,M$_{\odot}$ SN of the same metallicity would produce a sub-Solar Ca/Co-ratio.
This is in contrast with the results and interpretation of \citet{McWilliam2003, McWilliam2013} that imply the lack of such massive SN ($>20$\,M$_{\odot}$). 
Note that we here use the nomenclature 'massive' for SN heavier than 15\,M$_{\odot}$ as these have clear differences in their explosion mechanism and physics related to compactness, $\nu$-mechanism, and possible magnetic fields compared to the SN with masses below 12\,M$_{\odot}$ collapsing onto a O-Ne-Mg core. A cut at 30\,M$_{\odot}$ makes less physical sense and we furthermore note that the $\alpha-$elements are already enhanced in 15-20\,M$_{\odot}$ SN compared to the lower mass ones  \citep[][and references therein]{Janka2017}.

\begin{figure*}[htbp]
\begin{center}
\includegraphics[scale=0.6]{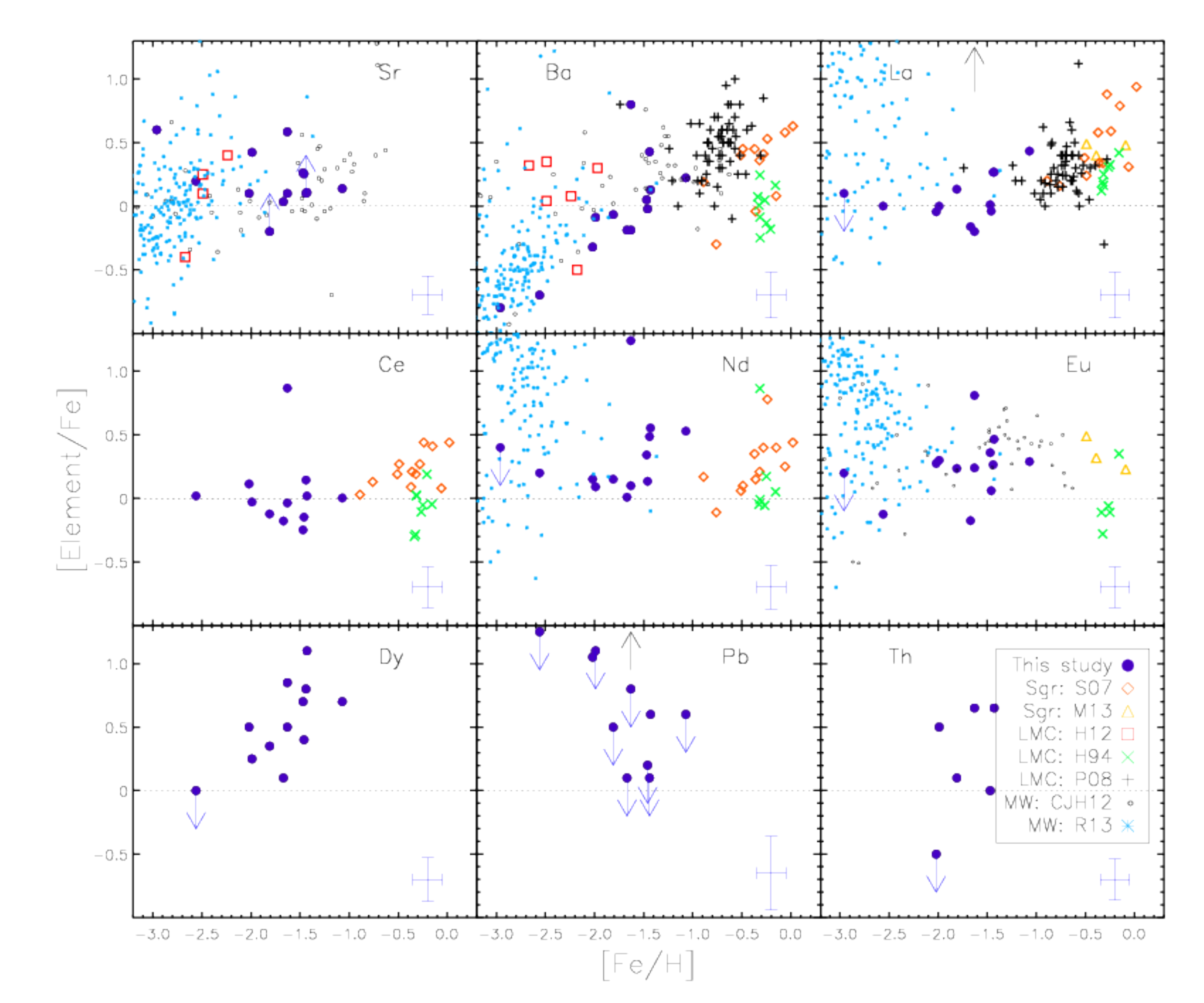}
\caption{Our Sr, Ba, La, Ce, Nd, Eu, Dy, Pb, and Th abundances for Sgr compared to literature studies of Sgr: \citet{Sbordone2007, McWilliam2013}; LMC: \citet{Hill1994,Pompeia2008,Haschke2012} and the MW halo: \citet{Hansen2012,Roederer2013}. The enhanced outlier is Sgr J190651.47-320147.23 where [La/Fe]=1.48 and [Pb/Fe]=2.05 are outside the plotting range (as indicated by the black arrows).The symbols are explained in the legend box.}
\label{Fig:multiheavy}
\end{center}
\end{figure*}

In the following we split the heavy neutron-capture elements into two groups - those that are predominantly formed by the s-process in the Solar system (Sr (weak s), Ba, La, Ce, Pb (main s)) while the r-process forms Eu, Dy, Th, and finally Nd is formed in almost equal amounts by either of the two processes \citep{Bisterzo2014}.
Figure~\ref{Fig:multiheavy} shows our 1D, LTE derived abundances compared to literature studies of Sgr, LMC, and the MW halo (as our stars are more metal-poor than the average MW disk stars and seem to show a chemical composition resembling the MW halo rather than that of its disk). 

\begin{table}[htp]
\caption{Average and standard deviation for the 12 elements studied. Values in parenthesis exclude upper limits.}
\begin{center}
\begin{tabular}{lcc}
\hline
Abundance & Mean & St.dev.\\
\hline
{[}C/Fe] &     $-0.71$    & 0.28 \\
{[}Ca/Fe] &     0.29   &  0.17 \\
{[}Co/Fe] &   $ -0.18$   &  0.17 \\ 
{[}Sr/Fe] &     0.21    & 0.22 \\
{[}Ba/Fe] &    $-0.06$    &  0.42 \\
{[}La/Fe] &     0.17    & 0.43 \\
{[}Ce/Fe] &     0.03   &  0.29 \\
{[}Nd/Fe] &     0.34   &  0.33 \\
{[}Eu/Fe] &     0.25   &  0.25 \\
{[}Dy/Fe] &     0.52   &   0.32 \\
{[}Pb/Fe] &   0.76 (1.23)   &   0.58 (0.74) \\
{[}Th/Fe] &   0.33 (0.38)  &  0.49 (0.31)\\
\hline
\end{tabular}
\end{center}
\label{tab:mean}
\end{table}%

Starting with the lightest n-capture element, Sr, we see a sparse trend of data clustered around [Sr/Fe]$=0.18$ (see Table~\ref{tab:mean}) making this the first sample probing the nature of Sr in Sgr. The stellar abundances of our giant sample agree well with those of the old, metal-poor RR lyrae stars from the LMC \citep{Haschke2012} as well as Sr from the MW halo \citep{Hansen2012, Roederer2013}.

Both Ba and La show increasing trends as a function of [Fe/H] with La abundances slightly higher than the Ba ones. This is in good agreement with \citet{Sbordone2007} and \citet{McWilliam2013}. The [La/Fe] trend is remarkably clean and consistently growing in both Sgr and the LMC, whereas Ba shows a slightly larger scatter (considering all samples or when excluding \object{Sgr J190651.47-320147.23}). Part of the explanation might be related to the Ba lines being stronger and possibly close to or saturated in the more metal-rich samples at [Fe/H]$=-1.0$. The [Ba/Fe] star-to-star scatter in Sgr and the LMC is slightly larger than what is found at that metallicity in the MW. This could indicate a large degree of inhomogeneity in the smaller dwarf galaxies \citep{Venn2012,Koch2013}. A difference in timescale and evolution might play a deciding factor here where the increased s-process level and mass loss of the dwarf galaxies could explain such trends.
However, a large star-to-star abundance scatter is expected for n-capture elements \citep[e.g.,][]{Francois2007, Hansen2012, Roederer2013}.

Despite Ce mainly being formed by the s-process, it shows a flat trend around 0 similar to that of Eu albeit $\langle[$Eu/Fe$]\rangle = 0.25$ is predominantly formed by the r-process.  
This is puzzling and could indicate that our Ce abundances are low or that the rare earth elements share formation processes at some level. While Ce shows a small star-to-star scatter, Eu is widely spread in the abundance diagnostics figure.
The Nd abundances are slightly increasing as a function of [Fe/H] \citep[in agreement with][]{Sbordone2007} but grow slower than the s-process elements Ba and La. Part of the explanation could be that Nd is produced in equal amounts by the r-process (creating Eu) and by the s-process forming Ba and La (see Sect.~\ref{sec:discussion}).

Another oddball is Dy which should be formed mainly by the r-process. However, Dy exhibits a clean growing trend as a function of [Fe/H] just like a boosted La (s-process) trend. We note that the Dy lines were blended and that the abundances therefore may be slightly high (despite conducting de-blending and spectrum synthesis). 

The two heaviest elements, Pb and Th, are presented here for the first time in Sgr. Most of the Pb abundances are only upper limits (owing to severe blends, see Fig.~\ref{Fig:specheavy}), but they support the high level of s-process enrichment both from the detections and the limits.  The slowly decaying Th is found at two levels - one around Solar and one just above 0.5\,dex. This might indicate that we are looking at two different population or groups with different ages (see Sect.~\ref{Sec:Th}). This would make sense considering the large [Fe/H] span of our sample.

\section{Discussion}
\label{sec:discussion}
The chemical composition of the very metal-poor stars of Sgr dSph provided surprising new results, from which we constrain the formation processes and objects that enriched this accreted, disrupted dwarf galaxy early on. From Fig.~\ref{Fig:multiheavy} the anticipated behaviour of a few elements seemed at odds with what we expect from their classical s- or r-production channels, and we explore their trends in more detail using absolute abundances.
We fit trends (lines) using `ladfit' in IDL which is a least absolute deviation method to obtain linear fits that are robust against outliers. Compared to a straight line fit (using, e.g., a minimum $\chi^2$) the changes are negligible (about 0.04, i.e., changes on the second digit). 
Here we also note that changes in the fitted lines/trends originating from using our weighted means or straight means are small (changes on the second or third digit on the slope and intersect with the y-axis \comm{-- see also Sect.~\ref{sec:abun}}). These are insignificant compared to the uncertainty in stellar abundances ($\sim 0.15$\,dex).

\begin{figure*}[htbp]
\begin{center}
\includegraphics[scale=0.19]{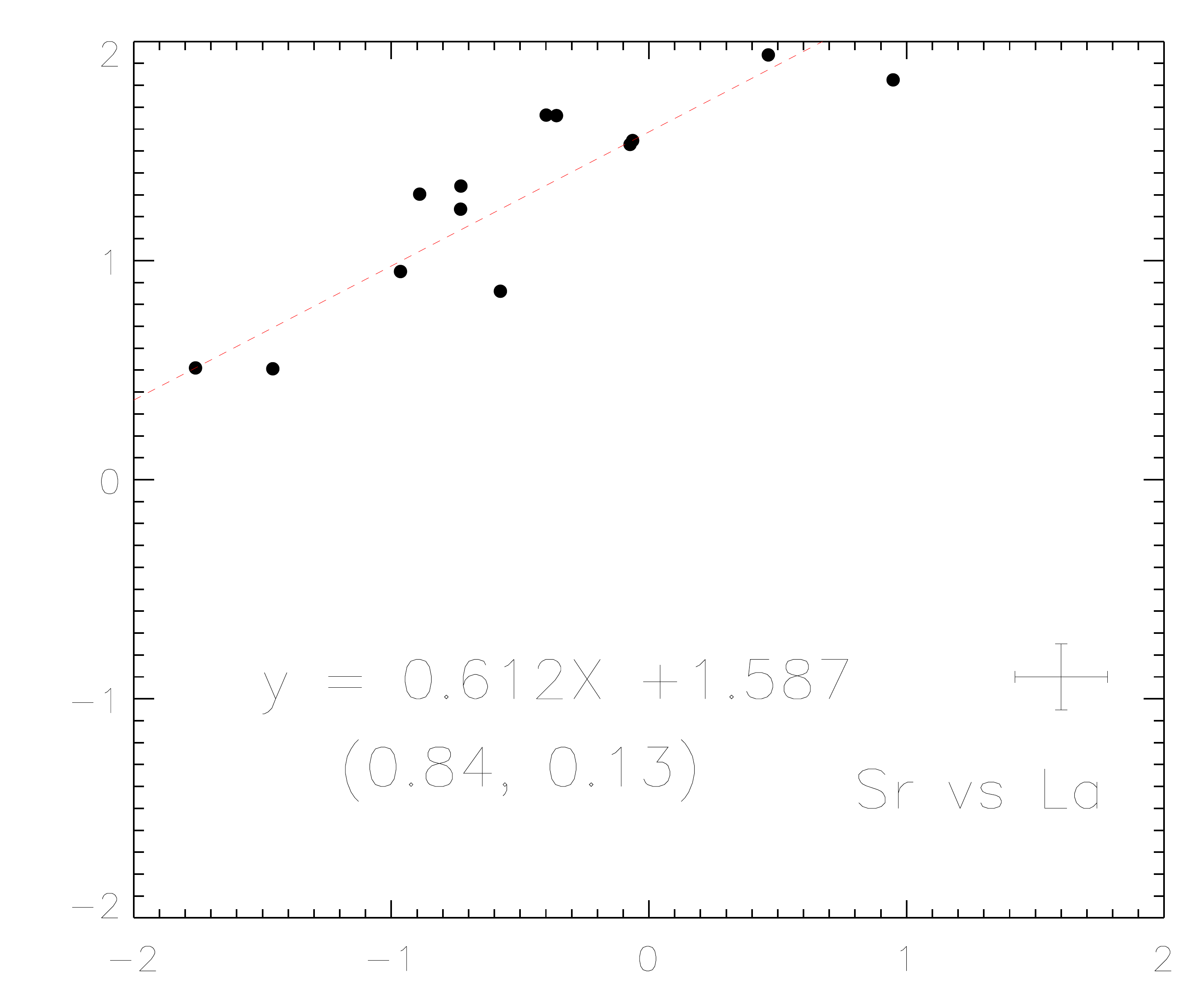}
\includegraphics[scale=0.19]{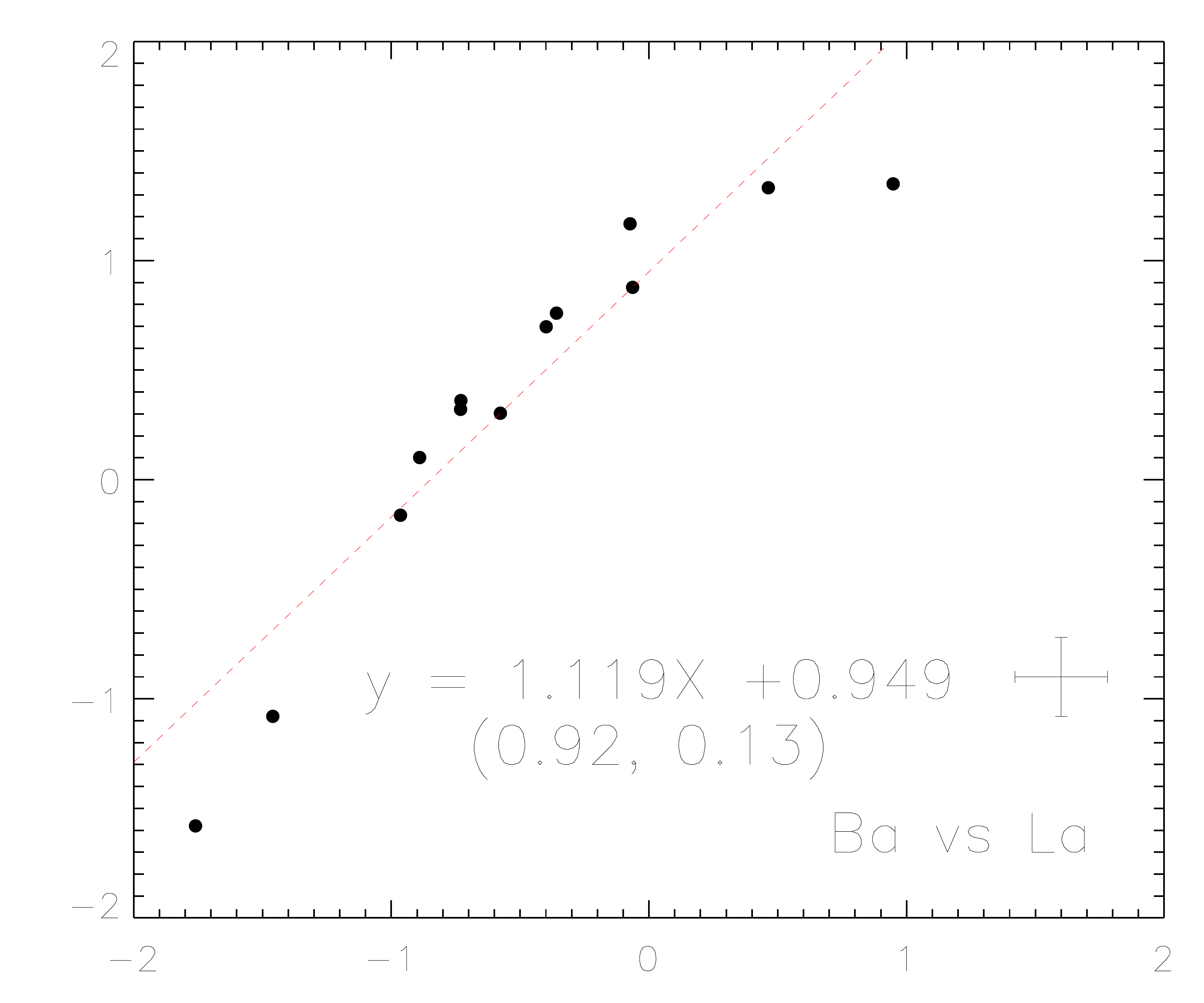}

\includegraphics[scale=0.19]{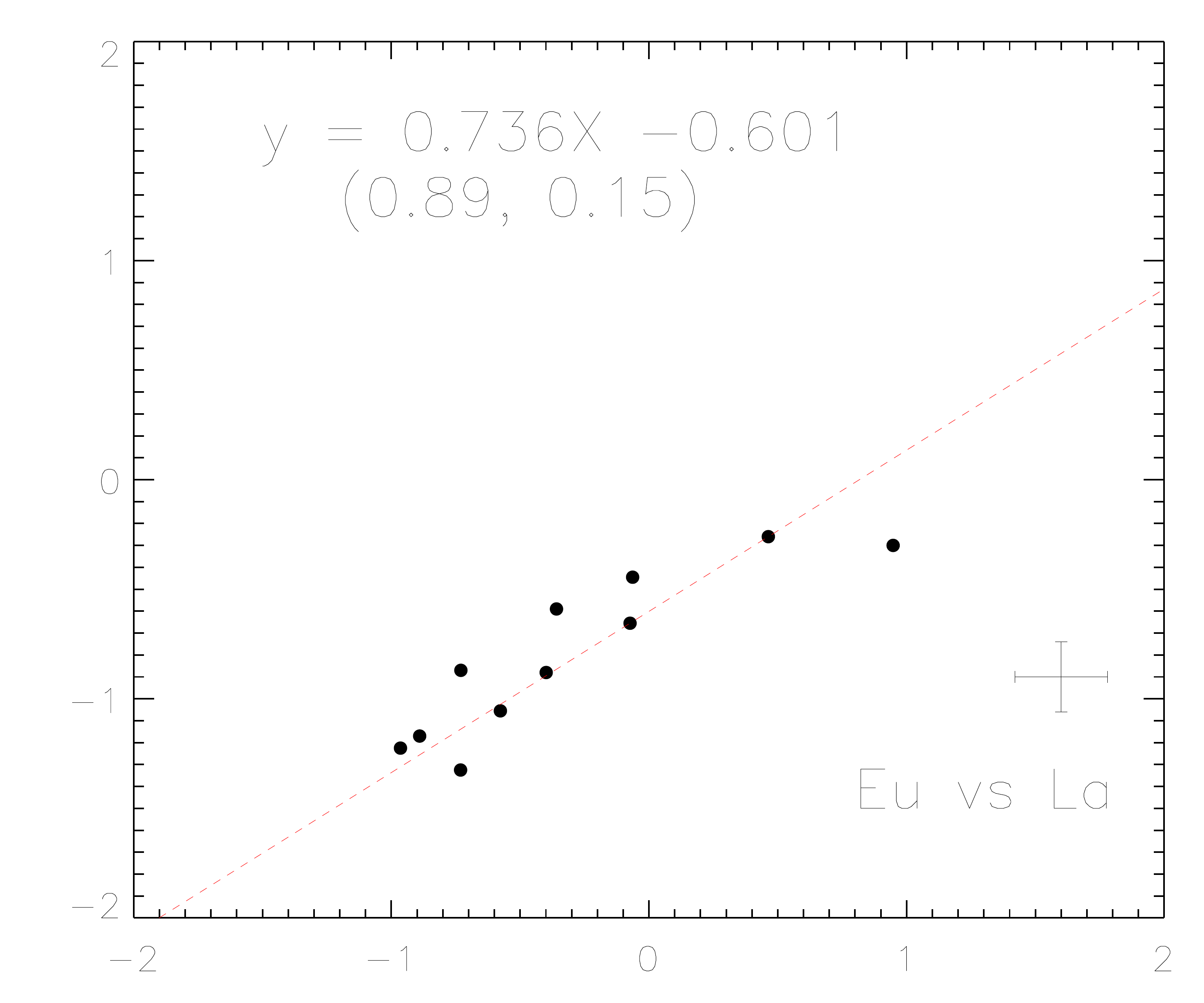}
\includegraphics[scale=0.19]{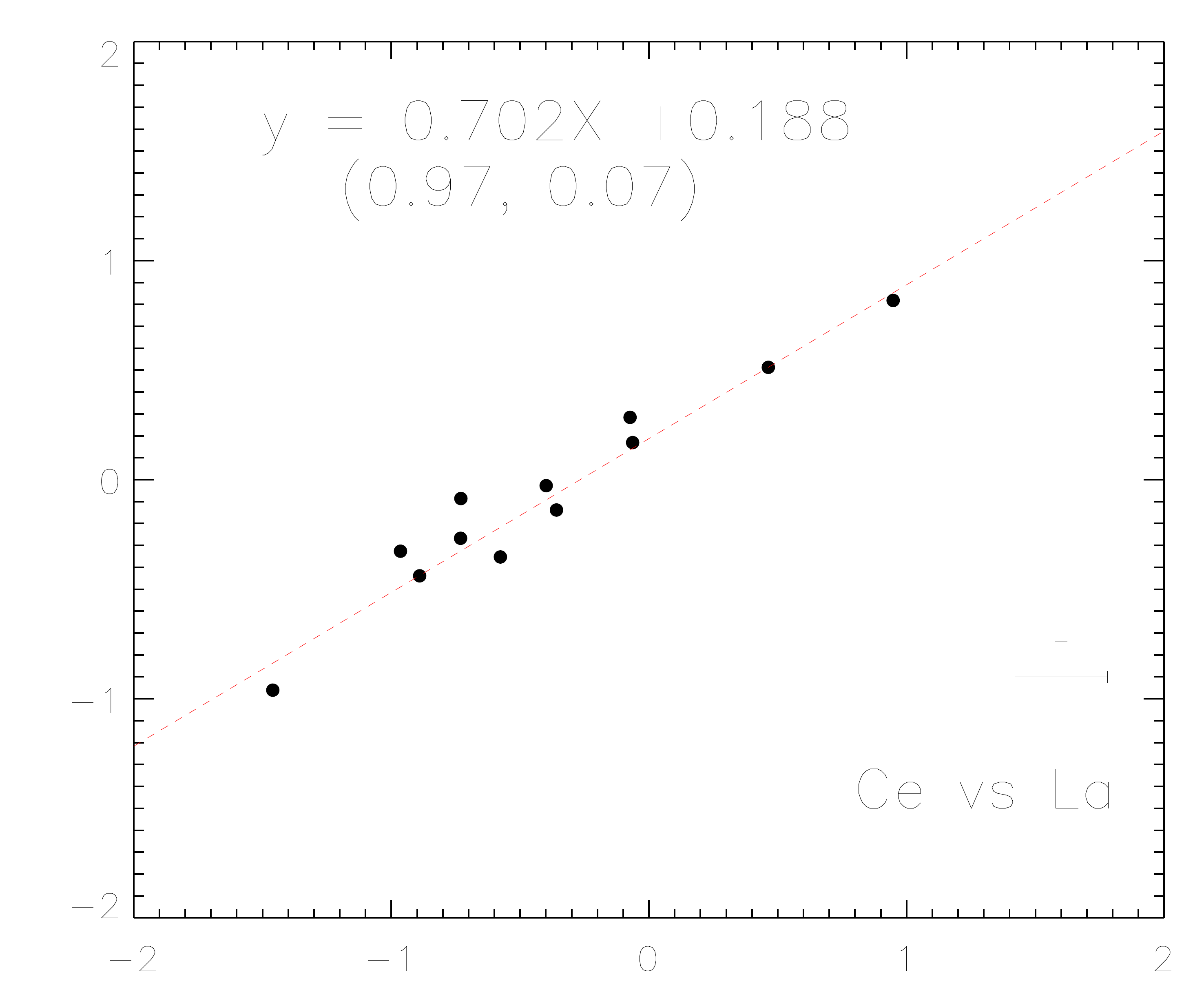}
\includegraphics[scale=0.19]{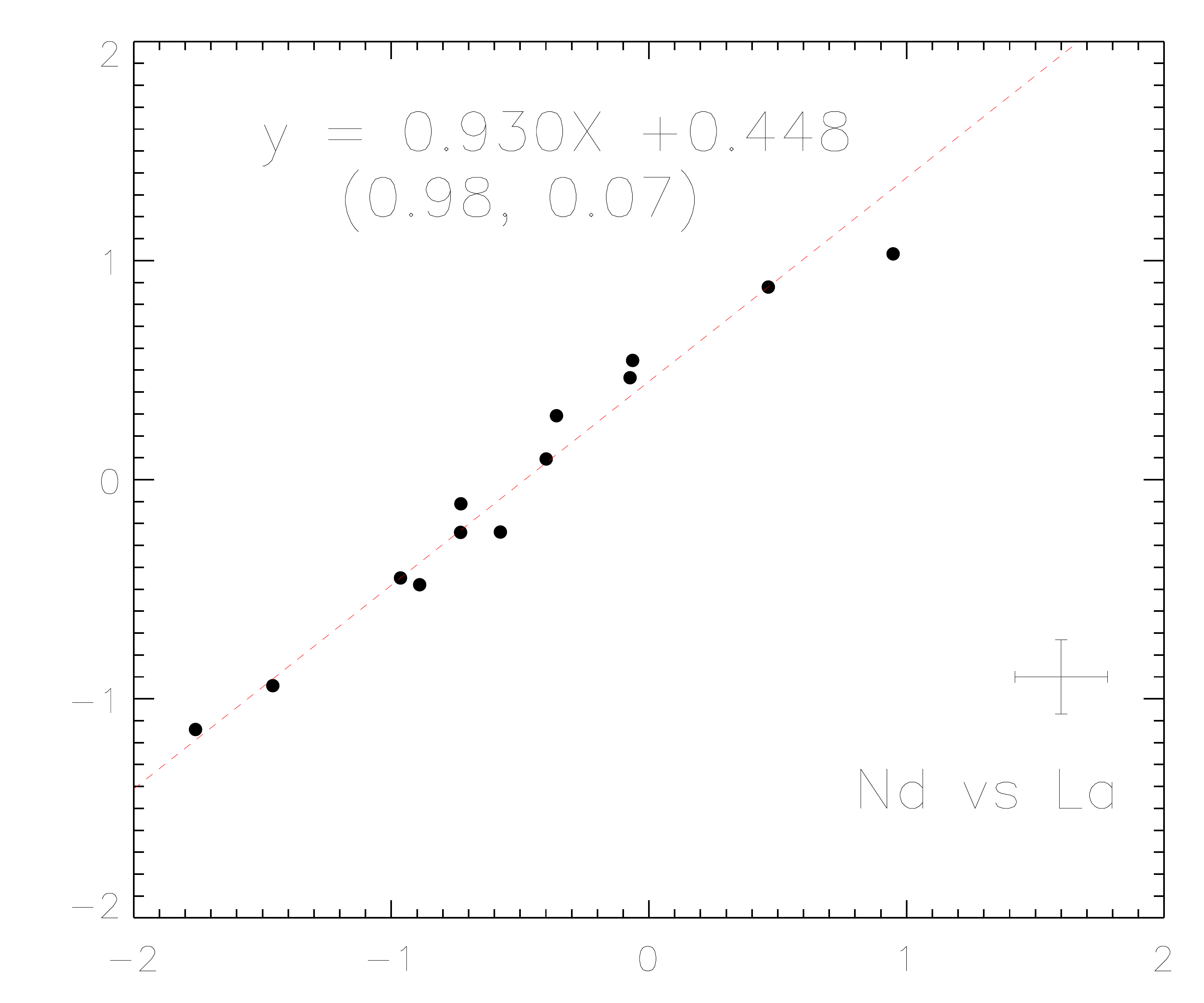}
\includegraphics[scale=0.19]{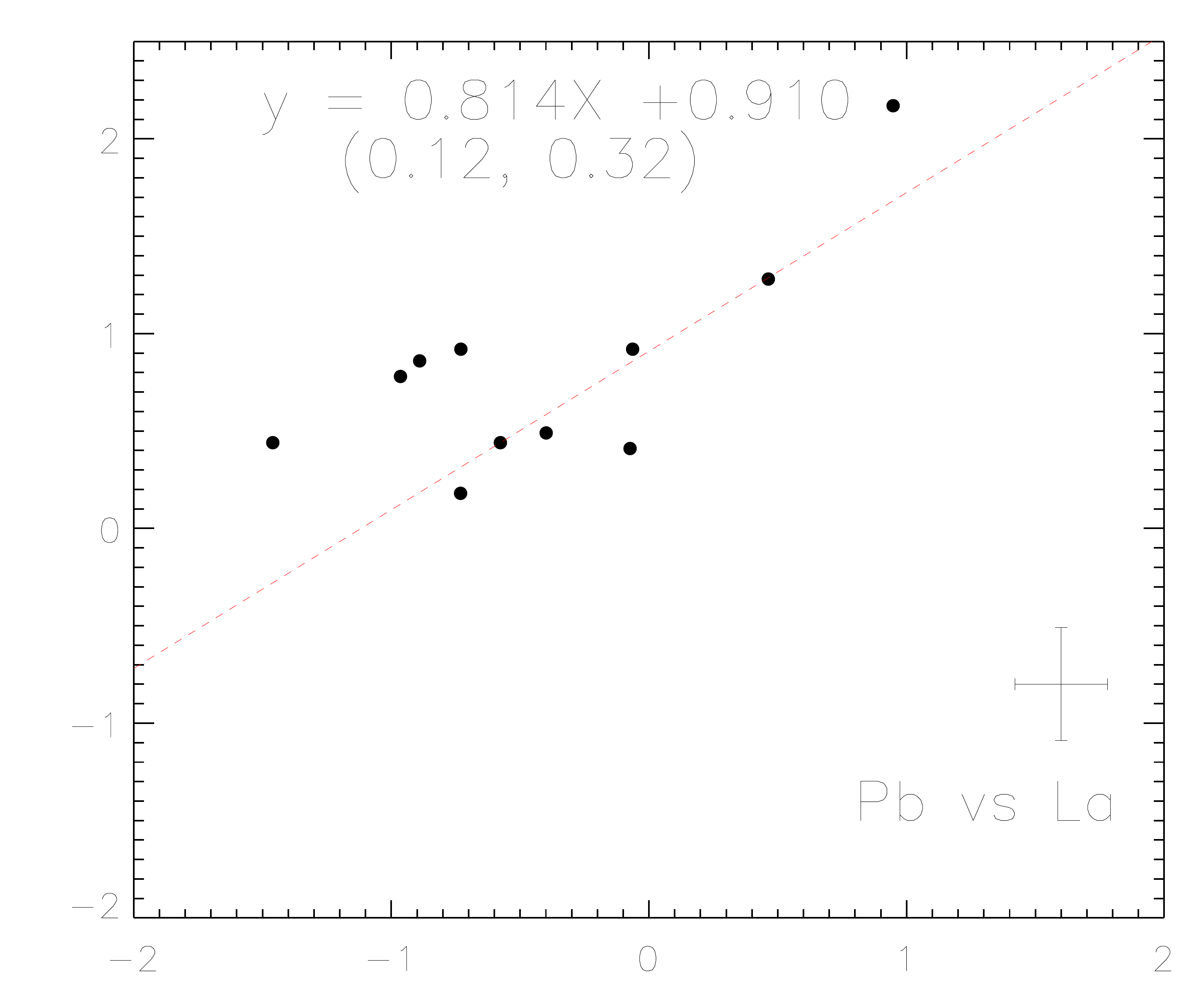}
\includegraphics[scale=0.19]{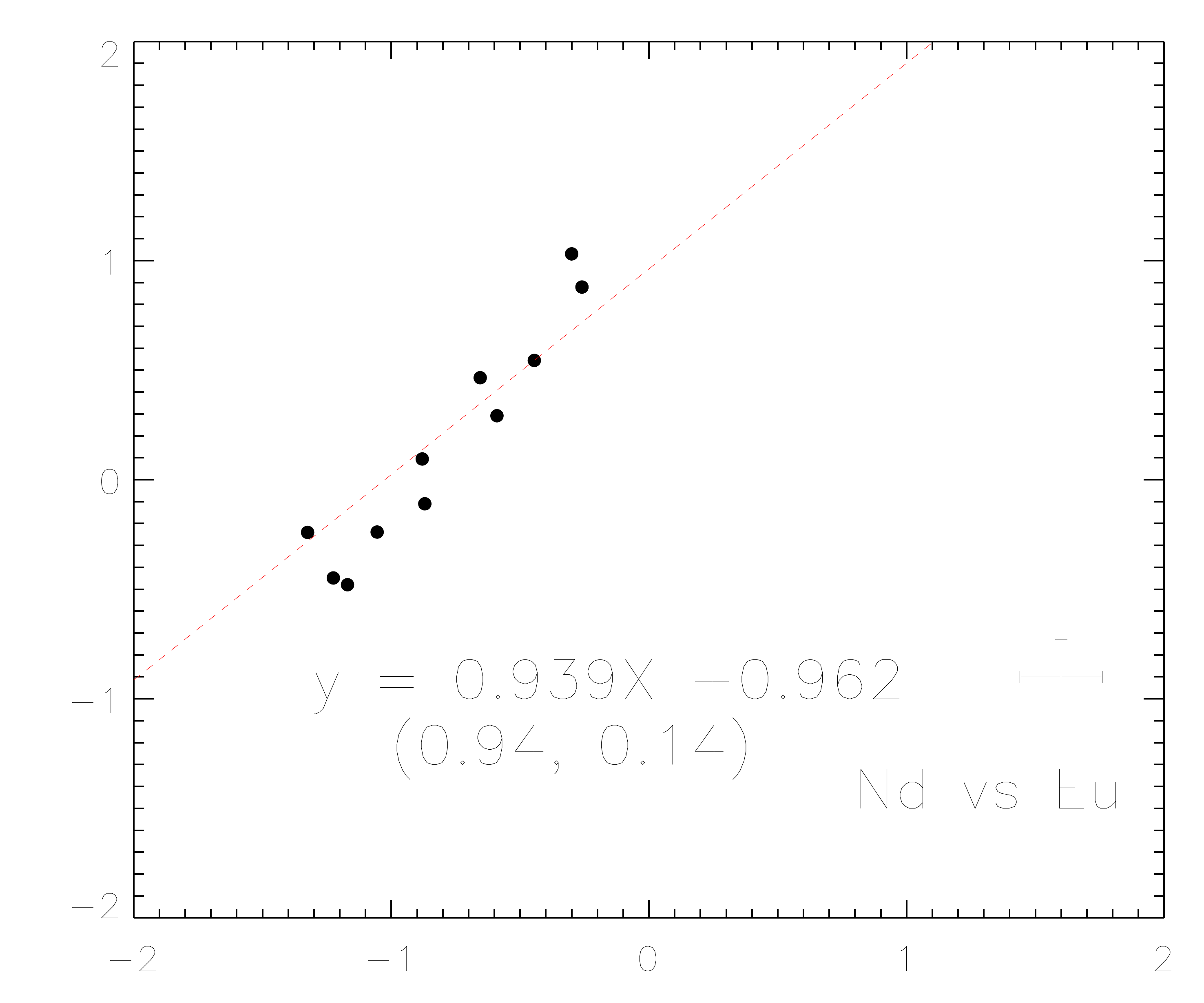}
\includegraphics[scale=0.19]{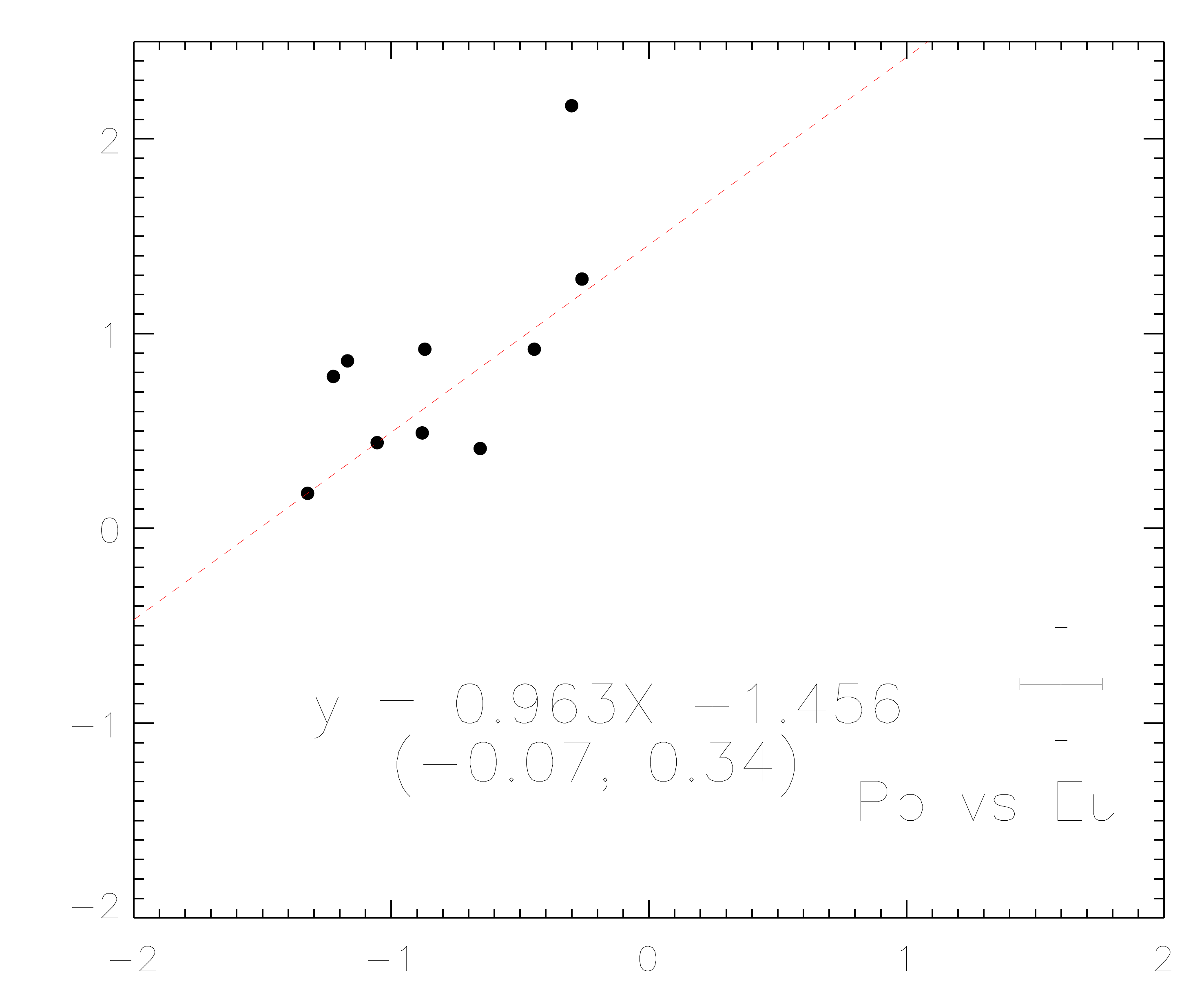}
\caption{Absolute stellar abundances of La (x-axis) vs Sr, Ba, Eu, Ce, Nd, Pb (y-axis), and Eu vs Nd, Pb (listed from top to bottom panel). Lines fitted to the data are indicted in the figures. In parenthesis the Pearson's r-value and the standard deviation on the distance to the line is provided.}
\label{Fig:trends}
\end{center}
\end{figure*}

\begin{figure*}[htp!]
\begin{center}
\includegraphics[scale=0.55]{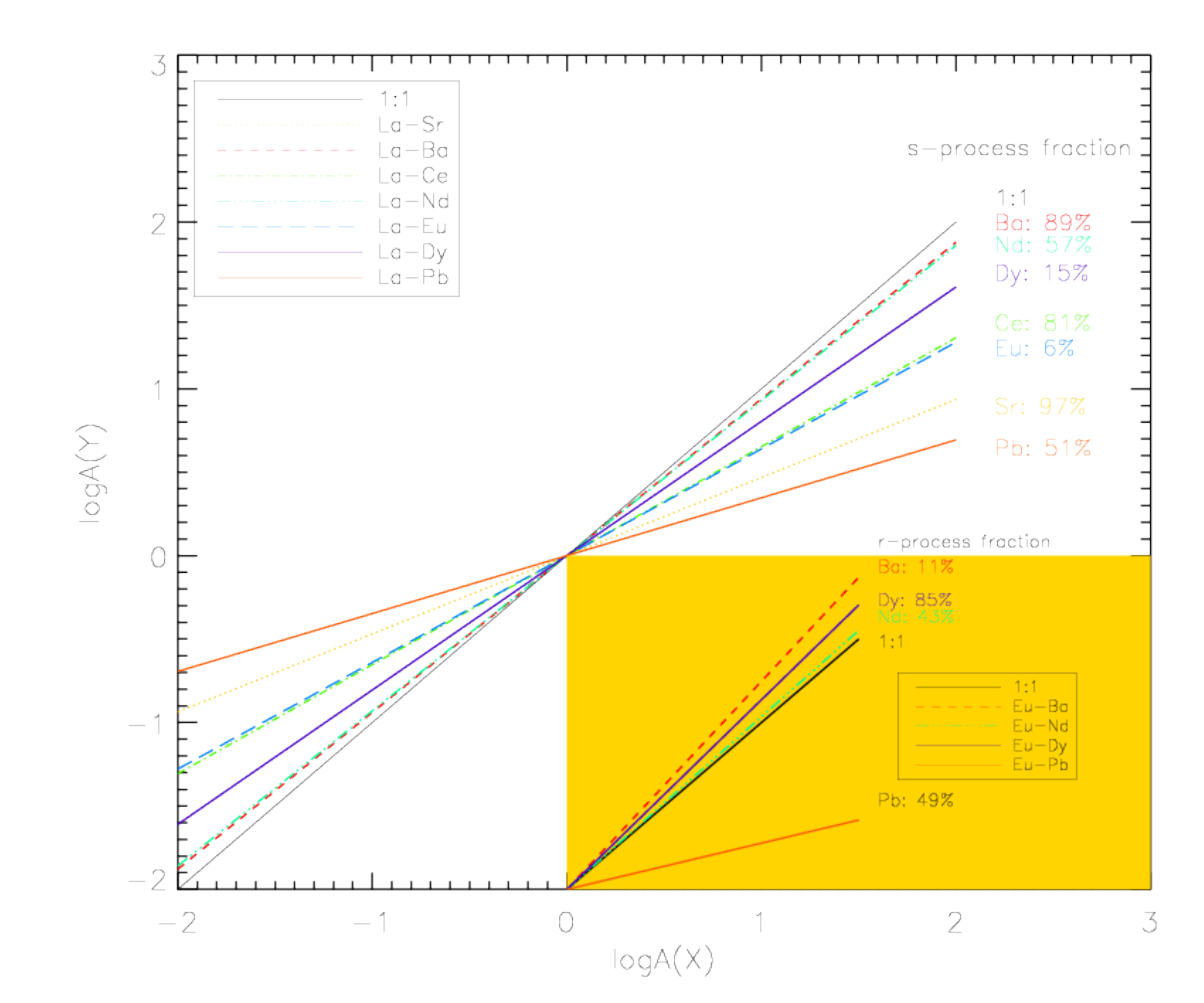}
\caption{Slopes fitted to Sr, Ba, Ce, Nd, Eu, Dy, and Pb with respect to La is shown along side with the fraction of main s-process material \citep[according to][]{Bisterzo2014}. The lower right corner shows selected linear trends between Eu and Ba, Nd, Dy, and Pb and their r-process fraction in percent.}
\label{Fig:slopes}
\end{center}
\end{figure*}

\begin{figure}[htbp]
\begin{center}
\includegraphics[scale=0.3]{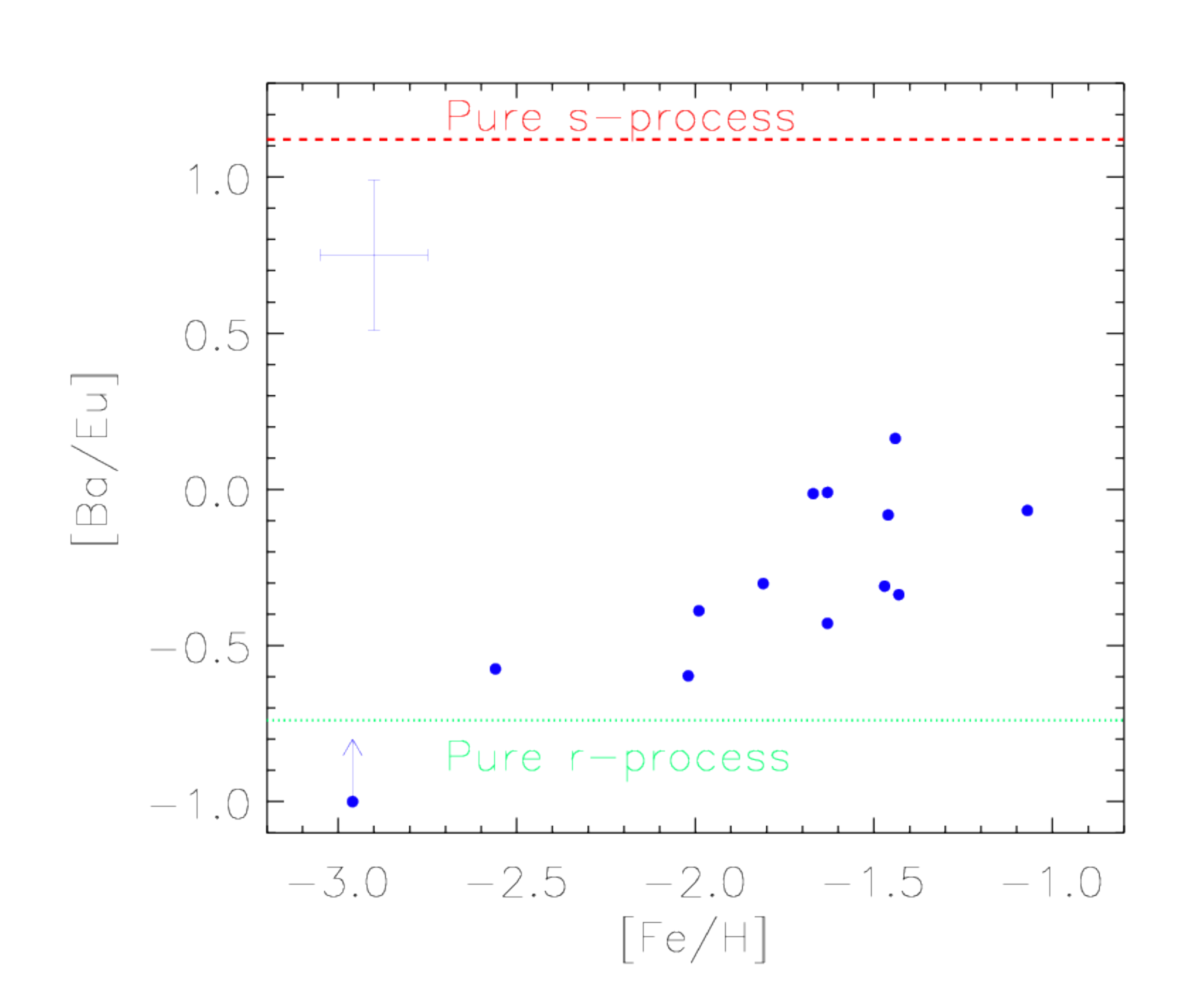}
\caption{[Ba/Eu] vs [Fe/H] for the sample stars. A pure r- and s-process content \citep{Arlandini1999} is indicated by dashed lines.}
\label{Fig:BaEu}
\end{center}
\end{figure}

As expected, two s-process elements like La and Ba show an almost perfect 1:1 correlation on an absolute abundance scale (see Figs.~\ref{Fig:trends}, \ref{Fig:slopes} within the standard deviation around the line), and La (s) vs Eu (r) show a linear trend clearly deviating from 1:1 with a slightly larger star-to-star scatter and a poorer Pearson's correlation coefficient (see Fig.~\ref{Fig:trends}, \ref{Fig:slopes}). Deviation from 1:1 and abundance scatter are clear indications of differing formation processes/origin as shown in \citet{Hansen2012,Hansen2014a}. In this regard it is puzzling that a typical s-process element like Ce does not correlate 1:1 with La, but good that the Pearson's r is 0.97 and the spread around the line is low (0.07). Neodymium ($\sim$50/50\% s/r) shows clear and almost equally good correlations with both Eu and La with high Pearson's r values which is encouraging. 
This could indicate that equal amounts of s- and r-process material have been mixed into these metal-poor stars (see also Fig.~\ref{Fig:BaEu}). Moreover, the light element Sr shows a very different origin compared to La, as does the heavy s element Pb compared to La. This indicates that different s-process formation sites are at work in Sgr forming different amounts of Sr, La, and Pb (see Fig.~\ref{Fig:slopes}). Alternatively, it could be an expression of different (evolving) physics in the same environment/object, e.g., different neutron density as a function of time. The poor Pearson's r and large scatter (standard deviation around the line) in lower panels of Fig.~\ref{Fig:trends} show that neither detections nor upper limits of Pb can be explained by the same formation channel creating Eu (which is reassuring and illustrates that the `trends method' works).

Figure \ref{Fig:slopes} shows how the abundances and elements studied here correlate with the s-process element La, and in the inserted lower right corner trends versus the r-process element Eu is shown for comparison. Indicated in the figure are the relative fractions of r- and s-process taken from the main s contribution from \citet{Bisterzo2014}. As indicated above a few surprises were found, the main one being Nd correlating stronger with La than Ce, despite La and Ce being closer in atomic number than La and Nd. Moreover, there is only a 10\% difference in main s-process contribution between La (75\%) and Ce (84\%) while Nd is only 57\% created by the s-process. This is hard to explain. We tested if there were differences in the weighted vs straight means but both resulted in very similar linear fits (a slight change on the second digit \comm{-- see also Sect.~\ref{sec:abun}}). The poorer agreement between La and Sr (a large scatter and a slope clearly different from 1) indicated that a larger fraction of Sr could be formed by the weak s-process than accounted for in the models. Similar observational indication of the weak s-process making a larger contribution to the production of Sr was also found in \citet{Hansen2012,Hansen2014a}.

\begin{figure*}[htbp]
\begin{center}
\includegraphics[scale=0.5]{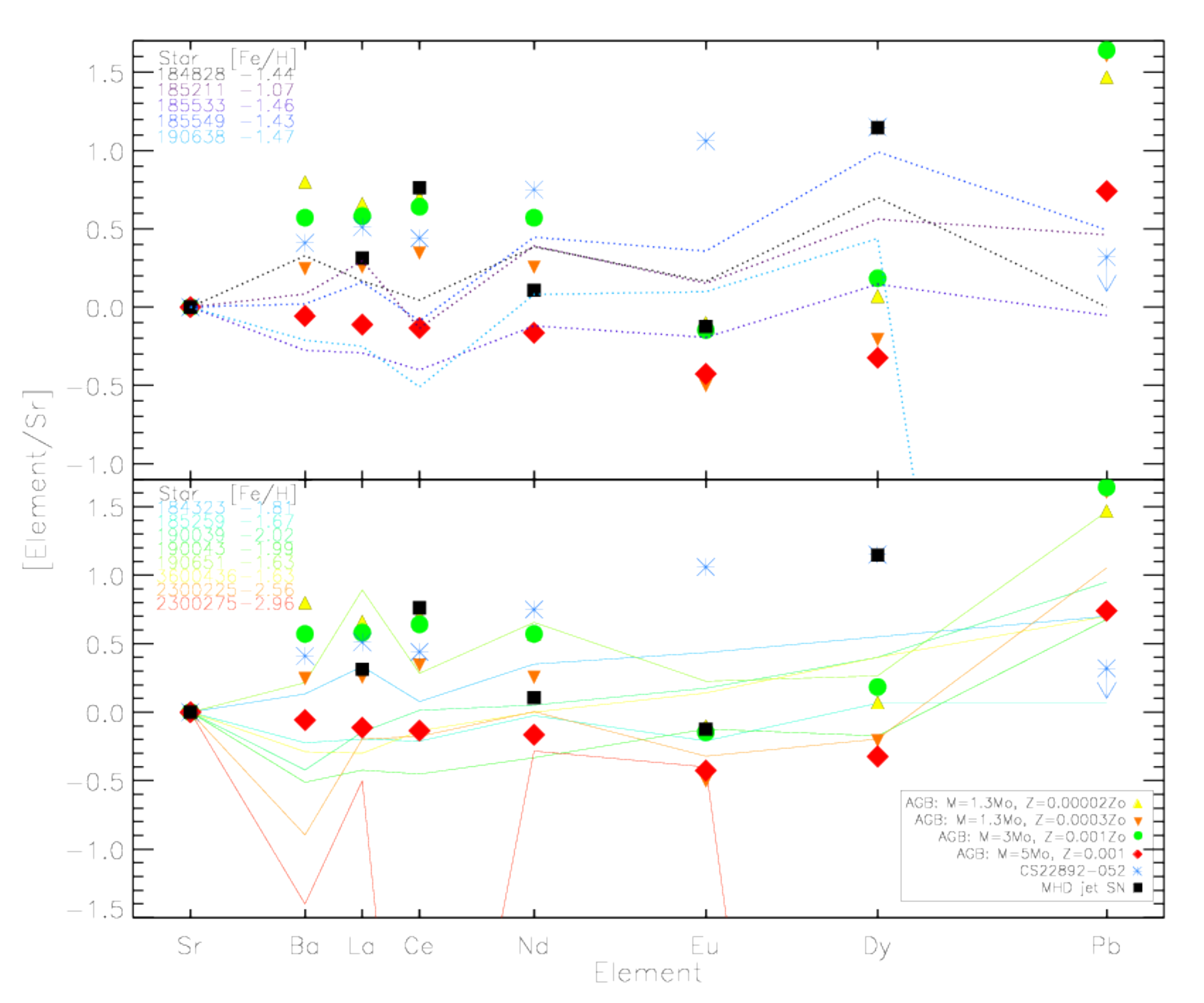}
\caption{Abundance patterns of our Sgr stars normalised to Sr and compared to AGB \citep{Cristallo2011} and MHD jet supernova model \citep{Winteler2012,Heger2005} predictions as well as the r-rich star CS22892-052 \citep{Sneden2003}. (Symbol size increase as a function of mass. For details see legend.)}
\label{Fig:pattern}
\end{center}
\end{figure*}
To find the origin of the elements we compare the observationally derived abundances to model predictions from AGB stars (from the F.R.U.I.T.Y. database; \citealt{Cristallo2011}) and magneto-hydrodynamic driven supernovae of 15\,M$_{\odot}$ exploding with jets \citep{Winteler2012,Heger2005}. This kind of SN might host an r-process. Based on previous studies (e.g., \cite{Letarte2010,McWilliam2013}) the very metal-poor, low-mass AGB stars were predicted to enrich the more metal-rich Sgr stars, and we therefore select AGB model predictions with Z=0.00002Z$_{\odot}$, 0.0002Z$_{\odot}$, 0.0003Z$_{\odot}$ and 0.001Z$_{\odot}$ which corresponds to [Fe/H] =$-2.8, -1.8, -1.6$, and $-1.2$ to test this. The former is the most metal-poor model available in the database and the latter the highest value the AGB star could have had if its material should be incorporated into any star from our sample. The top panel of Fig.~\ref{Fig:pattern} shows the abundance pattern for our sample stars with [Fe/H] $>-1.5$ where both supernovae of type Ia and II can, in principle, have enriched the ISM. This panel shows a much lower star-to-star scatter than the bottom one illustrating more metal-poor stars. It hints that the ISM of Sgr was likely to be homogeneous at metallicities above $-1.5$, and inhomogeneous below this value. Surprisingly, both the metal-poor and the more metal-rich stars in our sample show a better agreement with the intermediate massive 3-5\,M$_{\odot}$ AGB stars than the lower mass AGB except for \object{Sgr J190651.47-320147.23} which agrees well with both of the 1.3-3\,M$_{\odot}$ AGB models (this is investigated further below and in Fig.~\ref{Fig:HSLS}). None of the stars provide a good fit to the most metal-poor, low-mass AGB star. Again in stark contrast to \citet{McWilliam2013} and \citet{Letarte2010}.
Moreover, some of the metal-poor stars  ($-2.5 <$[Fe/H]$<-1.5$) show some agreement with the MHD jet SN (except for the Dy abundances) indicating supernova or mild r-process enrichment but not to the extent of the well studied r-rich star, CS22892-052 \citep{Sneden2003}. A mixture of (r+s) formation sites  is clearly needed to explain the chemistry of metal-poor as well as metal-rich Sgr stars (see also Fig.~\ref{Fig:BaEu}).

\subsection*{More massive progenitor stars}
\begin{figure*}[htbp]
\begin{center}
\includegraphics[scale=0.5]{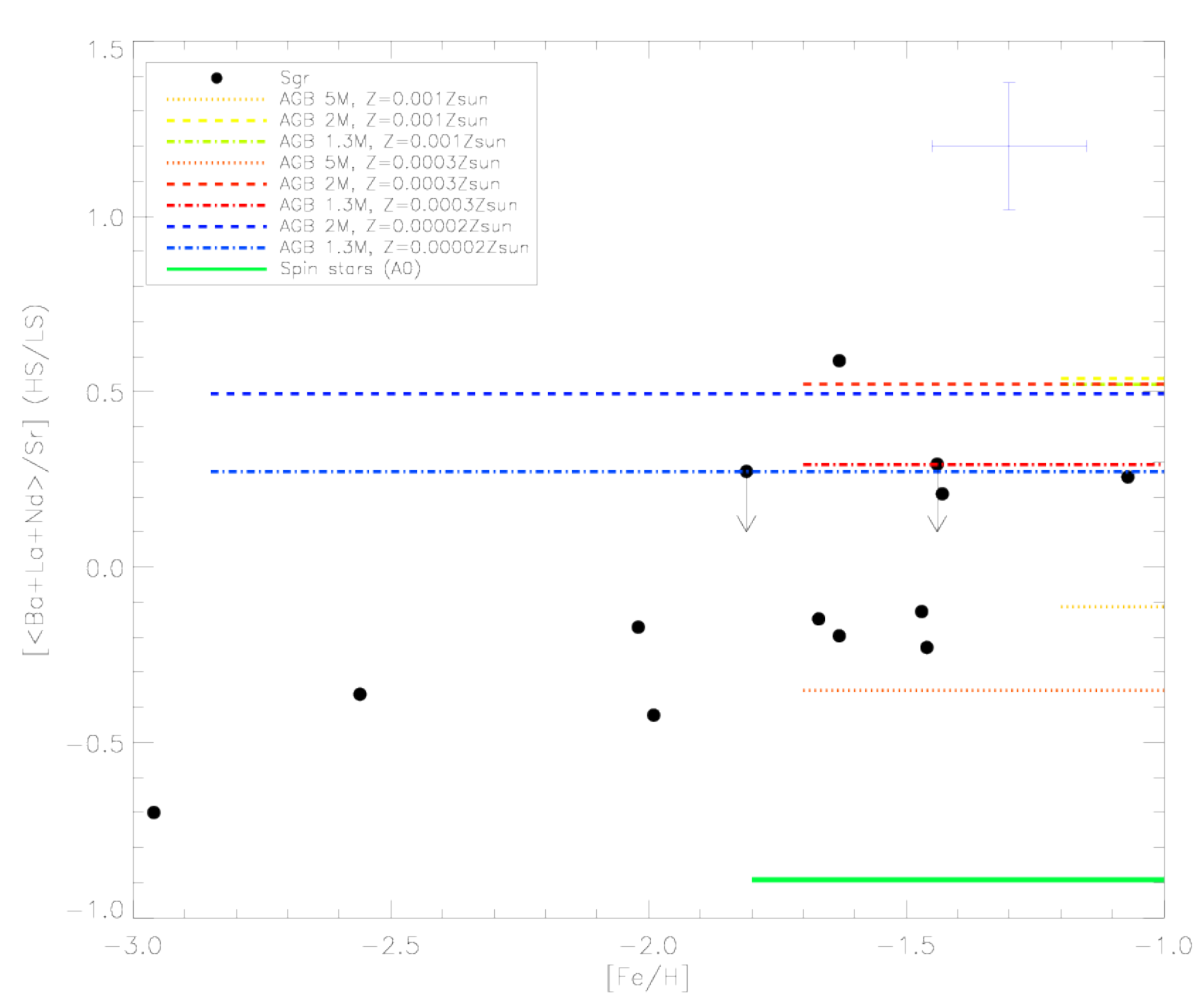}
\caption{[HS/LS] vs [Fe/H] for the sample stars compared to AGB yields \citep{Cristallo2011} with three different masses (1.3,2,5\,M$_{\odot}$) and metallicities corresponding to [Fe/H]=$-2.8, -1.6,-1.2$ as well as \comm{massive} stars \citep{Frischknecht2012}. The star with the highest HS/LS-ratio is Sgr J190651.47$-$320147.23. }
\label{Fig:HSLS}
\end{center}
\end{figure*}

As indicated by the higher $\alpha-$abundances and stellar abundance patterns, the enrichment in Sgr seems to have been of a more massive stellar progenitor population than previously believed. This includes both supernovae and AGB stars. The level of s-process enrichment depends on the mass of the AGB star, which we explore here using a standard $^{13}$C-pocket, no rotation, and the final yield composition of 1.3, 2, and 5\,M$_{\odot}$ AGB stars from the F.R.U.I.T.Y. database. We combine our Ba, La, and Nd into a heavy s-process (HS) tracer (in line with the approach of the database and literature) \comm{and use Sr to represent the light s-process (LS)}. We compare to the final elemental abundances from their model predictions using different metallicities (Z=0.00002Z$_{\odot}$ -- 0.001Z$_{\odot} \sim$ [Fe/H] = $-2.8$ to $-1.2$). In addition we include the yields from massive (25\,M$_{\odot}$, \comm{not} rotating) stars with a standard $^{17}$O reaction rate \citep{Frischknecht2012}. This means that fewer neutrons are available than if the  $^{17}$O-rate is reduced, since these reactions work as neutron poisons in reducing the number of neutrons available for creating s-process material. Moreover, \citet{Frischknecht2012} also showed that an increased rotation would  produce and burn slightly more $^{22}$Ne at the end of the convective He-core burning. Since the massive stars are hot enough to activate $^{22}$Ne($\alpha$,n)$^{25}$Mg reactions, this would lead to a larger production of s-process material than a non-rotation case. We select a non-rotating model with a normal (high) $^{17}$O reaction rate to get the lowest level of s-process enrichment from massive stars for this comparison \citep[this case corresponds to case A0 from ][see also Fig.~\ref{Fig:HSLS}]{Frischknecht2012}.  

As illustrated by Fig.~\ref{Fig:HSLS} only one star falls above the high heavy to light s-process (HS/LS) enrichment from a 2\,M$_{\odot}$, 0.001Z$_{\odot}$ AGB star. This is Sgr J190651.47-320147.23 which is discussed in more detail below (Sect.~\ref{sec:19065}). Four stars (two detections, two limits) agree with the low-mass 1.3\,M$_{\odot}$ AGB stars of different metallicities, while the remaining part of the sample lies between the bottom floor set by the massive (25\,M$_{\odot}$) non-rotating star and the intermediate mass (5\,M$_{\odot}$) AGB star. This shows that more massive (intermediate-mass) AGB stars and short lived $15-25$\,M$_{\odot}$ stars or even jet driven SN are needed to explain the chemical composition of the (very) metal-poor stars in our sample. Previous studies were limited to more metal-rich stars and could therefore, based on smaller (limited) samples, have drawn other conclusions \citep[e.g.,][]{McWilliam2013}.

\begin{figure}[htbp]
\begin{center}
\includegraphics[scale=0.2]{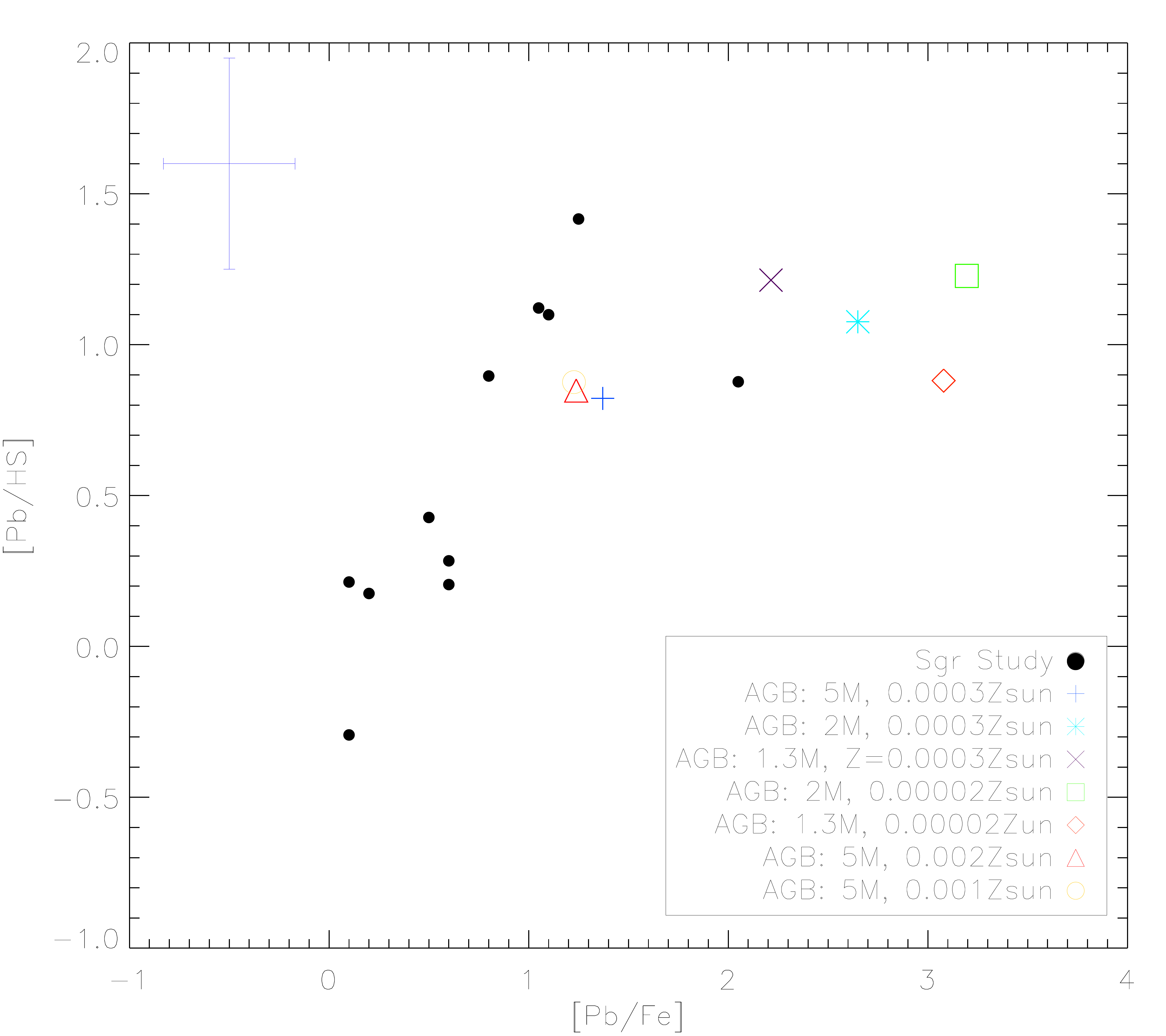}
\caption{[Pb/HS] vs [Pb/Fe] where HS = $\langle[$Ba,La,Nd/Fe$]\rangle$. AGB yields \citep{Cristallo2011} from stars with M=1.3, 2, 5\,M$_{\odot}$ and different metallicities in Solar units - see legend.}
\label{Fig:PbHS}
\end{center}
\end{figure}

Even though we mainly determine upper limits for Pb we compare our abundances to the same AGB predictions, and similarly find that only 1 star is in vicinity of the prediction from a 1.3\,M$_{\odot}$ AGB star, while the same four stars as shown in Fig.~\ref{Fig:PbHS} come closer to the Pb predictions from the 5\,M$_{\odot}$ AGB stars. Here we note that the metallicity of the AGB star might play a secondary role compared to its mass, as all the 5\,M$_{\odot}$ AGB stars regardless of their metallicity match (within the uncertainty) these four stars. 
In contrast to \citet{Letarte2010, McWilliam2013} who found high [Ba/Y] or [La/Y] (0 -- 1\,dex) we calculate [Ba/Sr] values in the range $-1.4$ to 0.3\,dex (or $-0.7< $[HS/LS]$<0.5$ cf. Fig.~\ref{Fig:HSLS}). For a few of the stars we find values above Solar, which is also in agreement with some degree of metal-poor AGB enrichment (with fewer seeds leading to more HS than LS). However, more than half of our sample show sub-Solar HS/LS values and therefore seem to need more seeds in a metal-rich AGB environment (or a more massive AGB star). This is in agreement with Figs.~\ref{Fig:pattern} -- \ref{Fig:PbHS}. 

The [Ba/Eu]-ratio we derive (typically below 0 -- see Fig.\ref{Fig:BaEu}) deviates from previous findings in both Sgr and Fornax  \citep{McWilliam2013,Letarte2010}. \comm{We} obtain an r-process fraction of $\ge50$\% which is much larger compared to the 10\% estimated by \citet{Letarte2010} in Fornax (which is the most massive dSph after Sgr). Restricted by the signal-to-noise ratio at the Eu lines, we also find the first pure r-process candidate at [Fe/H]$\sim -3$ namely Sgr 2300225 (see Sect.~\ref{sec:2300225} below).

\subsection{An r-process dominated star in Sgr: Sgr 2300225}
\label{sec:2300225}
The most metal-poor star with [Fe/H] =$-2.96$ in our sample shows a remarkable abundance pattern for a star in Sgr. Overall its chemical composition resembles that of MW halo stars, except from a low (uncertain) Ca abundance. One line is in the gap between the CCDs and the other line is rather noisy. It is the first star with an overabundant [Co/Fe]-ratio, and with its low barium abundance and upper limit of europium, it is well below the 'pure r-process' prediction in Fig.~\ref{Fig:BaEu}.
A very high [Sr/Ba] = 1.4 is rare and only two stars from the metal-poor sample presented in \citet{Francois2007} had [Sr/Ba]$>1$. They suggested an additional production site for Sr -- Ba at these low metallicities.
From Figs.~\ref{Fig:pattern} and \ref{Fig:HSLS} Sgr 2300225 is seen to have a very low Ba/Sr-ratio which could either indicate that it has been enriched by a massive, fast rotating star (it is slightly above the spin star A0 case, \citealt{Frischknecht2012}) albeit located at a very low metallicity. The exact origin of this star would need further high-resolution follow-up observation, but we speculate that, e.g., $\nu-$driven winds from a massive supernovae  may be possible formation site. An alternative would be winds from a massive (extremely metal-poor) AGB star. However, if an AGB star is responsible for the enrichment of Sgr 2300225, the low  [Ba/Fe]$=-0.8$ will be difficult to explain and the material would need to created and yielded below [Fe/H] $=-3$. We emphasise that none of the yield predictions in Fig.~\ref{Fig:pattern} provide a satisfactory explanation to the stellar abundances of Sgr 2300225.

\subsection{An s-process enhanced star: Sgr~J190651.47-320147.23}
\label{sec:19065}
\begin{figure}[htbp]
\begin{center}
\includegraphics[scale=0.55]{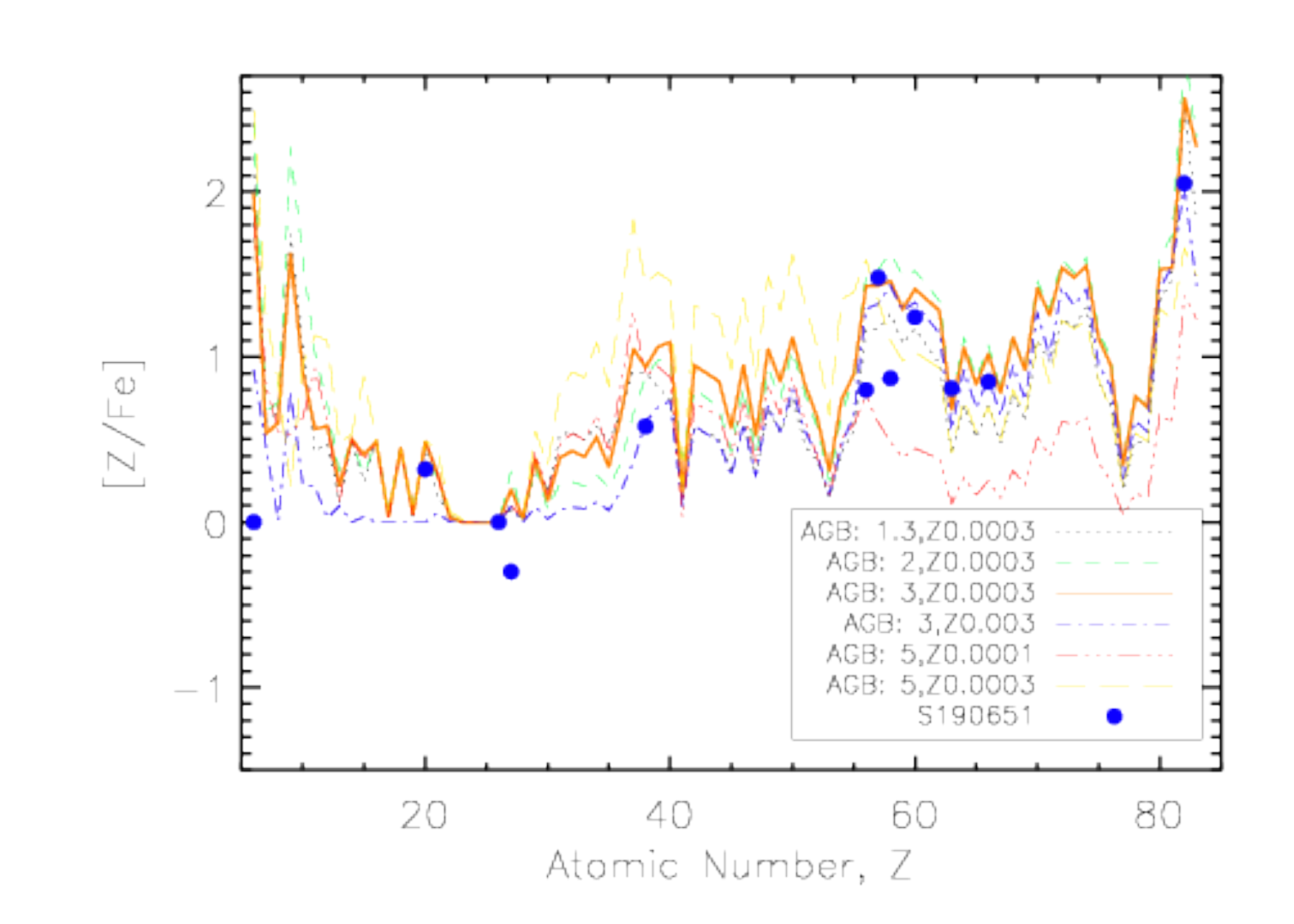}
\caption{Abundances of Sgr J190651.47-320147.23 compared to AGB yields \citep{Cristallo2011} of various mass and metallicity (see legend for values in M$_{\odot}$ and Z$_{\odot}$.)}
\label{Fig:19065}
\end{center}
\end{figure}

As seen from Table~\ref{tab:abun} and several figures above, one star sticks out as being the most s-process enriched star in notably Pb (just above 2\,dex), but also showing large La and Nd abundances. Therefore, we single out this star and compare it to AGB yields of varying mass and metallicity (see Fig.~\ref{Fig:19065}).
With our abundances spanning a broad range of atomic numbers we  have different ways of chemical tagging Sgr J190651.47-320147.23, either using single elements, or the stellar pattern. Starting from the lightest element, carbon, we find at  a [Fe/H]$=-1.63$ this star is a clear outlier in a [C/Fe] diagnostics diagram. It is not enhanced with respect to the Sun, but it is the most C-rich star in Sgr below [Fe/H] = $-1.0$ known to date. We have argued before that probably all of our stars have undergone some amount of internal mixing, resulting in the destruction of carbon. It seems very difficult to admit that this particular star has not undergone the same process as the other stars of similar luminosity.
To estimate the original C-composition of this evolved giant, we adopt an approximate correction from Figure 15 in \citet{Placco2014b} in which computations for a [Fe/H]=$-1.3$ star is presented. We assume the distance to Sgr (as outlined in Sect.~\ref{sec:results}) and use the apparent V magnitude (16.5) to estimate the $\log($L/L$_{\odot}) = 2.2$, resulting in a correction of $+0.4$\,dex. This would bring the current [C/Fe]=0 to 0.4, however, this is still not sufficiently high to classify this star as a carbon enhanced metal-poor (CEMP) star. For the star to be a CEMP it must be metal-poor ([Fe/H]$<-2$) and C-rich ([C/Fe]$>0.7$, see, e.g., \citealt{Hansen2016}). However, the star could belong to the more metal-rich counter parts, namely CH stars, but even for this class the C abundance is a bit low.
Interestingly, the largest sources of C are AGB and Wolf Rayet stars through stellar winds \citep{Kobayashi2006}. \citet{Hasselquist2017} state that the C yield from supernovae are mass sensitive, which would argue against a massive progenitor generation. However, Sgr is strongly polluted by AGB stars and only metal-rich, massive ($\sim6$\,M$_{\odot}$) AGB would produce some of the lowest C yields, which are still slightly above Solar in [C/Fe] \citep[in agreement with][]{Kobayashi2006}. This confirms a massive, metal-rich AGB progenitor for Sgr J190651.47-320147.23 in agreement with our results drawn from the other elements. We note that despite having derived C from the G-band using molecular equilibrium, the final [C/Fe] might still be uncertain and we therefore weigh results from other (heavy) elements higher in our conclusions.

Figure~\ref{Fig:19065} shows only the yields from a mass and metallicity combination that come close to the derived stellar abundances of Sgr J190651.47-320147.23. We see that all heavy elements but Ba and Ce fit the 3\,M$_{\odot}$  AGB model with a metallicity of [Fe/H] = $-1.7$ which is very close to the metallicity of Sgr J190651.47-320147.23 ($-1.63$). The close match in [Fe/H] does not allow for much delay time to incorporate the AGB ejecta into the following generation of low-mass stars (here Sgr J190651.47-320147.23). However, it indicates that this star was not enriched by very metal-poor AGB stars as only one of the other stars in our sample might be. Since Ce shows an odd behaviour most likely due to blends which we have not been able to remove completely, neither in the synthesis nor in our weighting scheme, it is not surprising that Ce is not well described by any of the yields for Sgr J190651.47-320147.23. However, the low Ba abundance is closer to the 5\,M$_{\odot}$  AGB model (Z=0.0001\,Z$_{\odot} \sim$ [Fe/H] = $-2.15$). For Sr and Pb, the more metal-rich 3\,M$_{\odot}$  AGB model provides a perfect fit to these two elements. This is however very unlikely as the AGB metallicity ($-0.63$) exceeds that of the low-mass, observed star. In any case, the best fit to Sgr J190651.47-320147.23 remains the 3\,M$_{\odot}$  AGB model at [Fe/H]=$-1.7$ as confirmed by the smallest $\chi^2$. The large Pb abundance of 2.05\,dex (an actual detection) is slightly overproduced by the preferred AGB model, which is a common problem \citep{Bisterzo2014,Cristallo2015}. The combination of mass and metallicity of the most likely progenitor AGB star, might have contributed to the special (enriched) s-process pattern of Sgr J190651.47-320147.23. Combining this with the radial velocities, we cannot reject that Sgr J190651.47-320147.23 could have resided in a binary system, and more follow-up observations would be needed to test this. For this star we also have upper limits for Th and we were therefore able to determine an approximate age for this star (see Sect.~\ref{Sec:Th} below).

\subsection{Thorium and ages}
\label{Sec:Th}
Nuclear cosmochronology has been conducted to derive stellar ages using a variety of elements for the past three decades \citep[e.g.][]{Butcher1987,Cowan1991,Sneden1996,Cowan1999,Cayrel2001,Truran2001,Hill2002,Kratz2007,Aoki2007}. We derive Th in five stars and upper limits in two stars belonging to the main body of Sgr. We focus on Th and ages from Th/Eu as we expect the stars to be old, and comment on the uncertainties associated with the abundances and ages derived from the 4019.1\,\AA~ line. We consider this line better suited for determining ages than the 4086\,\AA~line. Details on the blends and line list can be found in Sect.~\ref{sec:abun} and in Table~\ref{tab:online}. Owing to the blends and difficulty in placing continuum at $\sim4000$\,\AA~ uncertainties of 0.05 - 0.1\,dex in Th II are derived for all our stars. This translates into $\pm2$\,Gyr, which is in agreement with the findings of \citet{Ludwig2010}. However, in addition to the uncertainties from observations and model atmospheres, uncertainties in the nuclear prediction of the formation of the radioactive $^{232}$Th isotope with a half life of $\tau_{1/2} = 1.405\cdot 10^{10}$yrs \citep{Cowan1991,Kratz2007} arise. These nuclear uncertainties relate, e.g., to the $\beta-$ and $\alpha-$decay rates and $\beta-$delayed fission  and cause an uncertainty of $\sim2$\,Gyr \citep{Cowan1991,Cowan1999, Schatz2002,Otsuki2003}.

Earlier studies of 3D effects and corrections to Th II \citep{Caffau2008Th} found average values (corrections) for the Sun to be of the order $-0.1$\,dex. \citet{Mashonkina2012} calculated NLTE corrections of $\sim 0.1$\,dex for the same line but for giant stars with parameters closer to our stars. This could indicate that the corrections to the 1D, LTE abundances might cancel out for 3D, NLTE, but full calculations using the adequate stellar parameters would need to be carried out to test this (which is beyond the scope of this paper). Hence we continue using our 1D, LTE values to derive ages, which we consider accurate to within $2.5-2.8$\,Gyr (with the observational and nuclear uncertainties added in quadrature).

To calculate the ages ($\Delta$t) we start out with $Y_r(\Delta t) = Y_r(0) \cdot exp(-\Delta t/\tau_r)$ where Y is the yields/abundances, and we use the formula:
\begin{equation}
\Delta t = 46.7 [log(Th/Eu)_0 - log(Th/Eu)_{obs}]
\label{Eq:Th}
\end{equation}
where 46.7 is ln(10) times the mean lifetime (20.3\,Gyr) in Gyr ($\tau_{1/2}/ln(2)$), \comm{and} (Th/Eu)$_{0}$ is the initial produced Th ratio, where we adopt the predicted value of 0.507 from Table 4 in \citet{Cowan2002} resulting in a log(Th/Eu)$_0 = -0.295$. The last term in Eq.~\ref{Eq:Th} is the absolute abundances derived from the present-day, observed abundance. 
For three of the stars we derive 'realistic' ages ($\Delta T < 14$\,Gyr) and these are listed in Table~\ref{tab:Th}. These stars are the more metal-poor ones where the blends are less severe than in the more metal-rich stars in our sample. By adopting a different value for (Th/Eu)$_0$ using the meteoritic values as representation of the initial Solar system r-process fraction from \citet{AndersGrev}, the ages would be $\sim 1.6$\,Gyr lower, which is within the adopted uncertainty.

\begin{table}[htp]
\caption{Ages and metallicities for three sample stars}
\begin{center}
\begin{tabular}{lcc}
\hline
Star & [Fe/H] & age [Gyr] \\
\hline
\object{Sgr J184323.07-290337.64} & $-1.81$  & $9.6\pm 2.8$ \\
\object{Sgr J190043.03-311704.33} & $-1.99$ & $7.2\pm2.3$ \\
\object{Sgr J190651.47-320147.23} & $-1.63$ &$>8.2$\\
\hline
\end{tabular}
\end{center}
\label{tab:Th}
\end{table}%
  
We note that the ages listed in Table~\ref{tab:Th} are likely to be low (the stars are most likely older) owing to uncertainties from observations and nuclear physics. However, within the uncertainties these values are in agreement with \citet{deBoer2015}, predicting that the faint old stream stripped about 9\,Gyr ago from Sgr main body consists of stars with [Fe/H]$\leq-1.3$ coinciding with the $\alpha-$knee. It would therefore make sense if our sample stars from the main body, from which the stripped stars originate, with their slightly lower metallicities, would at least be around 9\,Gyr. Moreover, the ages we derive are slightly lower (i.e., the stars are younger) than the predictions by \citet{Siegel2007}, who found ages of 9-13\,Gyr for the metal-poor population of Sgr centred around [Fe/H]$=-1.3$. This agrees with our assessment, that since our stars are more metal-poor we should likely have found slightly higher ages (i.e., the stars might be older than predicted).
We note that we could not derive U from the spectra and since most Pb abundances are merely upper limits, we constrain our age determinations to the Th/Eu estimates despite the large gap in atomic number between Th and Eu leading to its slightly poorer value as stellar clock.
 
\section{Conclusion}
\label{sec:concl}
Dwarf galaxies are often studied to understand if they could be the building blocks in a hierarchical merger scheme, where smaller systems merge and build up larger systems like the MW \citep{Searle1978,Dekel1986,Bullock2005}. Moreover, the smaller dwarf galaxies are often thought to be simpler as a poorer gas reservoir and gas loss place constraints on the mass of the objects that facilitated the chemical evolution of such dwarf galaxies \citep{Hyde2012}. Here we have analysed one of the best study cases for chemical evolution and galaxy formation by studying Sagittarius, which has merged with the MW, and is thought to carry a unique chemical imprint based on previous studies. However, our study of the most metal-poor stars associated with Sgr dSph has shown a chemical composition somewhat more similar to that of the MW halo and not lending support to a top-light IMF. We find a high level of s-process enrichment as found in earlier studies, however, our Co-values despite (mainly) being sub-Solar are at the metal-poor end higher than previously found, but most notably we find a clear $\alpha-$enrichment and a strong contribution from a main r-process (with Sgr 2300225 at [Fe/H]$\sim-3$ indicating a pure r-process origin). This is the first time in a metal-poor sample that clear abundance enhancements are found in Sgr stars. However, still no extreme r-rich star like CS22982-052 has been discovered in Sgr.

The [Ba/Eu] ratio as a function of [Fe/H] show that the r-process contributed more than 50\% of the heavy elements to the most metal-poor stars of our sample. Combining this with the average Ca/Co-ratio and SN yield predictions a clear presence of massive stars (15-25\,M$_{\odot}$ SN) has been shown using two different abundance ratios and \comm{they} can explain both Ca ($\alpha$) and Eu (r-process).

Sgr seems to host stars with a broad range of s-process enhancements spanning from the s-rich star Sgr J190651.47-320147.23 to s-normal very metal-poor stars (such as Sgr 2300275). Sgr J190651.47-320147.23 could have been polluted by a 3\,M$_{\odot}$ with [Fe/H] $\sim-1.7$; more observations are needed to probe the binary nature of this star. Also stars with a much lower s-process content are found at metallicities below [Fe/H]=$-1.5$ pointing towards an inhomogeneous early ISM of Sgr. Most of our sample stars are best mimicked by AGB stars of intermediate mass ($\sim5$\,M$_{\odot}$), while the metallicity seems to be a secondary factor, yet with a tendency towards the higher metallicity variety. We find a few stars with high [Ba/Sr] or high [HS/LS] ($>0$), yet more than half of our sample stars have [Ba/Sr]$<0$ down to a record low $-1.4$. This shows that a range of AGB stars (with high and low metalliticy combined with masses of 3-5\,M$_{\odot}$) are needed to explain both the HS/LS-ratios as well at the abundance patterns (see Figs.\ref{Fig:BaEu}--\ref{Fig:19065}). Stars stripped from Sgr and similar dwarf galaxies could indeed be building blocks of the MW halo and possibly offer an explanation for s-rich stars in the Galactic halo.

We also calculated ages for three of the stars and found realistic, consistent values around $9\pm2.5$\,Gyr for those three. This is in agreement with previous predictions of the first stripping from the main body taking place around 9\,Gyr ago \citep{Fellhauer2006,Belokurov2014,Koposov2015, Hyde2015,deBoer2015}, and our calculated ages are therefore to be taken as lower limits.

\acknowledgments
CJH thanks Augustinus FONDEN (16-2068) for financial support and Andreas Koch for numerous discussions. SV gratefully acknowledges the support provided by Fondecyt reg. n. 1170518. A special thanks to arXiv for slowly stripping this paper of its catchy title (original: "Ages and heavy element abundances from very metal-poor stars in the Sagittarius dwarf galaxy: An ongoing (slow) strip tease"), which ensured a very generic and delayed appearance of this already accepted publication on arXiv.
Finally, we thank the anonymous referee for her/his comments.

\bibliographystyle{apj}
\bibliography{Sgr_heavy_ms_REsubm}

\begin{thebibliography}{103}
\expandafter\ifx\csname natexlab\endcsname\relax\def\natexlab#1{#1}\fi

\bibitem[{{Alonso} {et~al.}(1999){Alonso}, {Arribas}, \&
  {Mart{\'{\i}}nez-Roger}}]{Alonso1999}
{Alonso}, A., {Arribas}, S., \& {Mart{\'{\i}}nez-Roger}, C. 1999, \aaps, 140,
  261

\bibitem[{{Anders} \& {Grevesse}(1989)}]{AndersGrev}
{Anders}, E. \& {Grevesse}, N. 1989, \gca, 53, 197

\bibitem[{{Aoki} {et~al.}(2007){Aoki}, {Beers}, {Christlieb}, {Norris}, {Ryan},
  \& {Tsangarides}}]{Aoki2007}
{Aoki}, W., {Beers}, T.~C., {Christlieb}, N., {Norris}, J.~E., {Ryan}, S.~G.,
  \& {Tsangarides}, S. 2007, \apj, 655, 492

\bibitem[{{Arlandini} {et~al.}(1999){Arlandini}, {K{\"a}ppeler}, {Wisshak},
  {Gallino}, {Lugaro}, {Busso}, \& {Straniero}}]{Arlandini1999}
{Arlandini}, C., {K{\"a}ppeler}, F., {Wisshak}, K., {Gallino}, R., {Lugaro},
  M., {Busso}, M., \& {Straniero}, O. 1999, \apj, 525, 886

\bibitem[{{Asplund} {et~al.}(2009){Asplund}, {Grevesse}, {Sauval}, \&
  {Scott}}]{Asplund2009}
{Asplund}, M., {Grevesse}, N., {Sauval}, A.~J., \& {Scott}, P. 2009, \araa, 47,
  481

\bibitem[{{Barklem} {et~al.}(2005){Barklem}, {Christlieb}, {Beers}, {Hill},
  {Bessell}, {Holmberg}, {Marsteller}, {Rossi}, {Zickgraf}, \&
  {Reimers}}]{Barklem2005}
{Barklem}, P.~S., {Christlieb}, N., {Beers}, T.~C., {Hill}, V., {Bessell},
  M.~S., {Holmberg}, J., {Marsteller}, B., {Rossi}, S., {Zickgraf}, F.-J., \&
  {Reimers}, D. 2005, \aap, 439, 129

\bibitem[{{Bellazzini} {et~al.}(2008){Bellazzini}, {Ibata}, {Chapman},
  {Mackey}, {Monaco}, {Irwin}, {Martin}, {Lewis}, \&
  {Dalessandro}}]{Bellazzini2008}
{Bellazzini}, M., {Ibata}, R.~A., {Chapman}, S.~C., {Mackey}, A.~D., {Monaco},
  L., {Irwin}, M.~J., {Martin}, N.~F., {Lewis}, G.~F., \& {Dalessandro}, E.
  2008, \aj, 136, 1147

\bibitem[{{Belokurov} {et~al.}(2014){Belokurov}, {Koposov}, {Evans},
  {Pe{\~n}arrubia}, {Irwin}, {Smith}, {Lewis}, {Gieles}, {Wilkinson},
  {Gilmore}, {Olszewski}, \& {Niederste-Ostholt}}]{Belokurov2014}
{Belokurov}, V., {Koposov}, S.~E., {Evans}, N.~W., {Pe{\~n}arrubia}, J.,
  {Irwin}, M.~J., {Smith}, M.~C., {Lewis}, G.~F., {Gieles}, M., {Wilkinson},
  M.~I., {Gilmore}, G., {Olszewski}, E.~W., \& {Niederste-Ostholt}, M. 2014,
  \mnras, 437, 116

\bibitem[{{Bergemann} {et~al.}(2012){Bergemann}, {Hansen}, {Bautista}, \&
  {Ruchti}}]{Bergemann2012}
{Bergemann}, M., {Hansen}, C.~J., {Bautista}, M., \& {Ruchti}, G. 2012, \aap,
  546, A90

\bibitem[{{Bergemann} {et~al.}(2010){Bergemann}, {Pickering}, \&
  {Gehren}}]{Bergemann2010}
{Bergemann}, M., {Pickering}, J.~C., \& {Gehren}, T. 2010, \mnras, 401, 1334

\bibitem[{{Bisterzo} {et~al.}(2014){Bisterzo}, {Travaglio}, {Gallino},
  {Wiescher}, \& {K{\"a}ppeler}}]{Bisterzo2014}
{Bisterzo}, S., {Travaglio}, C., {Gallino}, R., {Wiescher}, M., \&
  {K{\"a}ppeler}, F. 2014, \apj, 787, 10

\bibitem[{{Bonifacio} {et~al.}(2000){Bonifacio}, {Hill}, {Molaro}, {Pasquini},
  {Di Marcantonio}, \& {Santin}}]{Bonifacio2000}
{Bonifacio}, P., {Hill}, V., {Molaro}, P., {Pasquini}, L., {Di Marcantonio},
  P., \& {Santin}, P. 2000, \aap, 359, 663

\bibitem[{{Bonifacio} {et~al.}(2009){Bonifacio}, {Spite}, {Cayrel}, {Hill},
  {Spite}, {Fran{\c c}ois}, {Plez}, {Ludwig}, {Caffau}, {Molaro}, {Depagne},
  {Andersen}, {Barbuy}, {Beers}, {Nordstr{\"o}m}, \& {Primas}}]{Bonifacio2009}
{Bonifacio}, P., {Spite}, M., {Cayrel}, R., {Hill}, V., {Spite}, F., {Fran{\c
  c}ois}, P., {Plez}, B., {Ludwig}, H.-G., {Caffau}, E., {Molaro}, P.,
  {Depagne}, E., {Andersen}, J., {Barbuy}, B., {Beers}, T.~C., {Nordstr{\"o}m},
  B., \& {Primas}, F. 2009, \aap, 501, 519

\bibitem[{{Bonifacio} {et~al.}(2006){Bonifacio}, {Zaggia}, {Sbordone},
  {Santin}, {Monaco}, {Monai}, {Molaro}, {Marconi}, {Girardi}, {Ferraro}, {di
  Marcantonio}, {Caffau}, \& {Bellazzini}}]{Bonifacio2006}
{Bonifacio}, P., {Zaggia}, S., {Sbordone}, L., {Santin}, P., {Monaco}, L.,
  {Monai}, S., {Molaro}, P., {Marconi}, G., {Girardi}, L., {Ferraro}, F., {di
  Marcantonio}, P., {Caffau}, E., \& {Bellazzini}, M. {Abundances in
  Sagittarius Stars}, ed. S.~{Randich} \& L.~{Pasquini}, 232

\bibitem[{{Bullock} \& {Johnston}(2005)}]{Bullock2005}
{Bullock}, J.~S. \& {Johnston}, K.~V. 2005, \apj, 635, 931

\bibitem[{{Butcher}(1987)}]{Butcher1987}
{Butcher}, H.~R. 1987, \nat, 328, 127

\bibitem[{{Caffau} {et~al.}(2008){Caffau}, {Sbordone}, {Ludwig}, {Bonifacio},
  {Steffen}, \& {Behara}}]{Caffau2008Th}
{Caffau}, E., {Sbordone}, L., {Ludwig}, H.-G., {Bonifacio}, P., {Steffen}, M.,
  \& {Behara}, N.~T. 2008, \aap, 483, 591

\bibitem[{{Carretta} {et~al.}(2010){Carretta}, {Bragaglia}, {Gratton},
  {Lucatello}, {Bellazzini}, {Catanzaro}, {Leone}, {Momany}, {Piotto}, \&
  {D'Orazi}}]{Carretta2010}
{Carretta}, E., {Bragaglia}, A., {Gratton}, R.~G., {Lucatello}, S.,
  {Bellazzini}, M., {Catanzaro}, G., {Leone}, F., {Momany}, Y., {Piotto}, G.,
  \& {D'Orazi}, V. 2010, \aap, 520, A95

\bibitem[{{Castelli} \& {Kurucz}(2003)}]{Castelli2003}
{Castelli}, F. \& {Kurucz}, R.~L. 2003, in IAU Symposium, Vol. 210, Modelling
  of Stellar Atmospheres, ed. N.~{Piskunov}, W.~W. {Weiss}, \& D.~F. {Gray},
  20P

\bibitem[{{Cayrel} {et~al.}(2004){Cayrel}, {Depagne}, {Spite}, {Hill}, {Spite},
  {Fran{\c c}ois}, {Plez}, {Beers}, {Primas}, {Andersen}, {Barbuy},
  {Bonifacio}, {Molaro}, \& {Nordstr{\"o}m}}]{Cayrel2004}
{Cayrel}, R., {Depagne}, E., {Spite}, M., {Hill}, V., {Spite}, F., {Fran{\c
  c}ois}, P., {Plez}, B., {Beers}, T., {Primas}, F., {Andersen}, J., {Barbuy},
  B., {Bonifacio}, P., {Molaro}, P., \& {Nordstr{\"o}m}, B. 2004, \aap, 416,
  1117

\bibitem[{{Cayrel} {et~al.}(2001){Cayrel}, {Hill}, {Beers}, {Barbuy}, {Spite},
  {Spite}, {Plez}, {Andersen}, {Bonifacio}, {Fran{\c c}ois}, {Molaro},
  {Nordstr{\"o}m}, \& {Primas}}]{Cayrel2001}
{Cayrel}, R., {Hill}, V., {Beers}, T.~C., {Barbuy}, B., {Spite}, M., {Spite},
  F., {Plez}, B., {Andersen}, J., {Bonifacio}, P., {Fran{\c c}ois}, P.,
  {Molaro}, P., {Nordstr{\"o}m}, B., \& {Primas}, F. 2001, \nat, 409, 691

\bibitem[{{Cohen}(2004)}]{Cohen2004}
{Cohen}, J.~G. 2004, \aj, 127, 1545

\bibitem[{{Cowan} {et~al.}(1999){Cowan}, {Pfeiffer}, {Kratz}, {Thielemann},
  {Sneden}, {Burles}, {Tytler}, \& {Beers}}]{Cowan1999}
{Cowan}, J.~J., {Pfeiffer}, B., {Kratz}, K.-L., {Thielemann}, F.-K., {Sneden},
  C., {Burles}, S., {Tytler}, D., \& {Beers}, T.~C. 1999, \apj, 521, 194

\bibitem[{{Cowan} {et~al.}(2002){Cowan}, {Sneden}, {Burles}, {Ivans}, {Beers},
  {Truran}, {Lawler}, {Primas}, {Fuller}, {Pfeiffer}, \& {Kratz}}]{Cowan2002}
{Cowan}, J.~J., {Sneden}, C., {Burles}, S., {Ivans}, I.~I., {Beers}, T.~C.,
  {Truran}, J.~W., {Lawler}, J.~E., {Primas}, F., {Fuller}, G.~M., {Pfeiffer},
  B., \& {Kratz}, K.-L. 2002, \apj, 572, 861

\bibitem[{{Cowan} {et~al.}(1991){Cowan}, {Thielemann}, \& {Truran}}]{Cowan1991}
{Cowan}, J.~J., {Thielemann}, F.-K., \& {Truran}, J.~W. 1991, \araa, 29, 447

\bibitem[{{Cristallo} {et~al.}(2015){Cristallo}, {Abia}, {Straniero}, \&
  {Piersanti}}]{Cristallo2015}
{Cristallo}, S., {Abia}, C., {Straniero}, O., \& {Piersanti}, L. 2015, \apj,
  801, 53

\bibitem[{{Cristallo} {et~al.}(2011){Cristallo}, {Piersanti}, {Straniero},
  {Gallino}, {Dom{\'{\i}}nguez}, {Abia}, {Di Rico}, {Quintini}, \&
  {Bisterzo}}]{Cristallo2011}
{Cristallo}, S., {Piersanti}, L., {Straniero}, O., {Gallino}, R.,
  {Dom{\'{\i}}nguez}, I., {Abia}, C., {Di Rico}, G., {Quintini}, M., \&
  {Bisterzo}, S. 2011, \apjs, 197, 17

\bibitem[{{de Boer} {et~al.}(2015){de Boer}, {Belokurov}, \&
  {Koposov}}]{deBoer2015}
{de Boer}, T.~J.~L., {Belokurov}, V., \& {Koposov}, S. 2015, \mnras, 451, 3489

\bibitem[{{Dekel} \& {Silk}(1986)}]{Dekel1986}
{Dekel}, A. \& {Silk}, J. 1986, \apj, 303, 39

\bibitem[{{Dekker} {et~al.}(2000){Dekker}, {D'Odorico}, {Kaufer}, {Delabre}, \&
  {Kotzlowski}}]{Dekker2000}
{Dekker}, H., {D'Odorico}, S., {Kaufer}, A., {Delabre}, B., \& {Kotzlowski}, H.
  2000, in \procspie, Vol. 4008, Optical and IR Telescope Instrumentation and
  Detectors, ed. M.~{Iye} \& A.~F. {Moorwood}, 534--545

\bibitem[{{Fellhauer} {et~al.}(2006){Fellhauer}, {Belokurov}, {Evans},
  {Wilkinson}, {Zucker}, {Gilmore}, {Irwin}, {Bramich}, {Vidrih}, {Wyse},
  {Beers}, \& {Brinkmann}}]{Fellhauer2006}
{Fellhauer}, M., {Belokurov}, V., {Evans}, N.~W., {Wilkinson}, M.~I., {Zucker},
  D.~B., {Gilmore}, G., {Irwin}, M.~J., {Bramich}, D.~M., {Vidrih}, S., {Wyse},
  R.~F.~G., {Beers}, T.~C., \& {Brinkmann}, J. 2006, \apj, 651, 167

\bibitem[{{Fran{\c c}ois} {et~al.}(2007){Fran{\c c}ois}, {Depagne}, {Hill},
  {Spite}, {Spite}, {Plez}, {Beers}, {Andersen}, {James}, {Barbuy}, {Cayrel},
  {Bonifacio}, {Molaro}, {Nordstr{\"o}m}, \& {Primas}}]{Francois2007}
{Fran{\c c}ois}, P., {Depagne}, E., {Hill}, V., {Spite}, M., {Spite}, F.,
  {Plez}, B., {Beers}, T.~C., {Andersen}, J., {James}, G., {Barbuy}, B.,
  {Cayrel}, R., {Bonifacio}, P., {Molaro}, P., {Nordstr{\"o}m}, B., \&
  {Primas}, F. 2007, \aap, 476, 935

\bibitem[{{Frischknecht} {et~al.}(2012){Frischknecht}, {Hirschi}, \&
  {Thielemann}}]{Frischknecht2012}
{Frischknecht}, U., {Hirschi}, R., \& {Thielemann}, F.-K. 2012, \aap, 538, L2

\bibitem[{{Fulbright}(2000)}]{Fulbright2000}
{Fulbright}, J.~P. 2000, \aj, 120, 1841

\bibitem[{{Gallagher} {et~al.}(2012){Gallagher}, {Ryan}, {Hosford},
  {Garc{\'{\i}}a P{\'e}rez}, {Aoki}, \& {Honda}}]{Gallagher2012}
{Gallagher}, A.~J., {Ryan}, S.~G., {Hosford}, A., {Garc{\'{\i}}a P{\'e}rez},
  A.~E., {Aoki}, W., \& {Honda}, S. 2012, \aap, 538, A118

\bibitem[{{Giuffrida} {et~al.}(2010){Giuffrida}, {Sbordone}, {Zaggia},
  {Marconi}, {Bonifacio}, {Izzo}, {Szeifert}, \& {Buonanno}}]{Giuffrida2010}
{Giuffrida}, G., {Sbordone}, L., {Zaggia}, S., {Marconi}, G., {Bonifacio}, P.,
  {Izzo}, C., {Szeifert}, T., \& {Buonanno}, R. 2010, \aap, 513, A62

\bibitem[{{Hansen} {et~al.}(2014){Hansen}, {Andersen}, \&
  {Christlieb}}]{Hansen2014a}
{Hansen}, C.~J., {Andersen}, A.~C., \& {Christlieb}, N. 2014, \aap, 568, A47

\bibitem[{{Hansen} {et~al.}(2013){Hansen}, {Bergemann}, {Cescutti}, {Fran{\c
  c}ois}, {Arcones}, {Karakas}, {Lind}, \& {Chiappini}}]{Hansen2013}
{Hansen}, C.~J., {Bergemann}, M., {Cescutti}, G., {Fran{\c c}ois}, P.,
  {Arcones}, A., {Karakas}, A.~I., {Lind}, K., \& {Chiappini}, C. 2013, \aap,
  551, A57

\bibitem[{{Hansen} {et~al.}(2016){Hansen}, {Nordstr{\"o}m}, {Hansen},
  {Kennedy}, {Placco}, {Beers}, {Andersen}, {Cescutti}, \&
  {Chiappini}}]{Hansen2016}
{Hansen}, C.~J., {Nordstr{\"o}m}, B., {Hansen}, T.~T., {Kennedy}, C.~R.,
  {Placco}, V.~M., {Beers}, T.~C., {Andersen}, J., {Cescutti}, G., \&
  {Chiappini}, C. 2016, \aap, 588, A37

\bibitem[{{Hansen} {et~al.}(2012){Hansen}, {Primas}, {Hartman}, {Kratz},
  {Wanajo}, {Leibundgut}, {Farouqi}, {Hallmann}, {Christlieb}, \&
  {Nilsson}}]{Hansen2012}
{Hansen}, C.~J., {Primas}, F., {Hartman}, H., {Kratz}, K.-L., {Wanajo}, S.,
  {Leibundgut}, B., {Farouqi}, K., {Hallmann}, O., {Christlieb}, N., \&
  {Nilsson}, H. 2012, \aap, 545, A31

\bibitem[{{Haschke} {et~al.}(2012){Haschke}, {Grebel}, {Frebel}, {Duffau},
  {Hansen}, \& {Koch}}]{Haschke2012}
{Haschke}, R., {Grebel}, E.~K., {Frebel}, A., {Duffau}, S., {Hansen}, C.~J., \&
  {Koch}, A. 2012, \aj, 144, 88

\bibitem[{{Hasselquist} {et~al.}(2017){Hasselquist}, {Shetrone}, {Smith},
  {Holtzman}, {McWilliam}, {Fern{\'a}ndez-Trincado}, {Beers}, {Majewski},
  {Nidever}, {Tang}, {Tissera}, {Fern{\'a}ndez Alvar}, {Allende Prieto},
  {Almeida}, {Anguiano}, {Battaglia}, {Carigi}, {Delgado Inglada},
  {Frinchaboy}, {Garc{\'{\i}}a-Hern{\'a}ndez}, {Geisler}, {Minniti}, {Placco},
  {Schultheis}, {Sobeck}, \& {Villanova}}]{Hasselquist2017}
{Hasselquist}, S., {Shetrone}, M., {Smith}, V., {Holtzman}, J., {McWilliam},
  A., {Fern{\'a}ndez-Trincado}, J.~G., {Beers}, T.~C., {Majewski}, S.~R.,
  {Nidever}, D.~L., {Tang}, B., {Tissera}, P.~B., {Fern{\'a}ndez Alvar}, E.,
  {Allende Prieto}, C., {Almeida}, A., {Anguiano}, B., {Battaglia}, G.,
  {Carigi}, L., {Delgado Inglada}, G., {Frinchaboy}, P.,
  {Garc{\'{\i}}a-Hern{\'a}ndez}, D.~A., {Geisler}, D., {Minniti}, D., {Placco},
  V.~M., {Schultheis}, M., {Sobeck}, J., \& {Villanova}, S. 2017, ArXiv
  e-prints

\bibitem[{{Heger} {et~al.}(2005){Heger}, {Woosley}, \& {Spruit}}]{Heger2005}
{Heger}, A., {Woosley}, S.~E., \& {Spruit}, H.~C. 2005, \apj, 626, 350

\bibitem[{{Hill} {et~al.}(1994){Hill}, {Andrievsky}, \& {Spite}}]{Hill1994}
{Hill}, V., {Andrievsky}, S., \& {Spite}, M. 1994, VizieR Online Data Catalog,
  329

\bibitem[{{Hill} {et~al.}(2002){Hill}, {Plez}, {Cayrel}, {Beers},
  {Nordstr{\"o}m}, {Andersen}, {Spite}, {Spite}, {Barbuy}, {Bonifacio},
  {Depagne}, {Fran{\c c}ois}, \& {Primas}}]{Hill2002}
{Hill}, V., {Plez}, B., {Cayrel}, R., {Beers}, T.~C., {Nordstr{\"o}m}, B.,
  {Andersen}, J., {Spite}, M., {Spite}, F., {Barbuy}, B., {Bonifacio}, P.,
  {Depagne}, E., {Fran{\c c}ois}, P., \& {Primas}, F. 2002, \aap, 387, 560

\bibitem[{{Hyde} {et~al.}(2015){Hyde}, {Keller}, {Zucker}, {Ibata}, {Siebert},
  {Lewis}, {Penarrubia}, {Irwin}, {Gilmore}, {Lane}, {Koch}, {Conn},
  {Diakogiannis}, \& {Martell}}]{Hyde2015}
{Hyde}, E.~A., {Keller}, S., {Zucker}, D.~B., {Ibata}, R., {Siebert}, A.,
  {Lewis}, G.~F., {Penarrubia}, J., {Irwin}, M., {Gilmore}, G., {Lane}, R.~R.,
  {Koch}, A., {Conn}, A.~R., {Diakogiannis}, F.~I., \& {Martell}, S. 2015,
  \apj, 805, 189

\bibitem[{{Hyde} {et~al.}(2012){Hyde}, {Zucker}, {Irwin}, {Pe{\~n}arrubia}, \&
  {Koch}}]{Hyde2012}
{Hyde}, E.~A., {Zucker}, D.~B., {Irwin}, M., {Pe{\~n}arrubia}, J., \& {Koch},
  A. 2012, in Astronomical Society of the Pacific Conference Series, Vol. 458,
  Galactic Archaeology: Near-Field Cosmology and the Formation of the Milky
  Way, ed. W.~{Aoki}, M.~{Ishigaki}, T.~{Suda}, T.~{Tsujimoto}, \&
  N.~{Arimoto}, 325

\bibitem[{{Ibata} {et~al.}(1994){Ibata}, {Gilmore}, \& {Irwin}}]{Ibata1994}
{Ibata}, R.~A., {Gilmore}, G., \& {Irwin}, M.~J. 1994, \nat, 370, 194

\bibitem[{{Ishigaki} {et~al.}(2014){Ishigaki}, {Aoki}, {Arimoto}, \&
  {Okamoto}}]{Ishigaki2014}
{Ishigaki}, M.~N., {Aoki}, W., {Arimoto}, N., \& {Okamoto}, S. 2014, \aap, 562,
  A146

\bibitem[{{Janka}(2017)}]{Janka2017}
{Janka}, H.-T. 2017, ArXiv e-prints

\bibitem[{{Johnson} {et~al.}(2014){Johnson}, {Rich}, {Kobayashi}, {Kunder}, \&
  {Koch}}]{Johnson2014}
{Johnson}, C.~I., {Rich}, R.~M., {Kobayashi}, C., {Kunder}, A., \& {Koch}, A.
  2014, \aj, 148, 67

\bibitem[{{Johnson} {et~al.}(2006){Johnson}, {Ivans}, \&
  {Stetson}}]{Johnson2006}
{Johnson}, J.~A., {Ivans}, I.~I., \& {Stetson}, P.~B. 2006, \apj, 640, 801

\bibitem[{{Kobayashi} {et~al.}(2006){Kobayashi}, {Umeda}, {Nomoto}, {Tominaga},
  \& {Ohkubo}}]{Kobayashi2006}
{Kobayashi}, C., {Umeda}, H., {Nomoto}, K., {Tominaga}, N., \& {Ohkubo}, T.
  2006, \apj, 653, 1145

\bibitem[{{Koch} \& {Edvardsson}(2002)}]{Koch2002}
{Koch}, A. \& {Edvardsson}, B. 2002, \aap, 381, 500

\bibitem[{{Koch} {et~al.}(2013){Koch}, {Feltzing}, {Ad{\'e}n}, \&
  {Matteucci}}]{Koch2013}
{Koch}, A., {Feltzing}, S., {Ad{\'e}n}, D., \& {Matteucci}, F. 2013, \aap, 554,
  A5

\bibitem[{{Koch} {et~al.}(2008){Koch}, {McWilliam}, {Grebel}, {Zucker}, \&
  {Belokurov}}]{Koch2008}
{Koch}, A., {McWilliam}, A., {Grebel}, E.~K., {Zucker}, D.~B., \& {Belokurov},
  V. 2008, \apjl, 688, L13

\bibitem[{{Koposov} {et~al.}(2015){Koposov}, {Belokurov}, {Zucker}, {Lewis},
  {Ibata}, {Olszewski}, {L{\'o}pez-S{\'a}nchez}, \& {Hyde}}]{Koposov2015}
{Koposov}, S.~E., {Belokurov}, V., {Zucker}, D.~B., {Lewis}, G.~F., {Ibata},
  R.~A., {Olszewski}, E.~W., {L{\'o}pez-S{\'a}nchez}, {\'A}.~R., \& {Hyde},
  E.~A. 2015, \mnras, 446, 3110

\bibitem[{{Korotin} {et~al.}(2015){Korotin}, {Andrievsky}, {Hansen}, {Caffau},
  {Bonifacio}, {Spite}, {Spite}, \& {Fran{\c c}ois}}]{Korotin2015}
{Korotin}, S.~A., {Andrievsky}, S.~M., {Hansen}, C.~J., {Caffau}, E.,
  {Bonifacio}, P., {Spite}, M., {Spite}, F., \& {Fran{\c c}ois}, P. 2015, \aap,
  581, A70

\bibitem[{{Kratz} {et~al.}(2007){Kratz}, {Farouqi}, {Pfeiffer}, {Truran},
  {Sneden}, \& {Cowan}}]{Kratz2007}
{Kratz}, K.-L., {Farouqi}, K., {Pfeiffer}, B., {Truran}, J.~W., {Sneden}, C.,
  \& {Cowan}, J.~J. 2007, \apj, 662, 39

\bibitem[{{Kurucz}(1970)}]{Kurucz1970}
{Kurucz}, R.~L. 1970, SAO Special Report, 309

\bibitem[{{Law} \& {Majewski}(2010)}]{Law2010}
{Law}, D.~R. \& {Majewski}, S.~R. 2010, \apj, 718, 1128

\bibitem[{{Lawler} {et~al.}(2001){Lawler}, {Bonvallet}, \&
  {Sneden}}]{Lawler2001La}
{Lawler}, J.~E., {Bonvallet}, G., \& {Sneden}, C. 2001, \apj, 556, 452

\bibitem[{{Lawler} {et~al.}(2015){Lawler}, {Sneden}, \& {Cowan}}]{Lawler2015}
{Lawler}, J.~E., {Sneden}, C., \& {Cowan}, J.~J. 2015, \apjs, 220, 13

\bibitem[{{Lawler} {et~al.}(2009){Lawler}, {Sneden}, {Cowan}, {Ivans}, \& {Den
  Hartog}}]{Lawler2009}
{Lawler}, J.~E., {Sneden}, C., {Cowan}, J.~J., {Ivans}, I.~I., \& {Den Hartog},
  E.~A. 2009, \apjs, 182, 51

\bibitem[{{Letarte} {et~al.}(2010){Letarte}, {Hill}, {Tolstoy}, {Jablonka},
  {Shetrone}, {Venn}, {Spite}, {Irwin}, {Battaglia}, {Helmi}, {Primas},
  {Fran{\c c}ois}, {Kaufer}, {Szeifert}, {Arimoto}, \&
  {Sadakane}}]{Letarte2010}
{Letarte}, B., {Hill}, V., {Tolstoy}, E., {Jablonka}, P., {Shetrone}, M.,
  {Venn}, K.~A., {Spite}, M., {Irwin}, M.~J., {Battaglia}, G., {Helmi}, A.,
  {Primas}, F., {Fran{\c c}ois}, P., {Kaufer}, A., {Szeifert}, T., {Arimoto},
  N., \& {Sadakane}, K. 2010, \aap, 523, A17

\bibitem[{{Ludwig} {et~al.}(2010){Ludwig}, {Caffau}, {Steffen}, {Bonifacio}, \&
  {Sbordone}}]{Ludwig2010}
{Ludwig}, H.-G., {Caffau}, E., {Steffen}, M., {Bonifacio}, P., \& {Sbordone},
  L. 2010, \aap, 509, A84

\bibitem[{{Majewski} {et~al.}(2003){Majewski}, {Skrutskie}, {Weinberg}, \&
  {Ostheimer}}]{Majewski2003}
{Majewski}, S.~R., {Skrutskie}, M.~F., {Weinberg}, M.~D., \& {Ostheimer}, J.~C.
  2003, \apj, 599, 1082

\bibitem[{{Marino} {et~al.}(2008){Marino}, {Villanova}, {Piotto}, {Milone},
  {Momany}, {Bedin}, \& {Medling}}]{Marino2008}
{Marino}, A.~F., {Villanova}, S., {Piotto}, G., {Milone}, A.~P., {Momany}, Y.,
  {Bedin}, L.~R., \& {Medling}, A.~M. 2008, \aap, 490, 625

\bibitem[{{Mashonkina} {et~al.}(2012){Mashonkina}, {Ryabtsev}, \&
  {Frebel}}]{Mashonkina2012}
{Mashonkina}, L., {Ryabtsev}, A., \& {Frebel}, A. 2012, \aap, 540, A98

\bibitem[{{Matteucci} \& {Brocato}(1990)}]{Matteucci1990}
{Matteucci}, F. \& {Brocato}, E. 1990, \apj, 365, 539

\bibitem[{{McConnachie}(2012)}]{McConnachie2012}
{McConnachie}, A.~W. 2012, \aj, 144, 4

\bibitem[{{McWilliam} {et~al.}(2003){McWilliam}, {Rich}, \&
  {Smecker-Hane}}]{McWilliam2003}
{McWilliam}, A., {Rich}, R.~M., \& {Smecker-Hane}, T.~A. 2003, \apjl, 592, L21

\bibitem[{{McWilliam} {et~al.}(2013){McWilliam}, {Wallerstein}, \&
  {Mottini}}]{McWilliam2013}
{McWilliam}, A., {Wallerstein}, G., \& {Mottini}, M. 2013, \apj, 778, 149

\bibitem[{{Monaco} {et~al.}(2005){Monaco}, {Bellazzini}, {Bonifacio},
  {Ferraro}, {Marconi}, {Pancino}, {Sbordone}, \& {Zaggia}}]{Monaco2005}
{Monaco}, L., {Bellazzini}, M., {Bonifacio}, P., {Ferraro}, F.~R., {Marconi},
  G., {Pancino}, E., {Sbordone}, L., \& {Zaggia}, S. 2005, \aap, 441, 141

\bibitem[{{Monaco} {et~al.}(2004){Monaco}, {Bellazzini}, {Ferraro}, \&
  {Pancino}}]{Monaco2004}
{Monaco}, L., {Bellazzini}, M., {Ferraro}, F.~R., \& {Pancino}, E. 2004,
  \mnras, 353, 874

\bibitem[{{Mucciarelli} {et~al.}(2017){Mucciarelli}, {Bellazzini}, {Ibata},
  {Romano}, {Chapman}, \& {Monaco}}]{Mucciarelli2017}
{Mucciarelli}, A., {Bellazzini}, M., {Ibata}, R., {Romano}, D., {Chapman},
  S.~C., \& {Monaco}, L. 2017, ArXiv e-prints

\bibitem[{{Mucciarelli} {et~al.}(2008){Mucciarelli}, {Carretta}, {Origlia}, \&
  {Ferraro}}]{Mucciarelli2008}
{Mucciarelli}, A., {Carretta}, E., {Origlia}, L., \& {Ferraro}, F.~R. 2008,
  \aj, 136, 375

\bibitem[{{Mucciarelli} {et~al.}(2009){Mucciarelli}, {Origlia}, {Ferraro}, \&
  {Pancino}}]{Mucciarelli2009}
{Mucciarelli}, A., {Origlia}, L., {Ferraro}, F.~R., \& {Pancino}, E. 2009,
  \apjl, 695, L134

\bibitem[{{Otsuki} {et~al.}(2003){Otsuki}, {Mathews}, \& {Kajino}}]{Otsuki2003}
{Otsuki}, K., {Mathews}, G.~J., \& {Kajino}, T. 2003, \nat, 8, 767

\bibitem[{{Placco} {et~al.}(2014){Placco}, {Frebel}, {Beers}, \&
  {Stancliffe}}]{Placco2014b}
{Placco}, V.~M., {Frebel}, A., {Beers}, T.~C., \& {Stancliffe}, R.~J. 2014,
  \apj, 797, 21

\bibitem[{{Pomp{\'e}ia} {et~al.}(2008){Pomp{\'e}ia}, {Hill}, {Spite}, {Cole},
  {Primas}, {Romaniello}, {Pasquini}, {Cioni}, \& {Smecker Hane}}]{Pompeia2008}
{Pomp{\'e}ia}, L., {Hill}, V., {Spite}, M., {Cole}, A., {Primas}, F.,
  {Romaniello}, M., {Pasquini}, L., {Cioni}, M.-R., \& {Smecker Hane}, T. 2008,
  \aap, 480, 379

\bibitem[{{Ram{\'{\i}}rez} \& {Mel{\'e}ndez}(2005)}]{Ramirez2005}
{Ram{\'{\i}}rez}, I. \& {Mel{\'e}ndez}, J. 2005, \apj, 626, 446

\bibitem[{{Reddy} {et~al.}(2006){Reddy}, {Lambert}, \& {Allende
  Prieto}}]{Reddy2006}
{Reddy}, B.~E., {Lambert}, D.~L., \& {Allende Prieto}, C. 2006, \mnras, 367,
  1329

\bibitem[{{Reddy} {et~al.}(2003){Reddy}, {Tomkin}, {Lambert}, \& {Allende
  Prieto}}]{Reddy2003}
{Reddy}, B.~E., {Tomkin}, J., {Lambert}, D.~L., \& {Allende Prieto}, C. 2003,
  \mnras, 340, 304

\bibitem[{{Roederer}(2013)}]{Roederer2013}
{Roederer}, I.~U. 2013, \aj, 145, 26

\bibitem[{{Sbordone} {et~al.}(2007){Sbordone}, {Bonifacio}, {Buonanno},
  {Marconi}, {Monaco}, \& {Zaggia}}]{Sbordone2007}
{Sbordone}, L., {Bonifacio}, P., {Buonanno}, R., {Marconi}, G., {Monaco}, L.,
  \& {Zaggia}, S. 2007, \aap, 465, 815

\bibitem[{{Schatz} {et~al.}(2002){Schatz}, {Toenjes}, {Pfeiffer}, {Beers},
  {Cowan}, {Hill}, \& {Kratz}}]{Schatz2002}
{Schatz}, H., {Toenjes}, R., {Pfeiffer}, B., {Beers}, T.~C., {Cowan}, J.~J.,
  {Hill}, V., \& {Kratz}, K.-L. 2002, \apj, 579, 626

\bibitem[{{Searle} \& {Zinn}(1978)}]{Searle1978}
{Searle}, L. \& {Zinn}, R. 1978, \apj, 225, 357

\bibitem[{{Shetrone} {et~al.}(2001){Shetrone}, {C{\^o}t{\'e}}, \&
  {Sargent}}]{Shetrone2001}
{Shetrone}, M.~D., {C{\^o}t{\'e}}, P., \& {Sargent}, W.~L.~W. 2001, \apj, 548,
  592

\bibitem[{{Siegel} {et~al.}(2007){Siegel}, {Dotter}, {Majewski}, {Sarajedini},
  {Chaboyer}, {Nidever}, {Anderson}, {Mar{\'{\i}}n-Franch}, {Rosenberg},
  {Bedin}, {Aparicio}, {King}, {Piotto}, \& {Reid}}]{Siegel2007}
{Siegel}, M.~H., {Dotter}, A., {Majewski}, S.~R., {Sarajedini}, A., {Chaboyer},
  B., {Nidever}, D.~L., {Anderson}, J., {Mar{\'{\i}}n-Franch}, A., {Rosenberg},
  A., {Bedin}, L.~R., {Aparicio}, A., {King}, I., {Piotto}, G., \& {Reid},
  I.~N. 2007, \apjl, 667, L57

\bibitem[{{Simmerer} {et~al.}(2004){Simmerer}, {Sneden}, {Cowan}, {Collier},
  {Woolf}, \& {Lawler}}]{Simmerer2004}
{Simmerer}, J., {Sneden}, C., {Cowan}, J.~J., {Collier}, J., {Woolf}, V.~M., \&
  {Lawler}, J.~E. 2004, \apj, 617, 1091

\bibitem[{{Sneden} {et~al.}(2016){Sneden}, {Cowan}, {Kobayashi}, {Pignatari},
  {Lawler}, {Den Hartog}, \& {Wood}}]{Sneden2016}
{Sneden}, C., {Cowan}, J.~J., {Kobayashi}, C., {Pignatari}, M., {Lawler},
  J.~E., {Den Hartog}, E.~A., \& {Wood}, M.~P. 2016, \apj, 817, 53

\bibitem[{{Sneden} {et~al.}(2003){Sneden}, {Cowan}, {Lawler}, {Ivans},
  {Burles}, {Beers}, {Primas}, {Hill}, {Truran}, {Fuller}, {Pfeiffer}, \&
  {Kratz}}]{Sneden2003}
{Sneden}, C., {Cowan}, J.~J., {Lawler}, J.~E., {Ivans}, I.~I., {Burles}, S.,
  {Beers}, T.~C., {Primas}, F., {Hill}, V., {Truran}, J.~W., {Fuller}, G.~M.,
  {Pfeiffer}, B., \& {Kratz}, K.-L. 2003, \apj, 591, 936

\bibitem[{{Sneden} {et~al.}(2014){Sneden}, {Lucatello}, {Ram}, {Brooke}, \&
  {Bernath}}]{Sneden2014}
{Sneden}, C., {Lucatello}, S., {Ram}, R.~S., {Brooke}, J.~S.~A., \& {Bernath},
  P. 2014, \apjs, 214, 26

\bibitem[{{Sneden} {et~al.}(1996){Sneden}, {McWilliam}, {Preston}, {Cowan},
  {Burris}, \& {Armosky}}]{Sneden1996}
{Sneden}, C., {McWilliam}, A., {Preston}, G.~W., {Cowan}, J.~J., {Burris},
  D.~L., \& {Armosky}, B.~J. 1996, \apj, 467, 819

\bibitem[{{Sneden}(1973)}]{Sneden1973}
{Sneden}, C.~A. 1973, PhD thesis, The University of Texas at Austin.

\bibitem[{{Tinsley}(1979)}]{Tinsley1979}
{Tinsley}, B.~M. 1979, \apj, 229, 1046

\bibitem[{{Trager} {et~al.}(1995){Trager}, {King}, \&
  {Djorgovski}}]{Trager1995}
{Trager}, S.~C., {King}, I.~R., \& {Djorgovski}, S. 1995, \aj, 109, 218

\bibitem[{{Truran} {et~al.}(2001){Truran}, {Burles}, {Cowan}, \&
  {Sneden}}]{Truran2001}
{Truran}, J.~W., {Burles}, S., {Cowan}, J.~J., \& {Sneden}, C. 2001, in
  Astronomical Society of the Pacific Conference Series, Vol. 245,
  Astrophysical Ages and Times Scales, ed. T.~{von Hippel}, C.~{Simpson}, \&
  N.~{Manset}, 226

\bibitem[{{Venn} {et~al.}(2004){Venn}, {Irwin}, {Shetrone}, {Tout}, {Hill}, \&
  {Tolstoy}}]{Venn2004}
{Venn}, K.~A., {Irwin}, M., {Shetrone}, M.~D., {Tout}, C.~A., {Hill}, V., \&
  {Tolstoy}, E. 2004, \aj, 128, 1177

\bibitem[{{Venn} {et~al.}(2012){Venn}, {Shetrone}, {Irwin}, {Hill}, {Jablonka},
  {Tolstoy}, {Lemasle}, {Divell}, {Starkenburg}, {Letarte}, {Baldner},
  {Battaglia}, {Helmi}, {Kaufer}, \& {Primas}}]{Venn2012}
{Venn}, K.~A., {Shetrone}, M.~D., {Irwin}, M.~J., {Hill}, V., {Jablonka}, P.,
  {Tolstoy}, E., {Lemasle}, B., {Divell}, M., {Starkenburg}, E., {Letarte}, B.,
  {Baldner}, C., {Battaglia}, G., {Helmi}, A., {Kaufer}, A., \& {Primas}, F.
  2012, \apj, 751, 102

\bibitem[{{Winteler} {et~al.}(2012){Winteler}, {K{\"a}ppeli}, {Perego},
  {Arcones}, {Vasset}, {Nishimura}, {Liebend{\"o}rfer}, \&
  {Thielemann}}]{Winteler2012}
{Winteler}, C., {K{\"a}ppeli}, R., {Perego}, A., {Arcones}, A., {Vasset}, N.,
  {Nishimura}, N., {Liebend{\"o}rfer}, M., \& {Thielemann}, F.-K. 2012, \apjl,
  750, L22

\bibitem[{{Zaggia} {et~al.}(2004){Zaggia}, {Bonifacio}, {Bellazzini}, {Caffau},
  {Di Marcantonio}, {Ferraro}, {Marconi}, {Monaco}, {Monai}, {Santin}, \&
  {Sbordone}}]{Zaggia2004}
{Zaggia}, S., {Bonifacio}, P., {Bellazzini}, M., {Caffau}, E., {Di
  Marcantonio}, P., {Ferraro}, F., {Marconi}, G., {Monaco}, L., {Monai}, S.,
  {Santin}, P., \& {Sbordone}, L. 2004, Memorie della Societa Astronomica
  Italiana Supplementi, 5, 291

\end{thebibliography}

\appendix
\section{Online Table}
\begin{table}[ht]
\caption{Line list for region around Th: Wavelength, atomic number and ionisation, excitation potential, log$gf$, dissociation energy.}
\label{tab:online}
\begin{center}
\begin{tabular}{lcccc}
\hline
\hline
Wavelength & Atom.Ion & Ex.pot. & log$gf$ & E$_{dis}$\\
$[$\AA\,$]$ &   & [eV] & [dex] & [eV] \\
\hline
4018.605 &     26.0 	  &    4.301  &     -3.877          &                     \\  
4018.737 &     69.1	  &    3.349  &     -3.250	    &   		  \\  
4018.738 &     23.0	  &    0.287  &     -6.805	    &   		  \\  
4018.789 &     24.0	  &    4.440  &     -2.822	    &   		  \\  
4018.812 &     24.0	  &    3.648  &     -2.629	    &   		  \\  
4018.820 &     58.1	  &    1.546  &     -0.960	    &   		  \\  
4018.820 &     60.1	  &    0.064  &     -0.850	    &   		  \\  
4018.887 &     26.0	  &    4.256  &     -2.604	    &   		  \\  
4018.900 &     58.1	  &    1.013  &     -1.220	    &   		  \\  
4018.927 &     58.1	  &    0.635  &     -1.680	    &   		  \\  
4018.929 &     23.0	  &    2.581  &     -0.651	    &   		  \\  
4018.963 &     59.1	  &    0.204  &     -1.030	    &   		  \\  
4018.986 &     92.1	  &    0.036  &     -1.391	    &   		  \\  
4018.999 &     25.0	  &    4.354  &     -1.497	    &   		  \\  
4019.003 &     26.0	  &    4.320  &     -1.793	    &   		  \\  
4019.036 &     23.1	  &    3.753  &     -2.704	    &   		  \\  
4019.042 &     25.0	  &    4.666  &     -0.561	    &   		  \\  
4019.042 &     26.0	  &    2.608  &     -2.780	    &   		  \\  
4019.057 &     58.1	  &    1.014  &     -0.530	    &   		  \\  
4019.058 &     28.0	  &    1.935  &     -3.174	    &   		  \\  
4019.067 &  	  28.0    &    1.934  &     	 -3.40      &       		  \\  	  
4019.090 &   106.00112    &    1.509  &     	-2.437      &       	 3.464    \\  	  
4019.103 &   106.00113    &    1.589  &     	-4.535      &       	 3.464    \\  	  
4019.114 &     27.0059    &    2.278  &     	-2.272      &       		  \\  	  
4019.114 &     27.0059    &    2.278  &     	-2.448      &       		  \\  	  
4019.119 &     27.0059    &    2.278  &     	-2.147      &       		  \\  	  
4019.119 &     27.0059    &    2.278  &     	-2.272      &       		  \\  	  
4019.126 &     27.0059    &    2.278  &     	-2.147      &       		  \\  	  
4019.126 &     27.0059    &    2.278  &     	-2.261      &       		  \\  	  
4019.126 &     27.0059    &    2.278  &     	-2.466      &       		  \\  	  
4019.126 &   106.00113    &    1.589  &     	-2.232      &       	 3.464    \\  	  
4019.129 &  	  26.0    &    4.317  &     	 -4.45      &       		  \\  	  
4019.129 &     90.1000    &    0.000  &     	-0.651      &       		  \\  	  
4019.136 &     27.0059    &    2.278  &     	-1.850      &       		  \\  	  
4019.136 &     27.0059    &    2.278  &     	-2.261      &       		  \\  	  
4019.138 &  	  23.0    &    1.802  &     	 -2.15      &       		  \\  	  
4019.143 &  	  42.0    &    3.396  &     	 -1.39      &       		  \\  	  
4019.144 &   106.00113    &    0.462  &     	-1.337      &       	 3.464    \\  	  
4019.213 &   106.00113    &    0.914  &     	-3.793      &       	 3.464    \\  	  
4019.228 &  	  74.0    &    0.412  &     	 -2.20      &       		  \\  	  
4019.229 &   106.00112    &    1.490  &     	-4.458      &       	 3.464    \\  	  
4019.245 &   606.01212    &    0.252  &     	-9.130      &       	  6.24    \\  	  
4019.255 &     27.0059    &    0.581  &     	-4.436      &       		  \\  	  
4019.261 &     27.0059    &    0.581  &     	-4.436      &       		  \\  	  
4019.261 &     27.0059    &    0.581  &     	-4.612      &       		  \\  	  
4019.264 &     27.0059    &    0.629  &     	-4.336      &       		  \\  	  
4019.270 &     27.0059    &    0.581  &     	-4.272      &       		  \\  	  
4019.270 &     27.0059    &    0.581  &     	-4.737      &       		  \\  	  
4019.270 &     27.0059    &    0.581  &     	-4.862      &       		  \\  	  
4019.283 &     27.0059    &    0.581  &     	-4.264      &       		  \\  	  
4019.283 &     27.0059    &    0.581  &     	-4.298      &       		  \\  	  
4019.283 &     27.0059    &    0.581  &     	-5.290      &       		  \\  	  
4019.288 &     27.0059    &    0.629  &     	-4.552      &       		  \\  	  
4019.289 &     24.1000    &    5.326  &     	-5.604      &       		  \\  	  
4019.289 &     27.0059    &    0.629  &     	-4.962      &       		  \\  	  
4019.298 &     27.0059    &    0.581  &     	-4.015      &       		  \\  	  
4019.298 &     27.0059    &    0.581  &     	-4.425      &       		  \\  	  
4019.308 &     27.0059    &    0.629  &     	-4.801      &       		  \\  	  
4019.308 &     27.0059    &    0.629  &     	-4.835      &       		  \\  	  
4019.309 &     27.0059    &    0.629  &     	-5.827      &       		  \\  	  
4019.316 &     27.0059    &    0.581  &     	-3.799      &       		  \\  	  
4019.324 &     27.0059    &    0.629  &     	-4.809      &       		  \\  	  
4019.324 &     27.0059    &    0.629  &     	-5.274      &       		  \\  	  
4019.324 &     27.0059    &    0.629  &     	-5.399      &       		  \\  	  
4019.336 &     27.0059    &    0.629  &     	-4.973      &       		  \\  	  
4019.336 &     27.0059    &    0.629  &     	-5.149      &       		  \\  	  
4019.344 &     27.0059    &    0.629  &     	-4.973      &       		  \\  	  
4019.357 &   106.00113    &    1.501  &     	-4.746      &       	 3.464    \\  	  
\hline
\end{tabular}
\end{center}
\label{default}
\end{table}%

\end{document}